\documentclass[
aps,
prx,
reprint,
groupedaddress
]{revtex4-2}

\usepackage{amsmath}
\usepackage{amsfonts}
\usepackage{graphicx}
\usepackage{amssymb}
\usepackage{textcomp}
\usepackage{bm}
\usepackage{color}
\usepackage{dsfont}

\usepackage{pgfplots}

\usepackage{siunitx}

\usepackage{mathtools}

\usetikzlibrary{arrows,shapes,backgrounds,calc,
    positioning,
    intersections}

\definecolor{Blue}{cmyk}{1,0.6,0,0.05}
\definecolor{Red}{cmyk}{0.04,0.87,0.89,0}
\definecolor{Green}{cmyk}{1,0.,1,0}
\definecolor{Yellow}{cmyk}{0,0.2,1,0}
\definecolor{Orange}{cmyk}{0.0,0.69,1,0}
\definecolor{SoftCyan}{cmyk}{0.48,0.,0.07,0.11}

\definecolor{LaserBlue}{rgb}{0.4,0,1}
\definecolor{LaserMOT}{rgb}{1,0.372,0}
\definecolor{LaserJseven}{rgb}{1,0,0}
\definecolor{LaserJeight}{rgb}{0.35,0,0}

\pgfplotsset{colormap={CM}{color=(white) color=(Blue!50!white) color=(Blue)  color=(Blue!75!black) color=(Blue!50!black)  color=(Blue!25!black) color=(black)}}

\pgfplotsset{colormap={RedToBlue}{color=(Red)  color=(white!50!black) color=(Blue)}}

\pgfplotsset{colormap={CM2}{color=(white) color=(Blue!33!white) color=(Blue!67!white) color=(Blue)  color=(Blue!75!black) color=(Blue!50!black)  color=(Blue!25!black) color=(black)}}
\pgfplotsset{colormap={CM3}{color=(white)  color=(Blue)  color=(black)}}
\pgfplotsset{colormap={CM4}{color=(white)  color=(Red)  color=(black)}}

\DeclareMathOperator{\Tr}{Tr}

\DeclareMathOperator{\sgn}{sgn}

\DeclareMathOperator{\acos}{acos}

\newcommand{\fig}[1]{Fig.\,\ref{#1}}
\newcommand{\refe}[1]{Ref.\,\cite{#1}}

\newcommand{\eq}[1]{Eq.\,\ref{#1}}

\newcommand{\bitem}{\begin{itemize}}
\newcommand{\eitem}{\end{itemize}}

\newcommand{\bti}{\begin{tikzpicture}}
\newcommand{\eti}{\end{tikzpicture}}

\newcommand{\ket}[1]{\left| #1 \right>} 
\newcommand{\bra}[1]{\left< #1 \right|} 

\newcommand{\I}{\mathrm{i}}

\newcommand{\E}{\mathrm{e}}

\newcommand{\dbold}{\mathbf{d}}
\newcommand{\ebold}{\mathbf{e}}

\newcommand{\nbold}{\mathbf{n}}

\newcommand{\bas}{\begin{align}}
\newcommand{\eas}{\end{align}}

\newcommand{\bc}{\begin{center}}
\newcommand{\ec}{\end{center}}

\DeclareSIUnit\gauss{G}

\newcommand{\Pabs}[1]{P_{\text{abs}}}
\newcommand{\rhopair}{\rho_{\text{pair}}}
\newcommand{\Qpair}{Q_{\text{pair}}}

\newcommand{\Smin}{S_{\infty}}
\newcommand{\lmax}{\lambda_{\text{max}}}

\begin{document}

 \title{
Partitioning dysprosium's electronic spin to reveal entanglement in non-classical states
 }

\author{Tanish Satoor}
\thanks{These two authors contributed equally.}
\author{Aur\'elien Fabre}
\thanks{These two authors contributed equally.}
\author{Jean-Baptiste Bouhiron}
\author{Alexandre Evrard}
\author{Raphael Lopes}
\author{Sylvain Nascimbene}
\email{sylvain.nascimbene@lkb.ens.fr}
 \affiliation{Laboratoire Kastler Brossel,  Coll\`ege de France, CNRS, ENS-PSL University, Sorbonne Universit\'e, 11 Place Marcelin Berthelot, 75005 Paris, France}
 \date{\today}

  \begin{abstract}
Quantum spins of mesoscopic size are a well-studied playground for engineering non-classical states. If the spin represents the collective state of an ensemble of qubits, its non-classical behavior is linked to entanglement between the qubits. In this work, we report on an experimental study of entanglement in  dysprosium's electronic spin. Its ground state, of angular momentum $J=8$, can formally be viewed as a set of $2J$ qubits symmetric upon exchange. To access entanglement properties, we partition the spin by optically coupling it to an excited state $J'=J-1$, which removes a pair of qubits in a state defined by the light polarization. Starting with the well-known W and squeezed states, we extract the concurrence of qubit pairs, which quantifies their non-classical character. We also directly demonstrate entanglement between the 14- and 2-qubit subsystems via an increase in entropy upon partition. In a complementary set of experiments, we probe decoherence of a state prepared in the excited level $J'=J+1$ and interpret spontaneous emission as a loss of a qubit pair in a random state. This allows us to contrast the robustness of pairwise entanglement of the W state with the fragility of the coherence involved in a Schrödinger cat state. Our findings open up the possibility to engineer novel types of entangled atomic ensembles, in which entanglement occurs within each atom's electronic spin as well as between different atoms. 
 \end{abstract}
 
 \maketitle
 
 \section{Introduction}\label{sec_intro}

Entanglement is a hallmark of non-classical behavior in compound quantum systems. Minimal entangled systems of qubit pairs, as realized with correlated photon pairs, play a central role in testing the foundations of quantum mechanics \cite{freedman_experimental_1972,aspect_experimental_1982}. Entanglement can also be engineered in many-particle systems \cite{amico_entanglement_2008}, such as an ensemble of interacting atoms  \cite{pezze_quantum_2018}. In this case, the atoms are not individually addressable, and quantum correlations are indirectly revealed by measuring global properties, such as a squeezed spin projection quadrature \cite{sorensen_many-particle_2001,sorensen_entanglement_2001,esteve_squeezing_2008,riedel_atom-chip-based_2010} or via the quantum enhancement of magnetic sensitivity \cite{gross_nonlinear_2010,hyllus_fisher_2012,toth_multipartite_2012}. 
State of the art experiments on photonic systems \cite{pan_multiphoton_2012}, superconducting qubits \cite{wendin_quantum_2017}, trapped ions \cite{blatt_entangled_2008} and Rydberg atom arrays \cite{saffman_quantum_2010} can now produce highly entangled states of tens of individually identifiable qubits, in which entanglement is more readily observable.

Besides quantum state tomography, a wide array of methods have been developed for the detection of entanglement \cite{horodecki_quantum_2009,guhne_entanglement_2009}. In two-qubit systems, the degree of entanglement is quantified by the concurrence
 \cite{hill_entanglement_1997,wootters_entanglement_1998}. 
Its direct measurement remains challenging since it requires non-linear operations on the prepared state \cite{bovino_direct_2005,walborn_experimental_2006,schmid_experimental_2008,islam_measuring_2015}, and it was so far only achieved for photon pairs in pure quantum states \cite{walborn_experimental_2006}.
In the case of multipartite systems, the study of entanglement is cumbersome due to the existence of distinct classes of entanglement~\cite{dur_three_2000}. It is often revealed using entanglement witnesses, by measuring the fidelity with respect to a given entangled state \cite{toth_detecting_2005} -- the method being limited to simple enough target states.


\begin{figure}[!t]
\includegraphics[
draft=false,scale=0.9,
trim={18mm 0mm 0mm 0.cm},
]{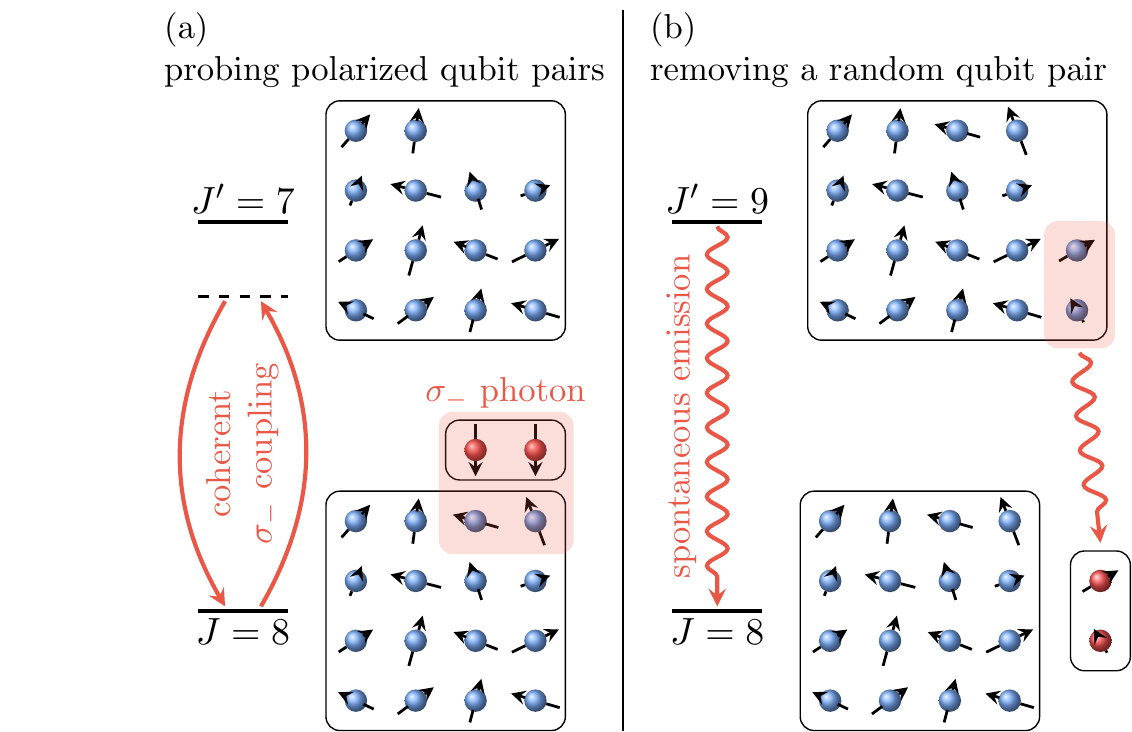}
\caption{
Scheme of the experiments manipulating qubit pairs in the electronic spin of dysprosium. An electronic spin of angular momentum $J$ can be viewed as a set of $2J$ virtual qubits symmetric upon exchange. (a) The coherent coupling to an excited state $J'=J-1$ with $\sigma_-$ polarized light probes the probability to find a qubit pair polarized in $\ket{\uparrow\uparrow}_z$. (b) The spontaneous emission from an excited state $J'=J+1$ removes a random pair of qubits. 
\label{fig_scheme}}
\end{figure}


In this work, we study the entanglement properties associated with non-classical states of the electronic spin of dysprosium atoms, of angular momentum $J=8$ in its ground state. Quantum states with non-classical correlations have been extensively studied in single large-spin systems, including two-photon states~\cite{lapkiewicz_experimental_2011}, ground state atomic spins~\cite{chaudhury_quantum_2007,fernholz_spin_2008}, molecules~\cite{gatteschi_quantum_2003}, and Rydberg atoms~\cite{facon_sensitive_2016}. In the formal analogy between a spin $J$ and a set of $2J$ qubits symmetric upon exchange~\cite{majorana_atomi_1932}, non-classicality goes hand in hand with entanglement between the virtual qubits. However, as long as the angular momentum $J$ is conserved, the qubit ensemble cannot be partitioned, and the relevance of entanglement is disputable. Here, we use an optical coupling to an excited electronic state of angular momentum $J'=J-1$ to partition the 16-qubit ensemble associated with the spin $J$, giving access to entanglement. The virtual absorption of a photon is interpreted as the annihilation of a qubit pair in a state defined by the light polarization, leaving a set of 14 qubits in the excited electronic level (see \fig{fig_scheme}a). We use this partition to probe entanglement in non-classical spin states, either by characterizing non-classical behavior of qubit pairs via the measurement of concurrence, or by revealing an increase of entropy upon partition. We extend this protocol to probe decoherence in states prepared in an excited electronic level $J'=J+1$ (see \fig{fig_scheme}b). There, the spontaneous emission of a photon drives the system to the electronic ground state $J$, which corresponds to  the removal of a qubit pair randomly drawn from the initial state. We reveal the robustness of pairwise entanglement with respect to qubit loss, as well as the fragility of coherence in Schr\"odinger cat states.

This paper is organized as follows. We present in section \ref{section_Qpair} the experimental protocol used to measure the properties of qubit pairs extracted from the electronic spin, based on the polarization dependence of the light-spin interaction. In section \ref{section_concurrence}, we investigate the non-classical character of these qubit pairs via the measurement of the concurrence of the reduced two-qubit density matrix, and apply it to a W state and a squeezed state. In section \ref{section_entropy}, we investigate the increase of entropy upon the $14|2$ partition as a proof of entanglement  for W and Schr\"odinger cat states, by  studying the mixed nature of the reduced two-qubit density matrix. Finally, in section \ref{section_decoherence}, we study the decoherence upon the loss of a qubit pair triggered by spontaneous emission. We show that non-classical pairwise correlations are robust with respect to the extraction of qubits. On the contrary,  the coherence of a Schrödinger cat state is completely destroyed upon qubit loss, due to the complete \emph{which path} information carried by the spontaneously emitted photon's polarization. In another superposition state, we show the existence of  a quantum jump leaving the path information hidden, such that maximal-order coherence remains visible.

\section{Pair Husimi function measurement\label{section_Qpair}}
\subsection{Probing pairs via light coupling}

The electronic ground state $J=8$ can be interpreted as the sum of $2J=16$ virtual spin-$1/2$s, in a state symmetric upon exchange.  We discuss here the partition of this qubit ensemble, prepared in a state $\rho$, through the coupling to an excited electronic level, of angular momentum $J'=7$. As sketched in \fig{fig_scheme}a, the coupling to the excited manifold is induced by light close to the optical transition, via the absorption of a photon. The photon polarization $\boldsymbol\epsilon$ defines an $L=1$ quantum state $\ket{\boldsymbol\epsilon}$ that can be considered as a symmetric 2-qubit state. We restrict here to the case of a circular polarization $\sigma_-$, which corresponds to qubits polarized in $\ket{\downarrow\downarrow}_z$. Since the excited state contains only $2J'=14$ qubits, two qubits are removed upon photon absorption. The conservation of angular momentum requires these removed qubits to be polarized in $\ket{\uparrow\uparrow}_z$, the time-reversed state of the absorbed photon's polarization. The excited state $\rho'$ can be then written as a projected state
$
\rho'=\bra{\uparrow\uparrow}_z\rho\ket{\uparrow\uparrow}_z.
$
The probability for a pair chosen from the 16 qubits to be polarized in  $\ket{\uparrow\uparrow}_z$ then reads
\[
\Qpair(\ebold_z)=\Tr\rho',
\]
defining the pair Husimi function along the direction $\ebold_z$.

To probe this behavior, we measure the light shift $V$ induced by an off-resonant light beam close to the considered optical transition. The light shift, being induced by virtual photon absorption processes, is proportional to the pair Husimi function, as 
\[
 V/V_0=\Qpair(\ebold_z),\quad V_0=\frac{(dE)^2}{\hbar\Delta},
\]
where $d=\langle{J-1}||\dbold|| J\rangle$ is the reduced dipole matrix element, $E$ is the light electric field amplitude, and $\Delta$ is the detuning from resonance.

\begin{figure}[!t]
\includegraphics[
draft=false,scale=0.9,
trim={2mm 2mm 0 0.cm},
]{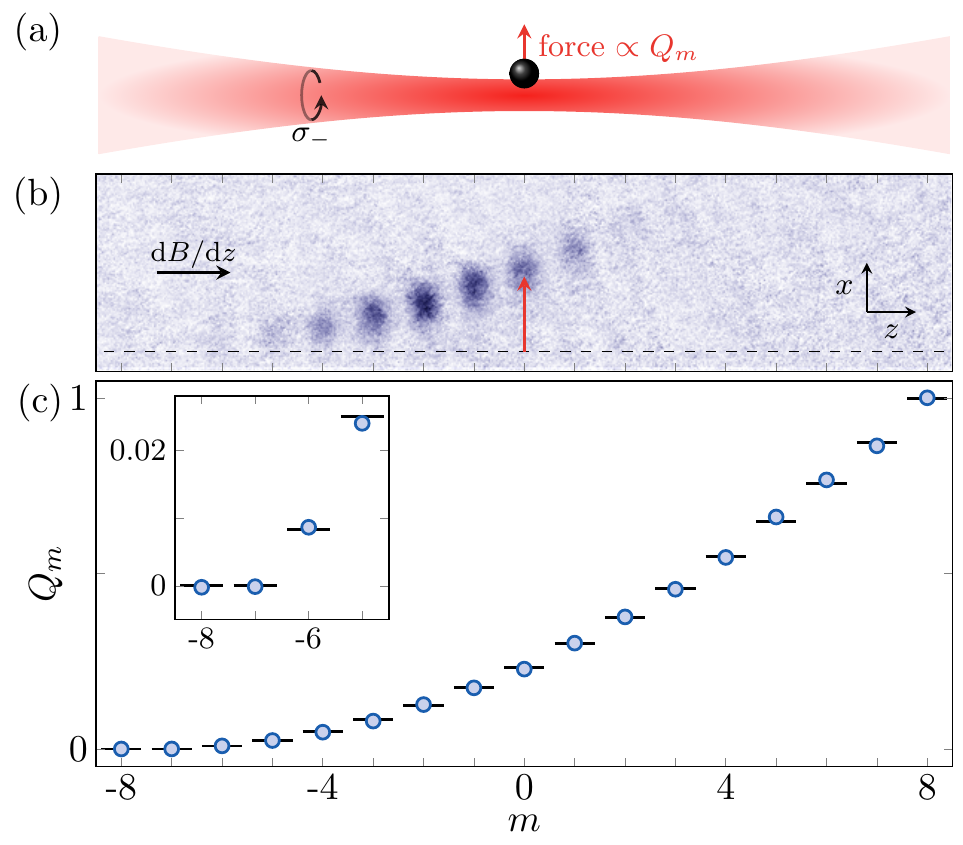}
\caption{
Husimi function measurement for Dicke states.
(a) Scheme of the light shift measurement. We measure the force induced on the atoms by an off-centered laser beam, blue detuned with respect to the optical resonance.
(b) Image of an atomic gas prepared in a coherent state of polar angle $\theta\simeq\SI{100}{\degree}$. The atoms are kicked along $x$ by the laser beam. Subsequently, we apply a magnetic field gradient separating the magnetic sublevels $\ket{m}$ along $z$ during time-of-flight. The dashed line indicates the mean $x$ position in the absence of the repulsive laser beam.
(c) Probability $Q_m$ for a qubit pair taken in the Dicke state $\ket{m}$ to be in $\ket{\uparrow\uparrow}_z$, deduced from the kick amplitudes. In all
figures, error bars represent the 1-$\sigma$ statistical uncertainty (here smaller than the blue dots).  The black lines are the theoretical values of \eq{Q_Dicke}.
\label{fig_Qm}}
\end{figure}

\subsection{Application to Dicke states}

We illustrate our method by measuring the value of the Husimi function $\Qpair(\ebold_z)$ for an arbitrary Dicke state $\ket{m}$ (with $-J\leq m\leq J$), that we denote $Q_m$ hereafter.  

All our experiments are performed on a cloud of $1.0(1)\times 10^5$ dysprosium atoms (of the bosonic isotope $^{162}$Dy), held in an optical dipole trap at a temperature $T=\SI{0.54(3)}{\micro\kelvin}$. 
The results described in this work can be understood by considering a single atom, with the ensemble acting as an averaging mechanism only.
The experimental scheme  for the $Q_m$ measurement is shown in  \fig{fig_Qm}a. We prepare the atoms in a coherent state $\ket{m=J}_{\nbold}$ polarized along a direction $\nbold$, parametrized by the spherical angles $(\theta,\phi)$. The polar angle $\theta$ determines the projection probabilities $\Pi_m$ along the Dicke states $\ket{m}$, which are significant for values of $m$ close to $J\cos\theta$. We then push the atomic cloud by applying an off-centered laser beam, with circular $\sigma_-$ polarization and blue detuning with respect to an optical transition at \SI{696}{\nano\metre}. The intensity gradient then leads to a force along $x$ proportional to the light shift (\fig{fig_Qm}a). After this kick, a magnetic field gradient is applied to spatially separate the different $m$-components along $z$, which allows us to retrieve the light shift experienced by each Dicke state independently. After a \SI{2.3}{\milli\second} time-of-flight, we image the atoms and measure the $x$-displacement for each Dicke state $\ket{m}$ that is significantly populated, and hence their values $Q_m$. A typical absorption image is shown in \fig{fig_Qm}b. Repeating this measurement for various angles $\theta$, we measure the light shifts for all projections $m$, and infer the $Q_m$ values shown in \fig{fig_Qm}c \cite{Note6}. 

Our measurements are consistent with an absence of light shift for the states $\ket{m=-J}$ and $\ket{m=-J+1}$, i.e. these states are dark with respect to the $J\rightarrow J'=J-1$ optical transition for $\sigma_-$ polarized light. In terms of the underlying qubits, the states $\ket{m=-J}$ only contains $\ket{\downarrow}_z$-polarized qubits, while the state $\ket{m=-J+1}$ has a single qubit in $\ket{\uparrow}_z$. In both cases, a qubit pair cannot be found polarized in $\ket{\uparrow\uparrow}_z$, hence $Q_{-J}=Q_{-J+1}=0$.

More generally, a Dicke state $\ket{m}$ is composed of $J-m$ qubits in $\ket\downarrow_z$ and $J+m$ qubits in $\ket\uparrow_z$ \cite{dicke_coherence_1954}. The probability to pick a pair $\ket{\uparrow\uparrow}_z$ simply reads 
\begin{equation}\label{Q_Dicke}
 Q_m  =\binom{J+m}{2}\bigg/\binom{2J}{2}=\frac{(J+m)(J+m-1)}{2J(2J-1)},
\end{equation}
in good agreement with our measurements.

We use these measurements to probe the Husimi function of states lacking $z$ rotation symmetry. For this,  we measure their projection probabilities $\Pi_m(\nbold)$ along $\nbold$ by combining a spin rotation and a Stern-Gerlach projective measurement along $z$. We then infer the Husimi function by weighting these probabilities with the $Q_m$ values, as
\begin{equation}\label{eq_Qpair}
 \Qpair(\nbold)=\sum_m Q_m\Pi_m(\nbold).
\end{equation}
In the following, we use the theoretical values  of \eq{Q_Dicke} rather than the measured ones to avoid propagating systematic errors.

\begin{figure*}[!t]
\includegraphics[
draft=false,scale=1,
trim={2mm 2mm 0 0.cm},
]{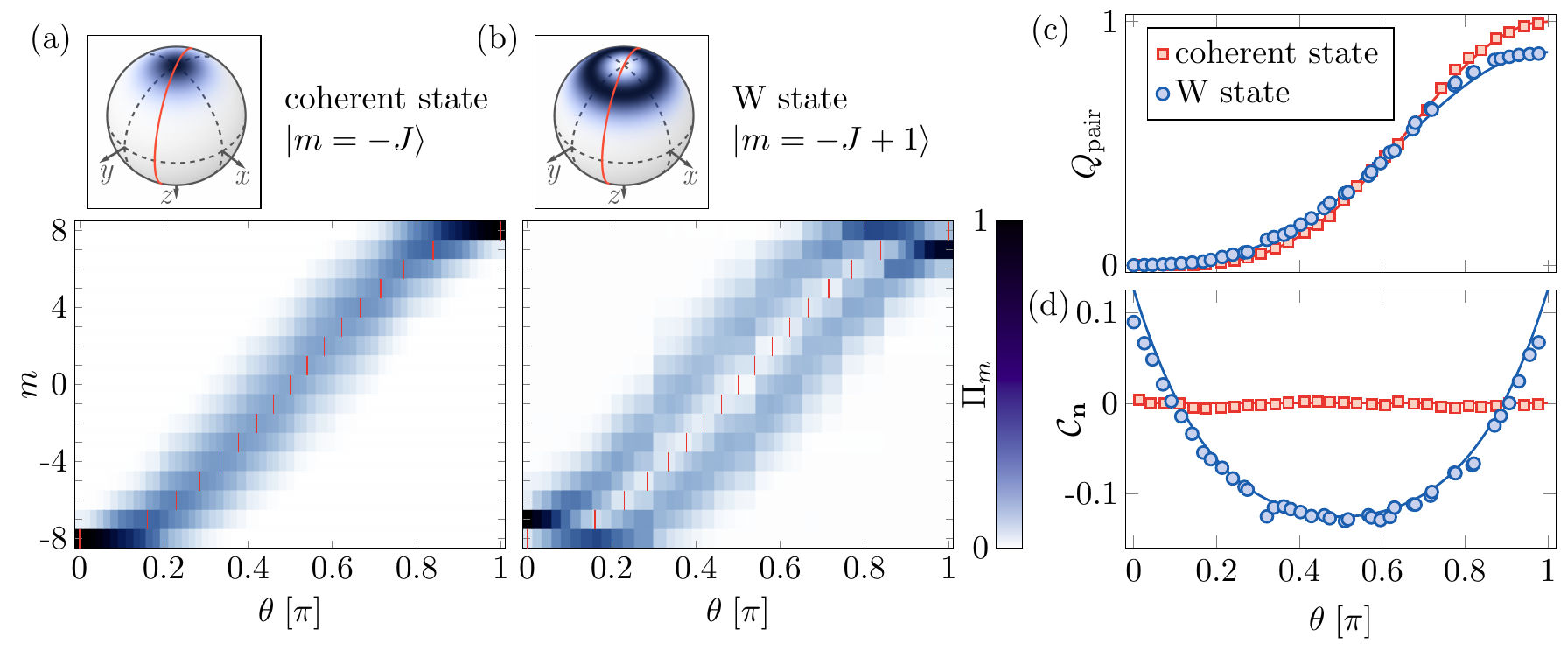}
\caption{
Qubit pair properties of coherent and W states.
(a,b) Measured spin projection probabilities $\Pi_m$ as a function of the polar angle $\theta$, for a coherent spin state (a) and for the W state (b). The red vertical lines indicate the expected maxima for the coherent state, also corresponding to minima for the W state. The top panels represent the considered spin-$J$ states on the Bloch sphere, the red circles indicate the spanned measurement projection axis.
(c) Pair Husimi function $\Qpair$ computed from the (a) and (b) data (blue disks and red squares, respectively).  The lines correspond to the expected functions $\Qpair(\theta)$ for the coherent and W states (red and blue lines).
(d) Distribution $\mathcal{C}_{\nbold}$ of non-classical correlations as a function of the polar angle $\theta$. The points $\mathcal{C}_{\nbold}>0$ measured for the W state evidence non-classicality.
\label{fig_Dicke}}
\end{figure*}

\subsection{Coherent and W states}

We first apply the above protocol to the quasi-classical coherent spin state $\ket{m=-J}$ and the W state  $\ket{m=-J+1}$. The coherent state can be viewed as a set of $2J$ qubits polarized in $\ket{\downarrow}_z$, forming a non-entangled product state. The W state, which hosts a single qubit in  $\ket{\uparrow}_z$, is a paradigmatic state of a fundamental class of entanglement \cite{dur_three_2000}, which  has been realized and studied in various settings \cite{choi_entanglement_2010,haas_entangled_2014,mcconnell_entanglement_2015,ebert_coherence_2015,
zeiher_microscopic_2015,haffner_scalable_2005,frowis_experimental_2017,pu_experimental_2018}.

In our experiment, the atoms are initially spin-polarized in the coherent state $\ket{m=-J}$. To produce the W state, we  apply a resonant radiofrequency $\pi$-pulse towards $\ket{m=-J+1}$, while the other Dicke states are made off-resonant using a quadratic light shift. This light shift is produced using the \SI{696}{\nano\metre} laser beam with a $\sigma_-$ polarization, leading to positive energy shifts for all Dicke states $\ket{m}$, except for $m=-J$ and $-J+1$. We reach a maximum W state fidelity of 0.91(1), with residual overlaps on other Dicke states below 4\% \cite{Note7}.

We report in \fig{fig_Dicke}a,b the measured projection probabilities $\Pi_m(\theta)$ for these two states. For a given projection $m$, the coherent state probabilities feature a single peak centered on the expected maximum at $\theta_m=\acos(m/J)$, shown as red lines. For the W state probabilities, we observe a double-peaked distribution for all non-stretched states $m\neq\pm J$. This behavior results from the interference between two processes, depending on whether the spin $\ket{\uparrow}_z$ is projected on $\ket{\uparrow}_\theta$ or $\ket{\downarrow}_\theta$. The first (second) process dominates for $\theta\simeq0$ ($\theta\simeq\pi$), and the two processes destructively interfere at $\theta_m$, as observed in our data.

We combine these measurements to infer the pair Husimi functions using \eq{eq_Qpair}, finding good agreement with theory for both states  (see \fig{fig_Dicke}c). In particular, for the coherent state our data matches well  the probability $\Qpair(\theta)=\sin^4(\theta/2)$ that two qubits in $\ket{\downarrow}_z$ are projected in $\ket{\uparrow}_\theta$. In the following sections we use these measurements to probe entanglement properties.

\section{Non-classicality of qubit pairs\label{section_concurrence}}

Our first characterization of entanglement of the $2J$-qubit state consists in revealing the non-classical character of qubits pairs extracted from it. 

\subsection{Measure of non-classicality via the concurrence\label{subsection_concurrence}}

The collective state $\rhopair$ of a qubit pair symmetric upon exchange can be written as the state of an angular momentum $L=1$. Drawing an analogy with quantum optics \cite{sudarshan_equivalence_1963,glauber_coherent_1963}, it will be called classical if it can be expressed as a statistical mixture of quasi-classical coherent states \cite{giraud_classicality_2008}, as 
\begin{equation}\label{eq_classical}
\rho_{\text{pair}}^{\text{(classical)}}=\sum_{\nbold} w_{\nbold} ||\nbold\rangle\langle\nbold||,
\end{equation}
where $||\nbold\rangle$ is a spin-1 coherent state pointing along $\nbold$, and $w_{\nbold}\geq0$, $\sum_{\nbold} w_{\nbold}=1$. Coherent states are the only pure states that satisfy the equality
\begin{equation}\label{eq_Z_coherent}
Z(\nbold)\equiv 2\langle L_{\nbold}^2 \rangle-\langle L_{\nbold} \rangle^2-1= 0
\end{equation}
for arbitrary measurement axis $\nbold$. Then it follows by convexity that $
Z(\nbold)\geq 0
$ for  classical states.
As shown in \cite{giraud_classicality_2008}, the existence of a strictly negative value $Z(\nbold)$ constitutes a necessary and sufficient criterion of non-classicality.

To apply this criterion to our system, we use the connection between the mean values of spin projection and the Husimi function, 
\begin{align*}
\langle L_{\nbold} \rangle&=\Qpair(\nbold)-\Qpair(-\nbold),\\
\langle L_{\nbold}^2 \rangle&=\Qpair(\nbold)+\Qpair(-\nbold),
\end{align*}
leading to the expression $Z(\nbold)=\alpha\,\mathcal{C}_{\nbold}$, where we introduce the coefficient \mbox{$\alpha=(\sqrt{\Qpair(-\nbold)}-\sqrt{\Qpair(\nbold)})^2-1$} and the distribution
\[
\mathcal{C}_{\nbold}=1-\left(\sqrt{\Qpair(-\nbold)}+\sqrt{\Qpair(\nbold)}\right)^2.
\]
The parameter $\alpha$ being negative, non-classicality is characterized by the existence of a direction $\nbold$ for which  $\mathcal{C}_{\nbold}$ is strictly positive. This criterion is equivalent to the bipartite entanglement witness established in \cite{korbicz_spin_2005}. 

We show in \fig{fig_Dicke}d the distribution $\mathcal{C}_{\nbold}$ computed from the measured Husimi functions, for the coherent and W states. For these states, symmetric upon rotations around $z$, we expect $\mathcal{C}_{\nbold}$ to only depend on the  polar angle $\theta$ of the measurement axis \cite{Note8}. For the coherent state, the measured $\mathcal{C}_{\nbold}$ remains close to zero for all angles $\theta$. Indeed, qubits pairs drawn from this state form themselves a spin-1 coherent state, for which $\mathcal{C}_{\nbold}$ vanishes according to Eq.\;(\ref{eq_Z_coherent}). For the W state, $\mathcal{C}_{\nbold}$ takes significantly positive values for $\theta$ close to 0 and $\pi$, showing a non-classical character.

We now show that the distribution $\mathcal{C}_{\nbold}$ can be used to quantify the degree of non-classicality of a quantum state, defined by its distance to the set of non-classical states  \cite{hillery_nonclassical_1987}. For a system of two qubits, this geometrical measure can be directly expressed in terms of the concurrence $\mathcal{C}$ \cite{wei_geometric_2003}, the most common measure of pairwise entanglement \cite{hill_entanglement_1997,wootters_entanglement_1998}. The concurrence can be explicitly written in terms of the density matrix, but it does not correspond to a directly accessible physical observable. Remarkably, the distribution $\mathcal{C}_{\nbold}$ can be used to retrieve the concurrence, as
\[
 \mathcal{C}=\max[0,\max_{\nbold}\mathcal{C}_{\nbold}].
\]
This relation was conjectured and numerically checked for randomly generated states in \refe{vidal_concurrence_2006}.

For the W state realized in the experiment, the measured $\mathcal{C}_{\nbold}$ takes its maximum for $\theta=0$ leading to a concurrence  $\mathcal{C}=0.089(5)$. This value is about 71\% of the maximum possible value $\mathcal{C}=1/J=0.125$ in a system of $2J$  qubits symmetric upon exchange  \cite{koashi_entangled_2000}, which would be reached for the W state in the absence of experimental imperfections. In our system, the concurrence is limited by the residual population $\Pi_{-J+2}\simeq0.03$ in the Dicke state $\ket{m=-J+2}$ that originates from spin-changing collisions between atoms in $\ket{m=-J+1}$. 

\begin{figure}[!t]
\includegraphics[
draft=false,scale=0.95,
trim={4mm 2mm 0 0.cm,clip},
]{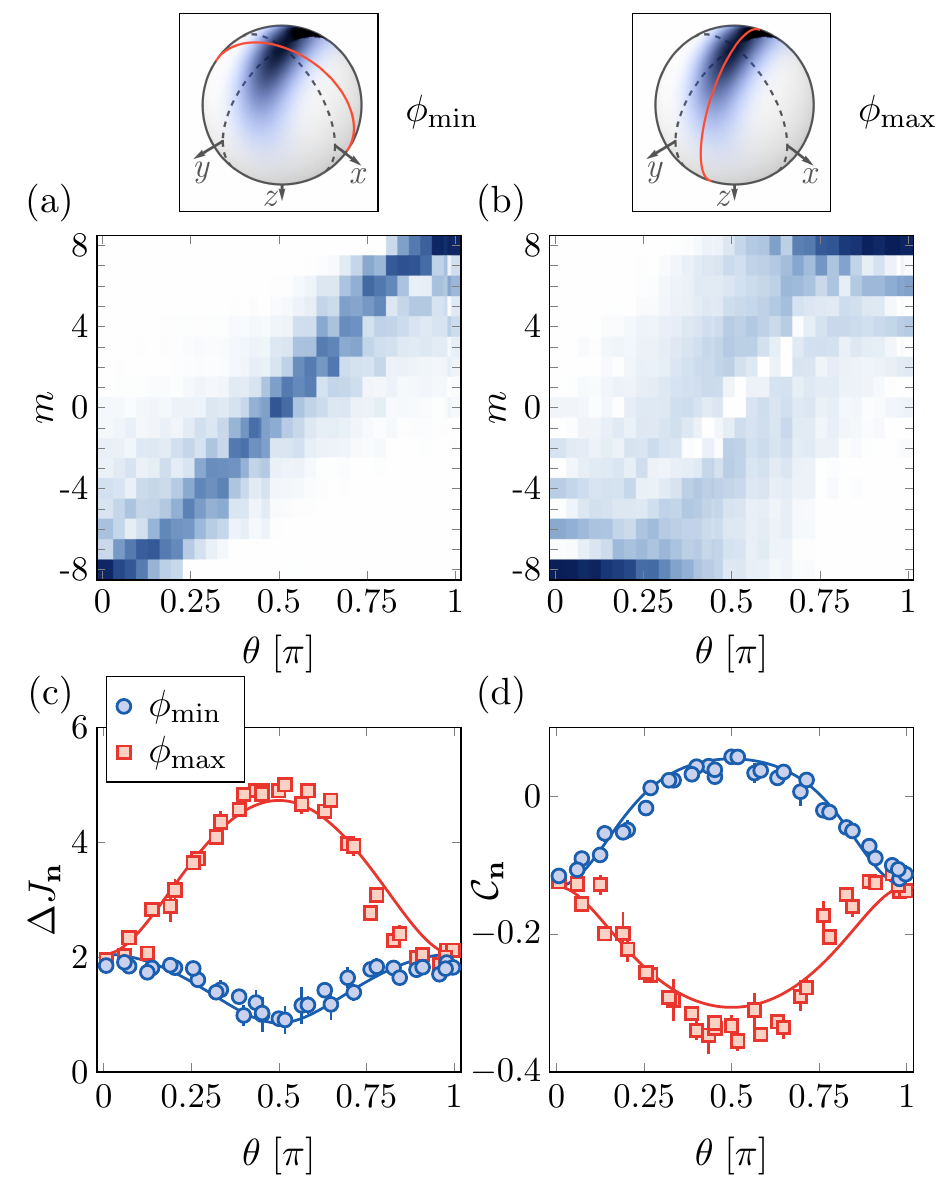}
\caption{
Qubit pair properties for a squeezed state.
(a,b) Measured spin projection probabilities $\Pi_m$ for a squeezed spin state, as a function of the polar angle $\theta$ with azimuthal angles  $\phi_{\text{min}}$ (a) and $\phi_{\text{max}}$(b).
(c) Spin projection uncertainty $\Delta J_{\nbold}$ computed from the (a) and (b) data (blue disks and red squares, respectively).  The lines correspond to the projection uncertainties  expected for the targeted spin state. 
(d) Distribution $\mathcal{C}_{\nbold}$ of non-classical correlations as a function of $\theta$. 
\label{fig_Squeezed}}
\end{figure}

\subsection{Pairwise correlations in a squeezed state}

Non-classical correlations between qubit pairs play a central role in the squeezing of a spin projection quadrature \cite{kitagawa_squeezed_1993}. In this section we extend the measurement of qubit pair properties to a squeezed spin state, that we  produce via a non-linear spin dynamics. We apply a $\hbar\chi J_x^2$ spin coupling, generated by the spin-dependent light shift of the \SI{696}{\nano\meter} laser beam, using a linear polarization $\ebold_x$ \cite{chalopin_quantum-enhanced_2018}.  This coupling induces a twisting of the spin distribution, leading to the squeezing of a spin projection quadrature \cite{kitagawa_squeezed_1993}, as first implemented in atomic Bose-Einstein condensates \cite{gross_nonlinear_2010,riedel_atom-chip-based_2010}. In our experiment, we apply a non-linear coupling of strength $\chi=2\pi\times\SI{32.1(4)}{kHz}$ for a duration $t\simeq\SI{700}{\nano\second}$, in the presence of a $z$ magnetic field $B=\SI{75(1)}{\milli G}$. 

Contrary to the Dicke states discussed above, the spin projection probabilities are no longer invariant around $z$. 
We show in \fig{fig_Squeezed}a,b the probabilities $\Pi_m(\theta,\phi)$ for two azimuthal angles $\phi_{\text{min}}=\SI{-0.4(2)}{\radian}$ and $\phi_{\text{max}}=\phi_{\text{min}}+\pi/2$, which feature minimal and maximal spin projection uncertainties, respectively. For $\theta=\pi/2$, a minimum spin projection uncertainty $\Delta J_{\text{min}}=0.92(16)$ is measured at $\phi_{\text{min}}$ (see \fig{fig_Squeezed}c), in agreement with the value $\Delta J_{\text{min}}=0.85$ expected for an optimally squeezed state (within the one-axis twisting dynamics). We report in \fig{fig_Squeezed}d the corresponding  distribution $\mathcal{C}_{\nbold}$. The measured $\mathcal{C}_{\nbold}$ takes its maximum for $\theta=\pi/2$ and $\phi=\phi_{\text{min}}$, i.e. along the squeezed quadrature direction. This maximum gives a value for the concurrence $\mathcal{C}=0.058(6)$, in agreement with the expected value of 0.055. 

Our measurements can be used to check the direct link between quadrature squeezing and pairwise entanglement \cite{wang_spin_2003}. Indeed, for the states reached via the one-axis twisting dynamics, one expects the concurrence to be expressed in terms of the minimum spin projection uncertainty, as 
\begin{equation}\label{eq_C}
 \mathcal{C}=\frac{1-2\Delta J_{\text{min}}^2/J}{2J-1}.
\end{equation}
From the measured projection quadrature, we calculate a value of 0.053(5) for the right hand side of \eq{eq_C}, in agreement with the direct measurement of the concurrence.

\section{Probing entanglement via the subsystem entropy\label{section_entropy}}

So far, we studied the entanglement of $2J$-qubit states via the non-classical character of their qubit pairs. In this section, we access entanglement more directly, by probing whether a given state of the spin $J=8$ is separable with respect to the $14|2$ partition performed by the photon absorption. For this, we use the fact that for a separable state, the global state is more disordered than its parts  \cite{horodecki_information-theoretic_1996}.  
More precisely, we quantify disorder via the Renyi entropy of infinite order (also called the min-entropy), defined as \cite{konig_operational_2009}
\[
\Smin(\rho)=-\log\lmax(\rho),
\]
where $\lmax$ is the maximum eigenvalue of the density matrix $\rho$. This eigenvalue corresponds to the maximum possible overlap of $\rho$ with a pure state. To reveal entanglement within a state $\rho$ of the collective spin $J$, it is thus sufficient to show that the entropy of the reduced pair state $\rhopair$ has a strictly higher entropy than the one of the original state $\rho$, i.e. if the conditional entropy satisfies \cite{horodecki_information-theoretic_1996}
\[
 \Smin(14|2)\equiv\Smin(\rho)-\Smin(\rhopair)<0.
\]

\subsection{Entanglement of the W state}

The evaluation of the pair state entropy $\Smin(\rhopair)$ is based on the tomography of the pair density matrix \cite{manko_spin_1997}. Full information on the density matrix is contained in the Husimi function $\Qpair(\nbold)$. We fit the measured Husimi function by a spherical harmonic expansion
\begin{equation}
 \Qpair(\nbold)=\frac{1}{3}+\sqrt{\frac{4\pi}{3}}\sum_{\ell=1}^2\sum_{m=-\ell}^\ell\lambda_{\ell,m}Y_{\ell}^m(\nbold),\label{eq_spherical_harmonics}
\end{equation}
and infer the density matrix as
\begin{equation}
 \rhopair=\frac{1}{3}\mathds{1}+\sum_{m=-1}^1\lambda_{1,m}\mathcal{L}_m+\sum_{m=-2}^2\lambda_{2,m}\mathcal{Q}_m,\label{eq_tomography}
\end{equation}
where the $\mathcal{L}_m$ and $\mathcal{Q}_m$ matrices correspond to the $L=1$ angular momentum components and quadrupole moments, respectively (see Appendix \ref{appendix_tomography}).

We apply this protocol to the W state, taking into account the slight variation of the Husimi function $\Qpair(\nbold)$ with respect to the azimuthal angle $\phi$ in the prepared state \cite{Note8}. We infer a density matrix
\[
 \rhopair\simeq\left(
\begin{array}{ccc}
 0.88 & 0.01+0.05\, \I & -0.01-0.01\, \I \\
 0.01-0.05\, \I & 0.12 & 0.01\, \I \\
 -0.01+0.01\, \I &  -0.01\, \I &0 \\
\end{array}
\right),
\]
with typically 1\% statistical uncertainty. 
The reconstructed density matrix matches well the expected one
\[
 \rhopair=\left(
\begin{array}{ccc}
 7/8 & 0 & 0 \\
 0 & 1/8 & 0 \\
 0 &  0 &0 
\end{array}
\right).
\]
Diagonalization of the reconstructed density matrix gives a maximum eigenvalue $\lmax(\rhopair)=0.882(5)$.  

We now consider the global spin-$J$ state. The projection probability $\Pi_{-J+1}=0.91(1)$ with the Dicke state $\ket{m=-J+1}$ provides a lower bound on the maximum overlap $\lmax(\rho)$ with pure states. 

Combining these results together, we obtain 
\[
 \Smin(14|2)<-0.03(1).
\]
Its negative value shows that the prepared state is not separable with respect to a $14|2$ partition, and is thus entangled.

\subsection{Entanglement of a Schr\"odinger cat state\label{section_cat}}
We now consider the case of a Schr\"odinger cat state, for which the effect of the $14|2$ partition is more striking. Schr\"odinger cat states, which constitute archetypal states with highly non-classical properties, have been realized in different types of experiments \cite{
monroe_schrodinger_1996,
brune_observing_1996,
friedman_quantum_2000,
sackett_experimental_2000,
leibfried_creation_2005,
ourjoumtsev_generating_2006,
neergaard-nielsen_generation_2006,
deleglise_reconstruction_2008,
monz_14-qubit_2011,
yao_observation_2012,
kirchmair_observation_2013,
facon_sensitive_2016,
degen_quantum_2017,
chalopin_quantum-enhanced_2018,
wang_18-qubit_2018,
dietsche_high-sensitivity_2019,
song_generation_2019,
omran_generation_2019,
wei_verifying_2020}.

\begin{figure*}[!t]
\includegraphics[
draft=false,scale=1,
trim={4mm 2mm 0 0.cm,clip},
]{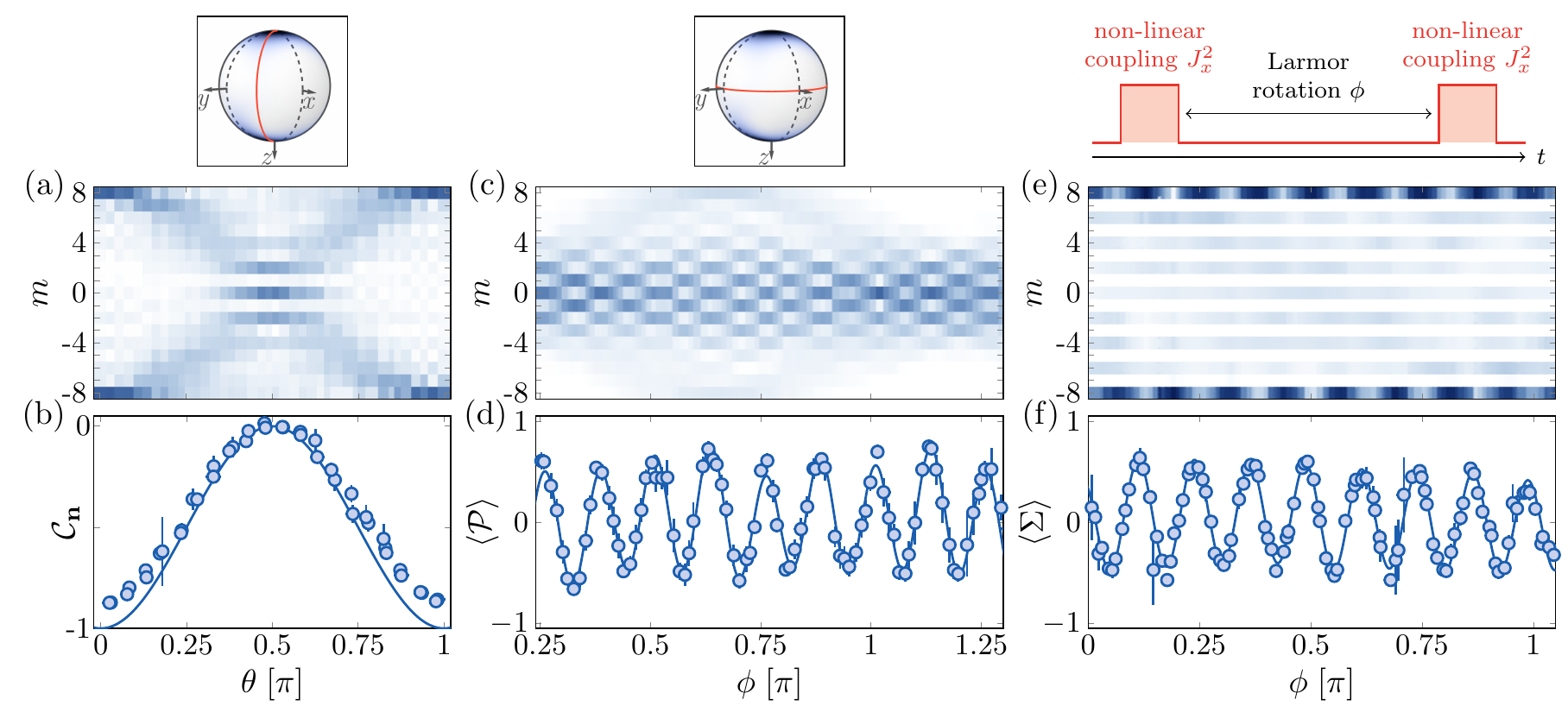}
\caption{
Characterization of entanglement in a Schr\"odinger cat state.
(a) Measured spin projection probabilities $\Pi_m$ for a cat state, as a function of the polar angle $\theta$.  The azimuthal angle $\phi=\SI{0.86(5)}{\radian}$ is chosen such that the two coherent state Husimi functions destructively interfere for odd $m$ values around $\theta=\pi/2$.
(b) Distribution $\mathcal{C}_{\nbold}$ inferred from the probabilities shown in (a) (blue disks). The solid line is the expected variation for a perfect cat state.
(c) Projection probabilities $\Pi_m$ measured along equatorial directions ($\theta=\pi/2$) parametrized by the azimuthal angle $\phi$. 
(d) Evolution of the mean parity $\langle\mathcal{P}\rangle$ deduced from (c). 
(e) Projection probabilities $\Pi_m$ measured after a Larmor rotation of angle $\phi$ followed by a second non-linear evolution. 
(f) Evolution of the mean sign of even projections $\langle\Sigma\rangle$ deduced from (e). The solid lines in (d,f) are fits with a Fourier series.
\label{fig_N00N}}
\end{figure*}

The cat state considered here is the coherent superposition of two quasi-classical spin states $\ket{m=\pm J}$ \cite{mermin_extreme_1990}. To produce it, we use the one-axis twisting dynamics discussed above, with a stronger non-linear coupling $\chi=2\pi\times\SI{1.25}{MHz}$ and a reduced magnetic field $B=\SI{53.7(1)}{\milli G}$. After showing quadrature squeezing at short times ($t\sim\SI{10}{ns}$), the spin quadratures collapse to a featureless spin distribution, before a revival at a time $t_{\text{cat}}=\pi/(2\chi)=\SI{200}{\nano\second}$, at which the system forms a coherent superposition of stretched states $\ket{m=\pm J}$ \cite{chalopin_quantum-enhanced_2018}. 

 In \fig{fig_N00N}a, we show the measured probabilities $\Pi_m(\nbold)$ for various polar angles $\theta$ with a fixed azimuthal angle $\phi$. For $\theta=0$, we confirm dominant population of the two stretched states, with $\Pi_{-J}=0.38(2)$ and $\Pi_{J}=0.42(2)$. When varying $\theta$, the distribution is a superposition of the contributions of each of the two coherent states forming the cat state. Interestingly, we observe an interference between the two distributions when they overlap, i.e. for $\theta\simeq\pi/2$. As shown in \fig{fig_N00N}c, the interference pattern depends on the azimuthal angle $\phi$, with an alternation between even- and odd-$m$ projections of period $2\pi/(2J)$ \cite{bollinger_optimal_1996}.

We first test whether a qubit pair extracted from this state features non-classical behavior. We expect the distribution $\mathcal{C}_{\nbold}$ to be rotationally invariant around $z$, and thus study its variation with the polar angle $\theta$ in \fig{fig_N00N}b \cite{Note9}. Our measurements are consistent with $\mathcal{C}_{\nbold}<0$ for all angles $\theta$, showing that the reduced two-qubit state is classical. This measurement highlights the well-known property of this state that any of its subsystems is classical.

We now extend the tomography protocol to the cat state, and obtain the reduced two-body density matrix
\[
 \rhopair\simeq\left(
\begin{array}{ccc}
 0.46 & -0.01\,\I & -0.03+0.05\,\I \\
 0.01\,\I & 0.05 & -0.01\,\I \\
 -0.03-0.05\,\I &  0.01\,\I & 0.49 
\end{array}
\right),
\]
that we compare to the expected matrix
\begin{equation}\label{eq_rhopair_cat}
 \rhopair=\left(
\begin{array}{ccc}
 1/2 & 0 & 0 \\
 0 & 0 & 0 \\
 0 &  0 &1/2 
\end{array}
\right)
\end{equation}
obtained for a perfect cat state. We compute the maximum eigenvalue $\lmax(\rhopair)=0.53(1)$ of the reconstructed matrix.

In order to reveal entanglement in the prepared state, we evaluate  its overlap with perfect cat states $\ket{\text{cat}(\alpha)}=(\ket{m=-J}+\E^{\I\alpha}\ket{m=J})/\sqrt{2}$, which constitute a family of pure quantum states. The simple form of these states in the Dicke basis allows us to express the overlap with a state $\rho$ as
\[
 \mathcal{O}_\alpha=\frac{\rho_{-J,-J}+\rho_{J,J}+2\,\text{Re}(\rho_{-J,J}\,\E^{\I\alpha})}{2},
\]
where the diagonal elements $\rho_{m,m}$ correspond to the spin projection probabilities $\Pi_m$. The overlap $\mathcal{O}_\alpha$ takes its maximum value $\mathcal{O}$ for $\alpha=-\arg\rho_{-J,J}$, with 
\[
 \mathcal{O}=\frac{\Pi_{-J}+\Pi_{J}+2|\rho_{-J,J}|}{2}.
\]
We present two protocols giving a lower bound on the extremal coherence $|\rho_{-J,J}|$, both based on the measurement of an observable $A$ defined on the spin $J$. We consider its mean value in a state obtained after the cat state preparation, followed by a Larmor rotation around $z$ of angle $\phi$, as
\[
 \langle A\rangle(\phi)=\sum_{m,m'}a_{m,m'}\rho_{m',m}\E^{\I(m'-m)\phi}.
\]
The extremal coherence can be singled out by measuring the Fourier coefficient $A_{2J}=|a_{J,-J}\rho_{-J,J}|$ at frequency $2J$ \cite{bollinger_optimal_1996,wei_verifying_2020}. We will use observables that can take values in the interval  $[-1,1]$ only, such that $|a_{J,-J}|\leq 1$. The coefficient  $A_{2J}$ then provides a lower bound on the extremal coherence $|\rho_{-J,J}|$.

The first observable we consider is the parity $\mathcal{P}$ of the spin projection along an equatorial direction $\nbold\perp\ebold_z$ -- an observable commonly used to characterize cat states \cite{sackett_experimental_2000,leibfried_creation_2005,monz_14-qubit_2011,yao_observation_2012,song_generation_2019,omran_generation_2019,wei_verifying_2020}. We fit its oscillation, shown in \fig{fig_N00N}d, with a Fourier series, from which we get the Fourier coefficient $\mathcal{P}_{2J}=0.26(1)$. The second observable uses a non-linear evolution, obtained by repeating the one-axis twisting evolution used to produce the cat state \cite{davis_approaching_2016,linnemann_quantum-enhanced_2016,frowis_detecting_2016,macri_loschmidt_2016,chalopin_quantum-enhanced_2018} (see the scheme of \fig{fig_N00N}e). In the absence of imperfections, the system is brought to a superposition $\sin(J\phi)\ket{m=-J}+\cos(J\phi)\ket{m=J}$, which allows us to extract the maximal coherence from the projection probabilities in stretched states only. The projection probabilities measured with this protocol are shown in \fig{fig_N00N}e. In practice, we observe residual probabilities in other projection values $m$, with $m$ even only, as expected from parity symmetry. We thus use an observable $\Sigma$ defined as the sign of the spin projection on even states, with 
\[
 \langle\Sigma\rangle=\sum_{m\;\text{even}}\sgn(m)\Pi_m.
\]
Its oscillation, shown in \fig{fig_N00N}f, gives a Fourier coefficient $\Sigma_{2J}=0.247(5)$. The advantage of the second method will become clear when we consider a more complex quantum state in the next section.  

The two protocols lead to comparable estimates of the extremal coherence. Using the measured probabilities $\Pi_{\pm J}$ quoted above, we infer a lower bound on the overlap $\mathcal{O}\geq0.66(2)$ and thus on the eigenvalue $\lmax(\rho)$. 
Together, these measurements provide a conditional entropy
\[
 \Smin(14|2)<-0.23(3),
\]
which proves entanglement more evidently than for the W state. We note that the requirement \mbox{$\mathcal{O}>\lmax(\rhopair)=0.53(1)$}, that we used to demonstrate the non-separability of the $14|2$ partition, is consistent with the entanglement witness $\mathcal{O}>0.5$ extensively used for cat states~\cite{toth_detecting_2005}.

\section{Decoherence upon qubit loss\label{section_decoherence}}

We now consider the removal of a pair of qubits randomly drawn from the electronic spin,  irrespective of its quantum state. For this purpose, we prepare a quantum state of interest $\rho'$ in an excited level of angular momentum $J'=9$, corresponding to a symmetric state of $2J'=18$ qubits (see~\fig{fig_scheme} b). The spontaneous emission of a photon drives the system to the  ground state $J=8$, which has two missing qubits. Since the emitted photon can carry an arbitrary polarization, the process allows for three independent quantum jumps associated with the polarizations  $\ebold_-,\ebold_z,\ebold_+$, with $\ebold_\pm=(\ebold_x\pm\I\ebold_y)/\sqrt2$. The ground-state density matrix then reads
\[
 \rho=\sum_{\ebold_u=\ebold_-,\ebold_z,\ebold_+}\bra{\ebold_u}\rho'\ket{\ebold_u},
\]
which can be simply  written as 
\[
 \rho=\Tr_2\rho',
\]
corresponding to the loss of an arbitrary qubit pair.

\begin{figure}[!t]
\includegraphics[
draft=false,scale=0.9,
trim={16mm 4mm 0 0.cm},
]{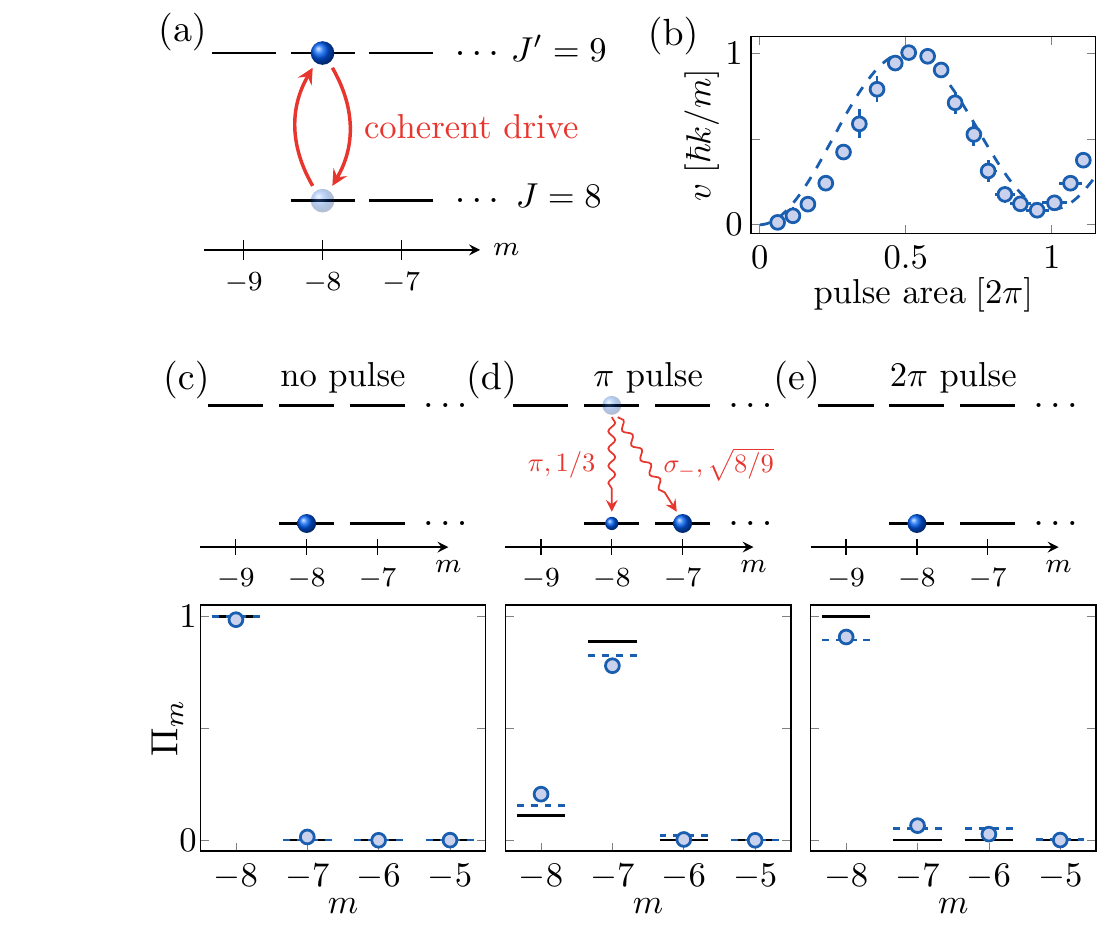}
\caption{
Loss of a qubit pair in a W state.
(a) Scheme for the preparation of the W state in the excited electronic level. (b) Evolution of the mean atom velocity  acquired due to the photon absorption recoil, as a function of the light pulse duration. The dashed line is a model taking into account spontaneous emission during the pulse. (c,d,e) Top panels: expected states, with a scheme of spontaneous emission in (d) showing the Clebsch-Gordan coefficients for the two possible quantum jumps. Bottom panels: spin projection probabilities in the absence of the resonant light pulse (c), for a $\pi$ pulse (d) and a $2\pi$ pulse (e). The solid lines are the probabilities  expected for a perfect W state, while the dashed lines use the same model as in (b). 
\label{fig_decoherence_W}}
\end{figure}

\subsection{Robustness of pairwise quantum correlations}

We first investigate the effect of particle loss on a W state prepared in an excited electronic level of angular momentum $J'=J+1$, coupled to the ground state with an optical transition of wavelength \SI{626}{\nano\meter}. To produce the state  $\ket{m^\prime=-J^\prime+1}$ in the excited level, we start in the coherent state $\ket{m=-J}$ of the lowest energy manifold, and use $\pi$ polarized resonant light to couple the system to the desired state (see \fig{fig_decoherence_W}a). As shown in \fig{fig_decoherence_W}b, we monitor the Rabi oscillation via the atom recoil upon light absorption. The comparison with a master equation model taking into account spontaneous emission during the Rabi flopping allows us to estimate a fidelity of 0.98 for a pulse duration  $t_{\text{pulse}}\simeq\SI{62}{\nano\second}$ -- the excited state lifetime being  $\tau_{\text{exc}}\simeq\SI{1.2}{\micro\second}$ \cite{gustavsson_lifetime_1979}.

Following the light pulse, we wait for spontaneous emission to occur before measuring the spin state in the ground level. We observe significant populations only in the states $\ket{m=-J}$ and $\ket{m=-J+1}$, as expected from the selection rule $|m'-m|\leq1$. The state $\ket{m=-J+1}$ is dominantly populated, showing that, in most cases, the $\ket{\uparrow}$ excitation of the W state is not removed upon the loss of a qubit pair. The projection probabilities, shown in \fig{fig_decoherence_W}d, are close to the expected values $\Pi_{-J+1}=1/(J+1)$ and $\Pi_{-J}=1-\Pi_{-J+1}$, with a residual difference mostly explained by the imperfect state preparation. 

The non-classicality of qubit pairs in the final state is probed via the distribution $\mathcal{C}_{\nbold}$ introduced in section \ref{subsection_concurrence}. We remind that $\mathcal{C}_{\nbold}$ is obtained from the spin projection probabilities along $\nbold$. Since its maximum value is expected to be reached along $z$, we only consider projections along this direction, and obtain $\mathcal{C}_{z}=0.104(3)$. This value provides a lower bound on the qubit pair concurrence, expected to be $\mathcal{C}=1/(J+1)\simeq0.111$ in the initial state. The proximity of the initial state concurrence and the measured one after decay  illustrates that losing qubits does not alter non-classicality of the remaining qubits pairs~\cite{dur_three_2000}. 

\subsection{Fragility of macroscopic coherence\label{section_decoherence_cat}}

\begin{figure*}[h!t]
\includegraphics[
draft=false,scale=0.9,
trim={15mm 4mm 0 0.cm},
]{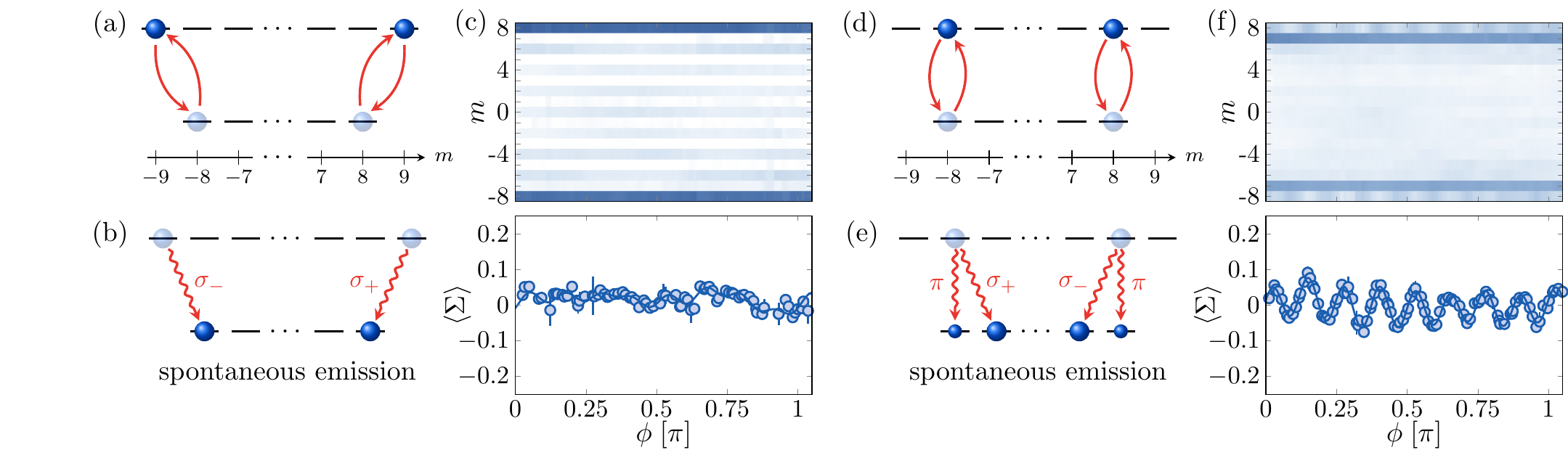}
\caption{
Loss of a qubit pair from superposition states.
(a) Preparation method for the Schr\"odinger cat state $\ket{\psi_1}$  in the excited electronic level. Given the small values of their Clebsch-Gordan coefficients, we neglect the couplings between $\ket{m=\pm8}$ to $\ket{m'=\pm7}$. (b) Scheme of the subsequent spontaneous emission.  (c) Top panel: spin projection probabilities measured in the $xy$ plane, as a function of the azimuthal angle $\phi$. The corresponding sign observable $\langle\Sigma\rangle$ is shown below, together with a fit with a Fourier series. The $y$ axis range has been reduced compared to \fig{fig_N00N}f to highlight the absence of oscillation. The panels (d,e,f) show the same information for the superposition state $\ket{\psi_2}=(\ket{m'=-8}+\ket{m'=8})/\sqrt2$.
\label{fig_decoherence_N00N}}
\end{figure*}

We contrast this behavior with the fragility of entanglement in coherent superpositions of states distant in phase space \cite{lee_quantification_2011}. 

We consider two examples, namely a cat state \mbox{$\ket{\psi_1}=(\ket{m^\prime=-J^\prime}+\ket{m^\prime=J^\prime})/\sqrt2$}, and the superposition \mbox{$\ket{\psi_2}=(\ket{m^\prime=-J^\prime+1}+\ket{m^\prime=J^\prime-1})/\sqrt2$}.
Their preparation consists in producing a cat state in the ground manifold $(\ket{m=-J}+\ket{m=J})/\sqrt2$ (see section \ref{section_cat}), and then applying resonant light to couple it to the excited manifold. The state $\ket{\psi_1}$ is produced using an $x$-linear polarization  $\ebold_x = (\ebold_++\ebold_-)/\sqrt{2} $, which dominantly couples the stretched states $\ket{m=\pm J}$ to states $\ket{m'=\pm J'}$ (see \fig{fig_decoherence_N00N}a). Couplings to states $\ket{m'=\pm(J'-2)}$ also occur, albeit with very small Clebsch-Gordan coefficients, such that these processes can be neglected \cite{Note10}. The state  $\ket{\psi_2}$ is obtained using a $z$-linear polarization (see \fig{fig_decoherence_N00N}d). In both cases, a coherent Rabi oscillation is observed when varying the pulse duration, and the fidelity of the preparation is limited by the one of the cat state in the ground level.
We show in Appendix \ref{appendix_cat_2pi} that the coherence of the superposition is maintained during Rabi flopping, by studying the states reached after $2\pi$ pulses.

We study the effect of qubit loss, triggered by spontaneous emission, on the superposition states $\ket{\psi_1}$ and $\ket{\psi_2}$. For the cat state $\ket{\psi_1}$, we only expect population of the stretched states $\ket{m=\pm J}$ (see \fig{fig_decoherence_N00N}b). To check the coherence between them, we measure the sign observable $\langle\Sigma\rangle$ as a function of the azimuthal angle $\phi$, as in section \ref{section_cat}. As shown in \fig{fig_decoherence_N00N}d, its oscillation is completely washed out, with a measured Fourier component $\Sigma_{2J}=0.006(10)$, indicating an absence of coherence. 
For the superposition state $\ket{\psi_2}$, we observe dominant projection probabilities in the states $\ket{m=\pm(J-1)}$, corresponding to the spontaneous emission of a $\sigma_\mp$ polarized photon, respectively (see \fig{fig_decoherence_N00N}f). We do not measure any significant variation of these probabilities with the azimuthal angle $\phi$, excluding coherence between them. We also measure residual projection probabilities in the stretched states $\ket{m=\pm J}$, which occur via the spontaneous emission of a $\pi$ polarized photon. The advantage of the sign observable $\Sigma$ becomes clear here: it allows one to test the coherence between the states $\ket{m=\pm J}$, without being perturbed by the atoms populating odd-$m$ states. The measured probabilities in stretched states coherently oscillate as a function of the angle $\phi$ (see \fig{fig_decoherence_N00N}h). More quantitatively, the sign observable, which involves even-$m$ only, evolves with a Fourier component $\Sigma_{2J}=0.024(1)$.

The complete loss of coherence when starting in the cat state $\ket{\psi_1}$ can be interpreted as follows. The spontaneous decay  involves two orthogonal polarizations, with a $\sigma_+$ polarized photon emitted when starting in the component $\ket{m'=-J'}$, while a $\sigma_-$ polarized photon is associated to the decay of the state $\ket{m'=J'}$ (see \fig{fig_decoherence_N00N}b). The photon polarization thus holds complete \emph{which path} information on the spin state polarization -- a term referring to Einstein's version of the double-slit interference experiment \cite{wootters_complementarity_1979,englert_fringe_1996}. In this case, the coherence between the different paths is erased after spontaneous emission. 

For the state $\ket{\psi_2}$, the most probable quantum jumps correspond to the emission of $\sigma_+$ and $\sigma_-$ polarized photons, which carry information about the state polarization (see \fig{fig_decoherence_N00N}e). On the contrary, the quantum jump associated with the emission of a $\pi$ polarized photon does not give this information, which explains the residual coherence. The measured Fourier coefficient $\Sigma_{2J}$ corresponds to 9.7(5)\% of the value measured in the absence of the excitation. This reduction is consistent with the probability $1/(J+1)\simeq11.1\%$ to scatter a $\pi$ polarized photon for the considered state, showing that this channel fully preserves coherence.

\section{Summary and outlook}

In this paper, we show that the $2J$-qubit ensemble associated with an atomic electronic spin $J$  can be partitioned  via the optical coupling to an excited level $J'=J-1$. Among these qubits, $2J-2$ of them constitute the excited level, and the remaining two are annihilated by the absorbed photon, in a state defined by the light polarization. We investigate this process using atomic dysprosium, and use it to probe entanglement in non-classical states of spin $J=8$. We fully characterize the non-classical character of its reduced two-qubit state, and study the increase of entropy upon partition as a smoking gun for entanglement. 

In a second set of experiments, we consider the partition of an angular momentum $J'=J+1$ of an excited electronic state. There, a random qubit pair is extracted by spontaneous emission towards the ground state $J$. We show that non-classical pairwise correlations are robust to particle loss. On the contrary, we observe that coherent superpositions of states distant in phase space are very fragile. 

In this work, the study of light-spin interaction is limited to measurements of the electronic spin. A first extension would be to collect the spontaneously emitted photon, whose polarization is entangled with the electronic spin, as for experiments performed with trapped ions, atoms in optical cavities or solid-state qubits \cite{blinov_observation_2004,volz_observation_2006,wilk_single-atom_2007,togan_quantum_2010}. One would thus explicitly access the \emph{which path} information carried by the photon upon spontaneous emission of a Schr\"odinger cat state. More generally, the photon would allow one to couple qubit pairs from the electronic spin $J=8$ to `flying qubits', which could then be manipulated to entangle distant atoms \cite{moehring_entanglement_2007}, and generalize quantum communication schemes to a mesoscopic degree of freedom \cite{cozzolino_high-dimensional_2019}. 

Another interesting perspective would be to place the atomic gas in an optical cavity. The electronic spin $J$ of a single atom would be coherently coupled to the cavity light mode, leading to a compound light-spin object \cite{brahms_spin_2010}. For a large set of atoms,   the cavity light would couple the electronic spins \cite{brennecke_cavity_2007,colombe_strong_2007} -- a process that can be interpreted as an exchange of qubits between collective spins. This mechanism could be used to create entangled states of many mesoscopic spins \cite{vitagliano_spin_2014}, which could feature a strong quantum enhancement of magnetic sensitivity \cite{norris_enhanced_2012-1}, or serve as a playground for studies of decoherence.

 \begin{acknowledgments}
 This work is supported by  European Union (ERC
 TOPODY 756722). We thank Jean Dalibard for stimulating discussions. 
 \end{acknowledgments}

\appendix

\section{Deviation to $z$ rotation symmetry\\in the W and cat states\label{appendix_phi_variation}}

\begin{figure}[!t]
\includegraphics[
draft=false,scale=1,
trim={2mm 2mm 0 0.cm},
]{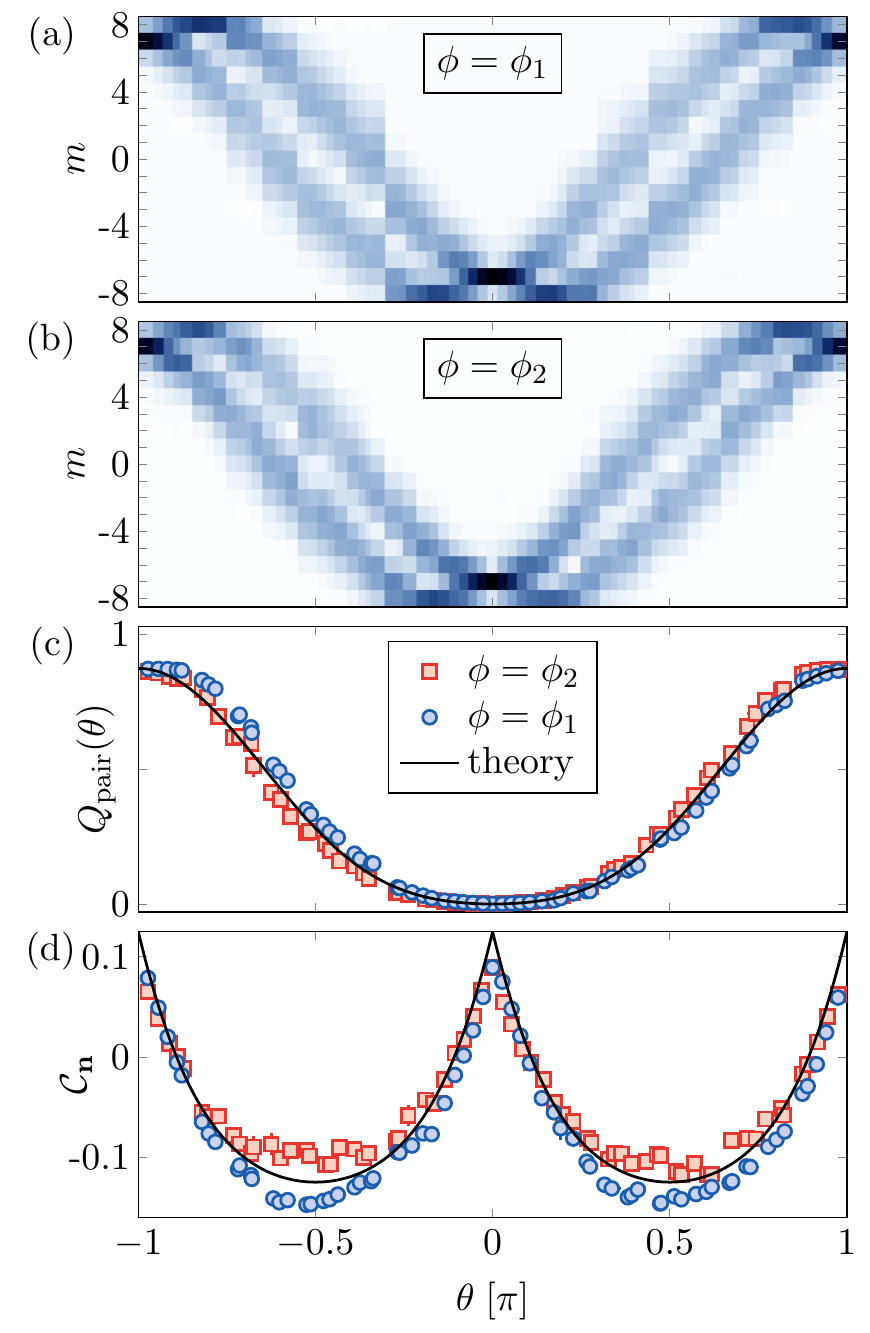}
\caption{
Deviation to $z$ rotation symmetry in the prepared W state.
(a,b) Projection probabilities $\Pi_m$ as a function of the polar angle $\theta$, for $\phi_1=\SI{0.36(5)}{\radian}$ and $\phi_2=\phi_1-\pi/2$.
(c) Pair Husimi functions $\Qpair$ inferred from the (a) and (b) data (blue disks and red squared, respectively). The error bars represent the statistical uncertainty from a bootstrap random sampling analysis. The line correspond to the expected variation for the  W state.
(d) Distribution $\mathcal{C}_{\nbold}$ as a function of $\theta$. The two azimuthal angles $\phi_1$ and $\phi_2$ are chosen to minimize and maximize the measured $\mathcal{C}_{\nbold}$, respectively.
\label{fig_W_phi_variation}}
\end{figure}

The $W$ state $\ket{m=-J+1}$ is invariant upon rotations around $z$, such that all observables should depend on the polar angle $\theta$ only. In practice, the state prepared close to the W state is not perfectly rotationally symmetric, because of a residual coherent admixture with other Dicke states. We measure a small $\phi$ variation of the measured probability distributions $\Pi_m(\nbold)$, as well as the pair Husimi function $\Qpair$ and distribution $\mathcal{C}_{\nbold}$ deduced from them. We show in \fig{fig_W_phi_variation} the measured data for two azimuthal angles $\phi_1=\SI{0.36(5)}{\radian}$ and $\phi_2=\phi_1-\pi/2$, for which  $\mathcal{C}_{\nbold}$ is minimized and maximized, respectively. The data shown in \fig{fig_Dicke} of the main text corresponds to an average over $\phi$, the error bars taking into account this dispersion.

\begin{figure}[!t]
\includegraphics[
draft=false,scale=1,
trim={2mm 2mm 0 0.cm},
]{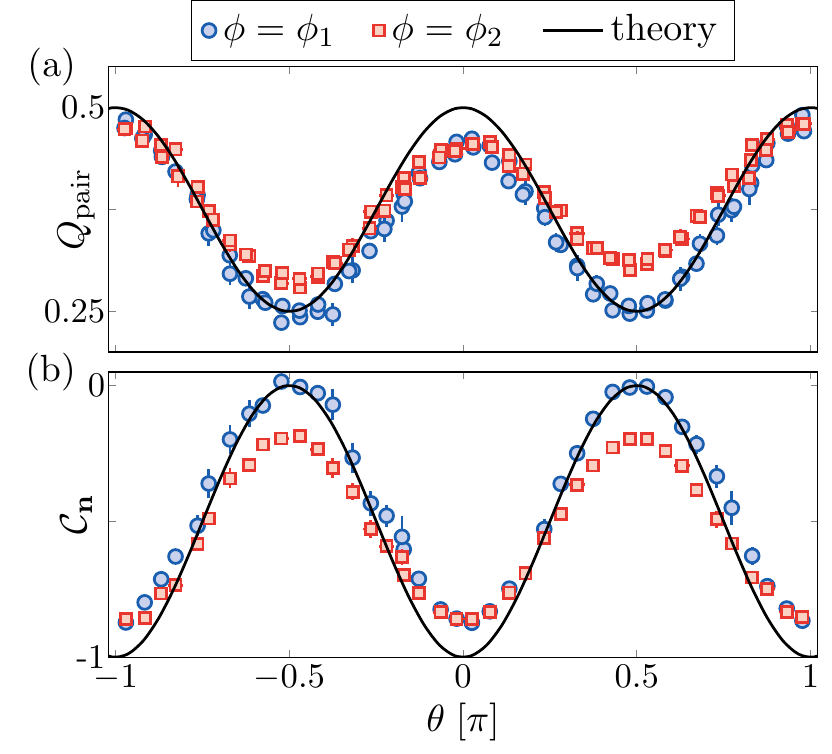}
\caption{
Deviation to $z$ rotation symmetry in the prepared Schr\"odinger cat state.
(a) Pair Husimi functions $\Qpair$ as a function of the polar angle $\theta$, for $\phi_1=\SI{3.3(1)}{\radian}$ and $\phi_2=\phi_1-\pi/2$ (blue disks and red squared, respectively). The line corresponds to the expected variation for a perfect cat state.
(b) Distribution $\mathcal{C}_{\nbold}$ as a function of $\theta$ deduced from the data in (a). 
\label{fig_N00N_phi_variation}}
\end{figure}

The cat state $\ket{m=-J}+\ket{m=J}$ is not rotationally invariant. Yet, its reduced two-body density matrix,  given by \eq{eq_rhopair_cat}, is invariant such that the pair Husimi function $\Qpair$ and distribution $\mathcal{C}_{\nbold}$ should depend on $\theta$ only. Like for the $W$ state, we measure a slight variation of these quantities with $\phi$, as shown in \fig{fig_N00N_phi_variation}. Since we focus on extracting the concurrence from the maximum of $\mathcal{C}_{\nbold}$, we show in the main text the data measured for an azimuthal angle $\phi_1=\SI{3.3(1)}{\radian}$ that maximizes  $\mathcal{C}_{\nbold}$.

\section{Spin-1 tomography\\using the pair Husimi function\label{appendix_tomography}}
The Husimi function of a spin-1 quantum state $\rho$ expands on the spherical harmonics $Y_{\ell}^m$ with $\ell=1,2$ and $|m|\leq \ell$, as written in \eq{eq_spherical_harmonics}. This linear decomposition allows us to retrieve the density matrix $\rho$, as given by \eq{eq_tomography}, where we introduce the operators
\begin{align}
 \mathcal{L}_{0}&=L_z,\\
 \mathcal{L}_{\pm 1}&=\mp(L_x\pm\I L_y)/\sqrt{2},\\
 \mathcal{Q}_0&=\sqrt{\frac{5}{3}}(3L_z^2-2),\\
 \mathcal{Q}_{\pm 1}&=\mp\sqrt{\frac{5}{2}}[(L_x\pm\I L_y)L_z+L_z(L_x\pm\I L_y)],\\
 \mathcal{Q}_{\pm 2}&=\sqrt{\frac{5}{2}}(L_x\pm\I L_y)^2.
\end{align}

\section{Coherence of superposition states during Rabi flopping\label{appendix_cat_2pi}}
The preparation of superposition states in the excited electronic state, as studied in section \ref{section_decoherence_cat},  uses coherent Rabi oscillations, starting in a Schr\"odinger cat state of the ground electronic level $(\ket{m=-J}+\ket{m=J})/\sqrt2$. To check that coherence is maintained during the Rabi oscillation, we study it after a $2\pi$ excitation, by measuring the oscillation of the sign observable $\langle\Sigma\rangle$.

\begin{figure}[!t]
\includegraphics[
draft=false,scale=0.88,
trim={2mm 2mm 0 0.cm},
]{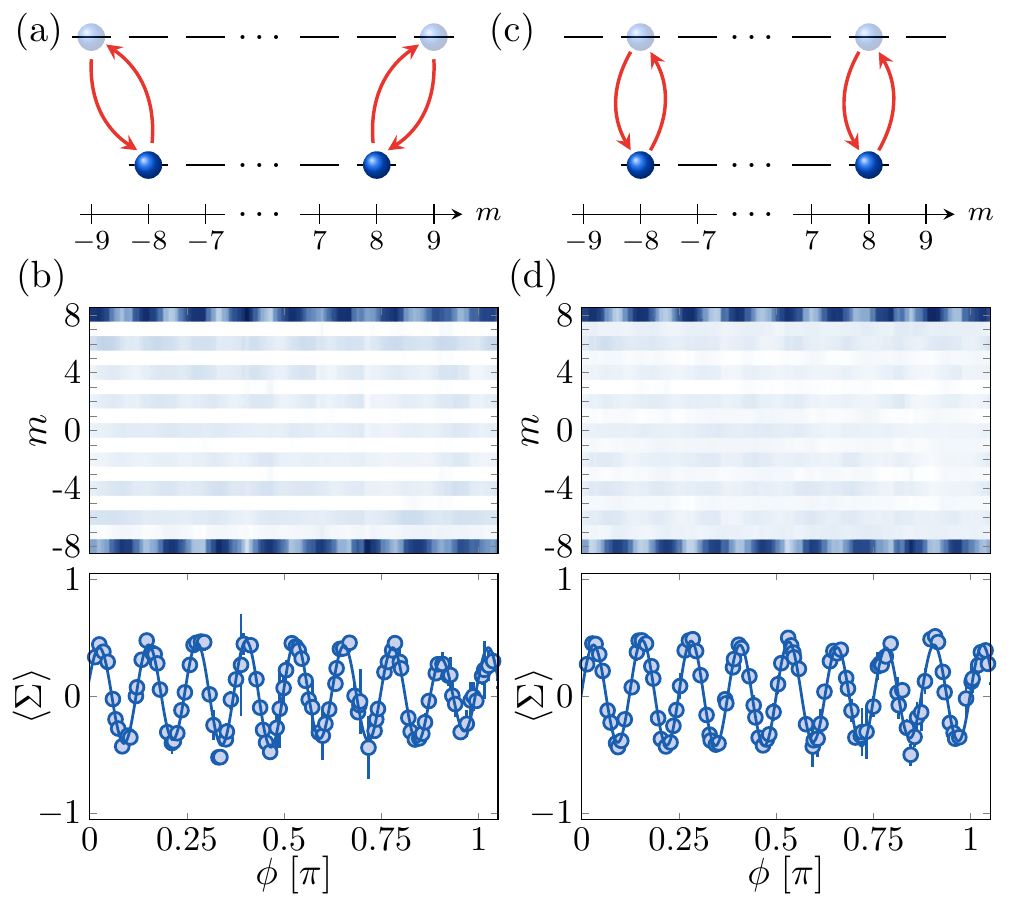}
\caption{
(a) Scheme of the $2\pi$ Rabi oscillation starting in a Schr\"odinger cat state of the electronic ground level, for an $x$-polarized laser excitation. (b)  Top panel: spin projection probabilities measured in the $xy$ plane, as a function of the azimuthal angle $\phi$. The corresponding sign observable $\langle\Sigma\rangle$ is shown below, together with a fit with a Fourier series. The panels (c,d) show the same information for a $z$-polarized laser excitation.
\label{fig_decoherence_cat_2pi}}
\end{figure}

As shown in \fig{fig_decoherence_cat_2pi}, we find that the coherence $|\rho_{-J,J}|$, estimated by the Fourier component $\Sigma_{2J}$, is reduced to 0.202(2) (0.211(6)) for the $x$-polarized ($z$-polarized) excitation, i.e. above 80\% of the value obtained  with no Rabi pulse. These measurements confirm that coherence is preserved during the Rabi oscillation.

\footnotetext[1]{See Supplemental Material  for  .}

\footnotetext[2]{We checked the lack of coherence between the states $\ket{-J}$ and $\ket{-J+1}$ by measuring an absence of Rabi oscillations when connecting these states by a resonant radiofrequency field.}

\footnotetext[3]{Other optical transitions contribute to an additional background light shift, that we characterized independently. For the detuning $\Delta=2\pi\times\SI{1}{\giga\hertz}$ used in our experiments, the background light shift is negligible compared to the near-resonant contribution given by \eq{eq_light_shift}.}

\footnotetext[4]{We also expect a coupling to other magnetic sublevels, but the corresponding Clebsch-Gordan coefficients appear to be much weaker, such that the fidelity of the $\ket{\text{N00N}'}$ state preparation is dominated by the finite fidelity of the initial $\ket{\text{N00N}}$ state. }

\footnotetext[5]{The preparation of the W state being imperfect, we observe a slight variation of the distribution $\mathcal{C}_{\nbold}$ with respect to the azimuthal angle $\phi$, see Appendix \ref{appendix_phi_variation}. The data shown in  \fig{fig_Dicke}c,d is averaged with respect to $\phi$, and the $\phi$ variation is taken into account in the error bars. The dependence on $\phi$ of $\mathcal{C}_{\nbold}$  does not play a role in the evaluation of its maximum value, which occurs at the north pole $\theta=0$.}

\footnotetext[6]{In practice, since the light shift amplitudes strongly vary with $m$, we vary the pulse duration in the range $10-\SI{100}{\micro\second}$ and the detuning in the range $\SI{50}{\mega\hertz}-\SI{1}{\giga\hertz}$, in order to keep similar displacements for all $m$ states (except the dark states). For the smallest detunings, we take into account the corrections to the second-order light shifts.
The uncertainties in the laser beam waist $w=\SI{40(5)}{\micro\meter}$ and on the excited state lifetime $\tau\simeq\SI{11}{\micro\second}$ \cite{dzuba_dynamic_2011} lead to a systematic error. We correct an overall 20\% error using the constraint $\sum_m Q_m=(2J+1)/3$, which states that in a completely undetermined state, a symmetric qubit pair has 1/3 chance to be in $\ket{\uparrow\uparrow}$.}

\footnotetext[7]{The W-state fidelity is mostly limited by inelastic collisions between atoms, which redistribute the spin among neighboring $m$ states.}

\footnotetext[8]{While the W state is rotationally symmetric around $z$, we observe in the prepared state a small but significant variation of the projection probabilities with the azimuthal angle $\phi$. We show in \fig{fig_Dicke}b,c,d the $\phi$-averaged values of our measurements. The variation with $\phi$ of the measured quantities is taken into account in the error bars, and is explicitly shown in Appendix \ref{appendix_phi_variation}. }

\footnotetext[9]{In practice, we observe a small variation of the  measured $\mathcal{C}_{\nbold}$ with the azimuthal angle $\phi$, discussed in Appendix \ref{appendix_phi_variation}. We show in \fig{fig_N00N}b the variation of $\mathcal{C}_{\nbold}$ with the polar angle $\theta$ with a fixed angle $\phi=\SI{3.3(1)}{\radian}$, which maximizes the value of $\mathcal{C}_{\nbold}$.}

\footnotetext[10]{The coupling amplitudes between $\ket{m=\pm J}$ and $\ket{m'=\pm(J'-2)}$ are $1/\sqrt{153}$ smaller than the couplings between $\ket{m=\pm J}$ and $\ket{m'=\pm J'}$. When the population of $\ket{m'=\pm J'}$ is maximized, we expect a residual population of the states $\ket{m'=\pm(J'-2)}$ of 3\% due to these small couplings.}


\begin{thebibliography}{98}%
\makeatletter
\providecommand \@ifxundefined [1]{%
 \@ifx{#1\undefined}
}%
\providecommand \@ifnum [1]{%
 \ifnum #1\expandafter \@firstoftwo
 \else \expandafter \@secondoftwo
 \fi
}%
\providecommand \@ifx [1]{%
 \ifx #1\expandafter \@firstoftwo
 \else \expandafter \@secondoftwo
 \fi
}%
\providecommand \natexlab [1]{#1}%
\providecommand \enquote  [1]{``#1''}%
\providecommand \bibnamefont  [1]{#1}%
\providecommand \bibfnamefont [1]{#1}%
\providecommand \citenamefont [1]{#1}%
\providecommand \href@noop [0]{\@secondoftwo}%
\providecommand \href [0]{\begingroup \@sanitize@url \@href}%
\providecommand \@href[1]{\@@startlink{#1}\@@href}%
\providecommand \@@href[1]{\endgroup#1\@@endlink}%
\providecommand \@sanitize@url [0]{\catcode `\\12\catcode `\$12\catcode
  `\&12\catcode `\#12\catcode `\^12\catcode `\_12\catcode `\%12\relax}%
\providecommand \@@startlink[1]{}%
\providecommand \@@endlink[0]{}%
\providecommand \url  [0]{\begingroup\@sanitize@url \@url }%
\providecommand \@url [1]{\endgroup\@href {#1}{\urlprefix }}%
\providecommand \urlprefix  [0]{URL }%
\providecommand \Eprint [0]{\href }%
\providecommand \doibase [0]{https://doi.org/}%
\providecommand \selectlanguage [0]{\@gobble}%
\providecommand \bibinfo  [0]{\@secondoftwo}%
\providecommand \bibfield  [0]{\@secondoftwo}%
\providecommand \translation [1]{[#1]}%
\providecommand \BibitemOpen [0]{}%
\providecommand \bibitemStop [0]{}%
\providecommand \bibitemNoStop [0]{.\EOS\space}%
\providecommand \EOS [0]{\spacefactor3000\relax}%
\providecommand \BibitemShut  [1]{\csname bibitem#1\endcsname}%
\let\auto@bib@innerbib\@empty
\bibitem [{\citenamefont {Freedman}\ and\ \citenamefont
  {Clauser}(1972)}]{freedman_experimental_1972}%
  \BibitemOpen
  \bibfield  {author} {\bibinfo {author} {\bibfnamefont {S.~J.}\ \bibnamefont
  {Freedman}}\ and\ \bibinfo {author} {\bibfnamefont {J.~F.}\ \bibnamefont
  {Clauser}},\ }\bibfield  {title} {\bibinfo {title} {Experimental {{Test}} of
  {{Local Hidden}}-{{Variable Theories}}},\ }\href@noop {} {\bibfield
  {journal} {\bibinfo  {journal} {Phys. Rev. Lett.}\ }\textbf {\bibinfo
  {volume} {28}},\ \bibinfo {pages} {938} (\bibinfo {year} {1972})}\BibitemShut
  {NoStop}%
\bibitem [{\citenamefont {Aspect}\ \emph {et~al.}(1982)\citenamefont {Aspect},
  \citenamefont {Dalibard},\ and\ \citenamefont
  {Roger}}]{aspect_experimental_1982}%
  \BibitemOpen
  \bibfield  {author} {\bibinfo {author} {\bibfnamefont {A.}~\bibnamefont
  {Aspect}}, \bibinfo {author} {\bibfnamefont {J.}~\bibnamefont {Dalibard}},\
  and\ \bibinfo {author} {\bibfnamefont {G.}~\bibnamefont {Roger}},\ }\bibfield
   {title} {\bibinfo {title} {Experimental {{Test}} of {{Bell}}'s
  {{Inequalities Using Time}}-{{Varying Analyzers}}},\ }\href@noop {}
  {\bibfield  {journal} {\bibinfo  {journal} {Phys. Rev. Lett.}\ }\textbf
  {\bibinfo {volume} {49}},\ \bibinfo {pages} {1804} (\bibinfo {year}
  {1982})}\BibitemShut {NoStop}%
\bibitem [{\citenamefont {Amico}\ \emph {et~al.}(2008)\citenamefont {Amico},
  \citenamefont {Fazio}, \citenamefont {Osterloh},\ and\ \citenamefont
  {Vedral}}]{amico_entanglement_2008}%
  \BibitemOpen
  \bibfield  {author} {\bibinfo {author} {\bibfnamefont {L.}~\bibnamefont
  {Amico}}, \bibinfo {author} {\bibfnamefont {R.}~\bibnamefont {Fazio}},
  \bibinfo {author} {\bibfnamefont {A.}~\bibnamefont {Osterloh}},\ and\
  \bibinfo {author} {\bibfnamefont {V.}~\bibnamefont {Vedral}},\ }\bibfield
  {title} {\bibinfo {title} {Entanglement in many-body systems},\ }\href@noop
  {} {\bibfield  {journal} {\bibinfo  {journal} {Rev. Mod. Phys.}\ }\textbf
  {\bibinfo {volume} {80}},\ \bibinfo {pages} {517} (\bibinfo {year}
  {2008})}\BibitemShut {NoStop}%
\bibitem [{\citenamefont {Pezz{\`e}}\ \emph {et~al.}(2018)\citenamefont
  {Pezz{\`e}}, \citenamefont {Smerzi}, \citenamefont {Oberthaler},
  \citenamefont {Schmied},\ and\ \citenamefont
  {Treutlein}}]{pezze_quantum_2018}%
  \BibitemOpen
  \bibfield  {author} {\bibinfo {author} {\bibfnamefont {L.}~\bibnamefont
  {Pezz{\`e}}}, \bibinfo {author} {\bibfnamefont {A.}~\bibnamefont {Smerzi}},
  \bibinfo {author} {\bibfnamefont {M.~K.}\ \bibnamefont {Oberthaler}},
  \bibinfo {author} {\bibfnamefont {R.}~\bibnamefont {Schmied}},\ and\ \bibinfo
  {author} {\bibfnamefont {P.}~\bibnamefont {Treutlein}},\ }\bibfield  {title}
  {\bibinfo {title} {Quantum metrology with nonclassical states of atomic
  ensembles},\ }\href@noop {} {\bibfield  {journal} {\bibinfo  {journal} {Rev
  Mod Phys}\ }\textbf {\bibinfo {volume} {90}},\ \bibinfo {pages} {035005}
  (\bibinfo {year} {2018})}\BibitemShut {NoStop}%
\bibitem [{\citenamefont {S{\o}rensen}\ \emph {et~al.}(2001)\citenamefont
  {S{\o}rensen}, \citenamefont {Duan}, \citenamefont {Cirac},\ and\
  \citenamefont {Zoller}}]{sorensen_many-particle_2001}%
  \BibitemOpen
  \bibfield  {author} {\bibinfo {author} {\bibfnamefont {A.}~\bibnamefont
  {S{\o}rensen}}, \bibinfo {author} {\bibfnamefont {L.-M.}\ \bibnamefont
  {Duan}}, \bibinfo {author} {\bibfnamefont {J.~I.}\ \bibnamefont {Cirac}},\
  and\ \bibinfo {author} {\bibfnamefont {P.}~\bibnamefont {Zoller}},\
  }\bibfield  {title} {\bibinfo {title} {Many-particle entanglement with
  {{Bose}}\textendash{{Einstein}} condensates},\ }\href@noop {} {\bibfield
  {journal} {\bibinfo  {journal} {Nature}\ }\textbf {\bibinfo {volume} {409}},\
  \bibinfo {pages} {63} (\bibinfo {year} {2001})}\BibitemShut {NoStop}%
\bibitem [{\citenamefont {S{\o}rensen}\ and\ \citenamefont
  {M{\o}lmer}(2001)}]{sorensen_entanglement_2001}%
  \BibitemOpen
  \bibfield  {author} {\bibinfo {author} {\bibfnamefont {A.~S.}\ \bibnamefont
  {S{\o}rensen}}\ and\ \bibinfo {author} {\bibfnamefont {K.}~\bibnamefont
  {M{\o}lmer}},\ }\bibfield  {title} {\bibinfo {title} {Entanglement and
  {{Extreme Spin Squeezing}}},\ }\href@noop {} {\bibfield  {journal} {\bibinfo
  {journal} {Phys. Rev. Lett.}\ }\textbf {\bibinfo {volume} {86}},\ \bibinfo
  {pages} {4431} (\bibinfo {year} {2001})}\BibitemShut {NoStop}%
\bibitem [{\citenamefont {Esteve}\ \emph {et~al.}(2008)\citenamefont {Esteve},
  \citenamefont {Gross}, \citenamefont {Weller}, \citenamefont {Giovanazzi},\
  and\ \citenamefont {Oberthaler}}]{esteve_squeezing_2008}%
  \BibitemOpen
  \bibfield  {author} {\bibinfo {author} {\bibfnamefont {J.}~\bibnamefont
  {Esteve}}, \bibinfo {author} {\bibfnamefont {C.}~\bibnamefont {Gross}},
  \bibinfo {author} {\bibfnamefont {A.}~\bibnamefont {Weller}}, \bibinfo
  {author} {\bibfnamefont {S.}~\bibnamefont {Giovanazzi}},\ and\ \bibinfo
  {author} {\bibfnamefont {M.}~\bibnamefont {Oberthaler}},\ }\bibfield  {title}
  {\bibinfo {title} {Squeezing and entanglement in a
  {{Bose}}\textendash{{Einstein}} condensate},\ }\href@noop {} {\bibfield
  {journal} {\bibinfo  {journal} {Nature}\ }\textbf {\bibinfo {volume} {455}},\
  \bibinfo {pages} {1216} (\bibinfo {year} {2008})}\BibitemShut {NoStop}%
\bibitem [{\citenamefont {Riedel}\ \emph {et~al.}(2010)\citenamefont {Riedel},
  \citenamefont {B{\"o}hi}, \citenamefont {Li}, \citenamefont {H{\"a}nsch},
  \citenamefont {Sinatra},\ and\ \citenamefont
  {Treutlein}}]{riedel_atom-chip-based_2010}%
  \BibitemOpen
  \bibfield  {author} {\bibinfo {author} {\bibfnamefont {M.~F.}\ \bibnamefont
  {Riedel}}, \bibinfo {author} {\bibfnamefont {P.}~\bibnamefont {B{\"o}hi}},
  \bibinfo {author} {\bibfnamefont {Y.}~\bibnamefont {Li}}, \bibinfo {author}
  {\bibfnamefont {T.~W.}\ \bibnamefont {H{\"a}nsch}}, \bibinfo {author}
  {\bibfnamefont {A.}~\bibnamefont {Sinatra}},\ and\ \bibinfo {author}
  {\bibfnamefont {P.}~\bibnamefont {Treutlein}},\ }\bibfield  {title} {\bibinfo
  {title} {Atom-chip-based generation of entanglement for quantum metrology},\
  }\href@noop {} {\bibfield  {journal} {\bibinfo  {journal} {Nature}\ }\textbf
  {\bibinfo {volume} {464}},\ \bibinfo {pages} {1170} (\bibinfo {year}
  {2010})}\BibitemShut {NoStop}%
\bibitem [{\citenamefont {Gross}\ \emph {et~al.}(2010)\citenamefont {Gross},
  \citenamefont {Zibold}, \citenamefont {Nicklas}, \citenamefont {Est{\`e}ve},\
  and\ \citenamefont {Oberthaler}}]{gross_nonlinear_2010}%
  \BibitemOpen
  \bibfield  {author} {\bibinfo {author} {\bibfnamefont {C.}~\bibnamefont
  {Gross}}, \bibinfo {author} {\bibfnamefont {T.}~\bibnamefont {Zibold}},
  \bibinfo {author} {\bibfnamefont {E.}~\bibnamefont {Nicklas}}, \bibinfo
  {author} {\bibfnamefont {J.}~\bibnamefont {Est{\`e}ve}},\ and\ \bibinfo
  {author} {\bibfnamefont {M.~K.}\ \bibnamefont {Oberthaler}},\ }\bibfield
  {title} {\bibinfo {title} {Nonlinear atom interferometer surpasses classical
  precision limit},\ }\href@noop {} {\bibfield  {journal} {\bibinfo  {journal}
  {Nature}\ }\textbf {\bibinfo {volume} {464}},\ \bibinfo {pages} {1165}
  (\bibinfo {year} {2010})}\BibitemShut {NoStop}%
\bibitem [{\citenamefont {Hyllus}\ \emph {et~al.}(2012)\citenamefont {Hyllus},
  \citenamefont {Laskowski}, \citenamefont {Krischek}, \citenamefont
  {Schwemmer}, \citenamefont {Wieczorek}, \citenamefont {Weinfurter},
  \citenamefont {Pezz{\'e}},\ and\ \citenamefont
  {Smerzi}}]{hyllus_fisher_2012}%
  \BibitemOpen
  \bibfield  {author} {\bibinfo {author} {\bibfnamefont {P.}~\bibnamefont
  {Hyllus}}, \bibinfo {author} {\bibfnamefont {W.}~\bibnamefont {Laskowski}},
  \bibinfo {author} {\bibfnamefont {R.}~\bibnamefont {Krischek}}, \bibinfo
  {author} {\bibfnamefont {C.}~\bibnamefont {Schwemmer}}, \bibinfo {author}
  {\bibfnamefont {W.}~\bibnamefont {Wieczorek}}, \bibinfo {author}
  {\bibfnamefont {H.}~\bibnamefont {Weinfurter}}, \bibinfo {author}
  {\bibfnamefont {L.}~\bibnamefont {Pezz{\'e}}},\ and\ \bibinfo {author}
  {\bibfnamefont {A.}~\bibnamefont {Smerzi}},\ }\bibfield  {title} {\bibinfo
  {title} {Fisher information and multiparticle entanglement},\ }\href@noop {}
  {\bibfield  {journal} {\bibinfo  {journal} {Phys. Rev. A}\ }\textbf {\bibinfo
  {volume} {85}},\ \bibinfo {pages} {022321} (\bibinfo {year}
  {2012})}\BibitemShut {NoStop}%
\bibitem [{\citenamefont {T{\'o}th}(2012)}]{toth_multipartite_2012}%
  \BibitemOpen
  \bibfield  {author} {\bibinfo {author} {\bibfnamefont {G.}~\bibnamefont
  {T{\'o}th}},\ }\bibfield  {title} {\bibinfo {title} {Multipartite
  entanglement and high-precision metrology},\ }\href@noop {} {\bibfield
  {journal} {\bibinfo  {journal} {Phys. Rev. A}\ }\textbf {\bibinfo {volume}
  {85}},\ \bibinfo {pages} {022322} (\bibinfo {year} {2012})}\BibitemShut
  {NoStop}%
\bibitem [{\citenamefont {Pan}\ \emph {et~al.}(2012)\citenamefont {Pan},
  \citenamefont {Chen}, \citenamefont {Lu}, \citenamefont {Weinfurter},
  \citenamefont {Zeilinger},\ and\ \citenamefont
  {{\.Z}ukowski}}]{pan_multiphoton_2012}%
  \BibitemOpen
  \bibfield  {author} {\bibinfo {author} {\bibfnamefont {J.-W.}\ \bibnamefont
  {Pan}}, \bibinfo {author} {\bibfnamefont {Z.-B.}\ \bibnamefont {Chen}},
  \bibinfo {author} {\bibfnamefont {C.-Y.}\ \bibnamefont {Lu}}, \bibinfo
  {author} {\bibfnamefont {H.}~\bibnamefont {Weinfurter}}, \bibinfo {author}
  {\bibfnamefont {A.}~\bibnamefont {Zeilinger}},\ and\ \bibinfo {author}
  {\bibfnamefont {M.}~\bibnamefont {{\.Z}ukowski}},\ }\bibfield  {title}
  {\bibinfo {title} {Multiphoton entanglement and interferometry},\ }\href@noop
  {} {\bibfield  {journal} {\bibinfo  {journal} {Rev. Mod. Phys.}\ }\textbf
  {\bibinfo {volume} {84}},\ \bibinfo {pages} {777} (\bibinfo {year}
  {2012})}\BibitemShut {NoStop}%
\bibitem [{\citenamefont {Wendin}(2017)}]{wendin_quantum_2017}%
  \BibitemOpen
  \bibfield  {author} {\bibinfo {author} {\bibfnamefont {G.}~\bibnamefont
  {Wendin}},\ }\bibfield  {title} {\bibinfo {title} {Quantum information
  processing with superconducting circuits: A review},\ }\href@noop {}
  {\bibfield  {journal} {\bibinfo  {journal} {Rep. Prog. Phys.}\ }\textbf
  {\bibinfo {volume} {80}},\ \bibinfo {pages} {106001} (\bibinfo {year}
  {2017})}\BibitemShut {NoStop}%
\bibitem [{\citenamefont {Blatt}\ and\ \citenamefont
  {Wineland}(2008)}]{blatt_entangled_2008}%
  \BibitemOpen
  \bibfield  {author} {\bibinfo {author} {\bibfnamefont {R.}~\bibnamefont
  {Blatt}}\ and\ \bibinfo {author} {\bibfnamefont {D.}~\bibnamefont
  {Wineland}},\ }\bibfield  {title} {\bibinfo {title} {Entangled states of
  trapped atomic ions},\ }\href@noop {} {\bibfield  {journal} {\bibinfo
  {journal} {Nature}\ }\textbf {\bibinfo {volume} {453}},\ \bibinfo {pages}
  {1008} (\bibinfo {year} {2008})}\BibitemShut {NoStop}%
\bibitem [{\citenamefont {Saffman}\ \emph {et~al.}(2010)\citenamefont
  {Saffman}, \citenamefont {Walker},\ and\ \citenamefont
  {M{\o}lmer}}]{saffman_quantum_2010}%
  \BibitemOpen
  \bibfield  {author} {\bibinfo {author} {\bibfnamefont {M.}~\bibnamefont
  {Saffman}}, \bibinfo {author} {\bibfnamefont {T.~G.}\ \bibnamefont
  {Walker}},\ and\ \bibinfo {author} {\bibfnamefont {K.}~\bibnamefont
  {M{\o}lmer}},\ }\bibfield  {title} {\bibinfo {title} {Quantum information
  with {{Rydberg}} atoms},\ }\href@noop {} {\bibfield  {journal} {\bibinfo
  {journal} {Rev. Mod. Phys.}\ }\textbf {\bibinfo {volume} {82}},\ \bibinfo
  {pages} {2313} (\bibinfo {year} {2010})}\BibitemShut {NoStop}%
\bibitem [{\citenamefont {Horodecki}\ \emph {et~al.}(2009)\citenamefont
  {Horodecki}, \citenamefont {Horodecki}, \citenamefont {Horodecki},\ and\
  \citenamefont {Horodecki}}]{horodecki_quantum_2009}%
  \BibitemOpen
  \bibfield  {author} {\bibinfo {author} {\bibfnamefont {R.}~\bibnamefont
  {Horodecki}}, \bibinfo {author} {\bibfnamefont {P.}~\bibnamefont
  {Horodecki}}, \bibinfo {author} {\bibfnamefont {M.}~\bibnamefont
  {Horodecki}},\ and\ \bibinfo {author} {\bibfnamefont {K.}~\bibnamefont
  {Horodecki}},\ }\bibfield  {title} {\bibinfo {title} {Quantum entanglement},\
  }\href@noop {} {\bibfield  {journal} {\bibinfo  {journal} {Rev Mod Phys}\
  }\textbf {\bibinfo {volume} {81}},\ \bibinfo {pages} {865} (\bibinfo {year}
  {2009})}\BibitemShut {NoStop}%
\bibitem [{\citenamefont {G{\"u}hne}\ and\ \citenamefont
  {T{\'o}th}(2009)}]{guhne_entanglement_2009}%
  \BibitemOpen
  \bibfield  {author} {\bibinfo {author} {\bibfnamefont {O.}~\bibnamefont
  {G{\"u}hne}}\ and\ \bibinfo {author} {\bibfnamefont {G.}~\bibnamefont
  {T{\'o}th}},\ }\bibfield  {title} {\bibinfo {title} {Entanglement
  detection},\ }\href@noop {} {\bibfield  {journal} {\bibinfo  {journal}
  {Physics Reports}\ }\textbf {\bibinfo {volume} {474}},\ \bibinfo {pages} {1}
  (\bibinfo {year} {2009})}\BibitemShut {NoStop}%
\bibitem [{\citenamefont {Hill}\ and\ \citenamefont
  {Wootters}(1997)}]{hill_entanglement_1997}%
  \BibitemOpen
  \bibfield  {author} {\bibinfo {author} {\bibfnamefont {S.}~\bibnamefont
  {Hill}}\ and\ \bibinfo {author} {\bibfnamefont {W.~K.}\ \bibnamefont
  {Wootters}},\ }\bibfield  {title} {\bibinfo {title} {Entanglement of a
  {{Pair}} of {{Quantum Bits}}},\ }\href@noop {} {\bibfield  {journal}
  {\bibinfo  {journal} {Phys. Rev. Lett.}\ }\textbf {\bibinfo {volume} {78}},\
  \bibinfo {pages} {5022} (\bibinfo {year} {1997})}\BibitemShut {NoStop}%
\bibitem [{\citenamefont {Wootters}(1998)}]{wootters_entanglement_1998}%
  \BibitemOpen
  \bibfield  {author} {\bibinfo {author} {\bibfnamefont {W.~K.}\ \bibnamefont
  {Wootters}},\ }\bibfield  {title} {\bibinfo {title} {Entanglement of
  {{Formation}} of an {{Arbitrary State}} of {{Two Qubits}}},\ }\href@noop {}
  {\bibfield  {journal} {\bibinfo  {journal} {Phys. Rev. Lett.}\ }\textbf
  {\bibinfo {volume} {80}},\ \bibinfo {pages} {2245} (\bibinfo {year}
  {1998})}\BibitemShut {NoStop}%
\bibitem [{\citenamefont {Bovino}\ \emph {et~al.}(2005)\citenamefont {Bovino},
  \citenamefont {Castagnoli}, \citenamefont {Ekert}, \citenamefont {Horodecki},
  \citenamefont {Alves},\ and\ \citenamefont {Sergienko}}]{bovino_direct_2005}%
  \BibitemOpen
  \bibfield  {author} {\bibinfo {author} {\bibfnamefont {F.~A.}\ \bibnamefont
  {Bovino}}, \bibinfo {author} {\bibfnamefont {G.}~\bibnamefont {Castagnoli}},
  \bibinfo {author} {\bibfnamefont {A.}~\bibnamefont {Ekert}}, \bibinfo
  {author} {\bibfnamefont {P.}~\bibnamefont {Horodecki}}, \bibinfo {author}
  {\bibfnamefont {C.~M.}\ \bibnamefont {Alves}},\ and\ \bibinfo {author}
  {\bibfnamefont {A.~V.}\ \bibnamefont {Sergienko}},\ }\bibfield  {title}
  {\bibinfo {title} {Direct {{Measurement}} of {{Nonlinear Properties}} of
  {{Bipartite Quantum States}}},\ }\href@noop {} {\bibfield  {journal}
  {\bibinfo  {journal} {Phys. Rev. Lett.}\ }\textbf {\bibinfo {volume} {95}},\
  \bibinfo {pages} {240407} (\bibinfo {year} {2005})}\BibitemShut {NoStop}%
\bibitem [{\citenamefont {Walborn}\ \emph {et~al.}(2006)\citenamefont
  {Walborn}, \citenamefont {Souto~Ribeiro}, \citenamefont {Davidovich},
  \citenamefont {Mintert},\ and\ \citenamefont
  {Buchleitner}}]{walborn_experimental_2006}%
  \BibitemOpen
  \bibfield  {author} {\bibinfo {author} {\bibfnamefont {S.~P.}\ \bibnamefont
  {Walborn}}, \bibinfo {author} {\bibfnamefont {P.~H.}\ \bibnamefont
  {Souto~Ribeiro}}, \bibinfo {author} {\bibfnamefont {L.}~\bibnamefont
  {Davidovich}}, \bibinfo {author} {\bibfnamefont {F.}~\bibnamefont
  {Mintert}},\ and\ \bibinfo {author} {\bibfnamefont {A.}~\bibnamefont
  {Buchleitner}},\ }\bibfield  {title} {\bibinfo {title} {Experimental
  determination of entanglement with a single measurement},\ }\href@noop {}
  {\bibfield  {journal} {\bibinfo  {journal} {Nature}\ }\textbf {\bibinfo
  {volume} {440}},\ \bibinfo {pages} {1022} (\bibinfo {year}
  {2006})}\BibitemShut {NoStop}%
\bibitem [{\citenamefont {Schmid}\ \emph {et~al.}(2008)\citenamefont {Schmid},
  \citenamefont {Kiesel}, \citenamefont {Wieczorek}, \citenamefont
  {Weinfurter}, \citenamefont {Mintert},\ and\ \citenamefont
  {Buchleitner}}]{schmid_experimental_2008}%
  \BibitemOpen
  \bibfield  {author} {\bibinfo {author} {\bibfnamefont {C.}~\bibnamefont
  {Schmid}}, \bibinfo {author} {\bibfnamefont {N.}~\bibnamefont {Kiesel}},
  \bibinfo {author} {\bibfnamefont {W.}~\bibnamefont {Wieczorek}}, \bibinfo
  {author} {\bibfnamefont {H.}~\bibnamefont {Weinfurter}}, \bibinfo {author}
  {\bibfnamefont {F.}~\bibnamefont {Mintert}},\ and\ \bibinfo {author}
  {\bibfnamefont {A.}~\bibnamefont {Buchleitner}},\ }\bibfield  {title}
  {\bibinfo {title} {Experimental {{Direct Observation}} of {{Mixed State
  Entanglement}}},\ }\href@noop {} {\bibfield  {journal} {\bibinfo  {journal}
  {Phys. Rev. Lett.}\ }\textbf {\bibinfo {volume} {101}},\ \bibinfo {pages}
  {260505} (\bibinfo {year} {2008})}\BibitemShut {NoStop}%
\bibitem [{\citenamefont {Islam}\ \emph {et~al.}(2015)\citenamefont {Islam},
  \citenamefont {Ma}, \citenamefont {Preiss}, \citenamefont {Eric~Tai},
  \citenamefont {Lukin}, \citenamefont {Rispoli},\ and\ \citenamefont
  {Greiner}}]{islam_measuring_2015}%
  \BibitemOpen
  \bibfield  {author} {\bibinfo {author} {\bibfnamefont {R.}~\bibnamefont
  {Islam}}, \bibinfo {author} {\bibfnamefont {R.}~\bibnamefont {Ma}}, \bibinfo
  {author} {\bibfnamefont {P.~M.}\ \bibnamefont {Preiss}}, \bibinfo {author}
  {\bibfnamefont {M.}~\bibnamefont {Eric~Tai}}, \bibinfo {author}
  {\bibfnamefont {A.}~\bibnamefont {Lukin}}, \bibinfo {author} {\bibfnamefont
  {M.}~\bibnamefont {Rispoli}},\ and\ \bibinfo {author} {\bibfnamefont
  {M.}~\bibnamefont {Greiner}},\ }\bibfield  {title} {\bibinfo {title}
  {Measuring entanglement entropy in a quantum many-body system},\ }\href@noop
  {} {\bibfield  {journal} {\bibinfo  {journal} {Nature}\ }\textbf {\bibinfo
  {volume} {528}},\ \bibinfo {pages} {77} (\bibinfo {year} {2015})}\BibitemShut
  {NoStop}%
\bibitem [{\citenamefont {D{\"u}r}\ \emph {et~al.}(2000)\citenamefont
  {D{\"u}r}, \citenamefont {Vidal},\ and\ \citenamefont
  {Cirac}}]{dur_three_2000}%
  \BibitemOpen
  \bibfield  {author} {\bibinfo {author} {\bibfnamefont {W.}~\bibnamefont
  {D{\"u}r}}, \bibinfo {author} {\bibfnamefont {G.}~\bibnamefont {Vidal}},\
  and\ \bibinfo {author} {\bibfnamefont {J.~I.}\ \bibnamefont {Cirac}},\
  }\bibfield  {title} {\bibinfo {title} {Three qubits can be entangled in two
  inequivalent ways},\ }\href@noop {} {\bibfield  {journal} {\bibinfo
  {journal} {Phys. Rev. A}\ }\textbf {\bibinfo {volume} {62}},\ \bibinfo
  {pages} {062314} (\bibinfo {year} {2000})}\BibitemShut {NoStop}%
\bibitem [{\citenamefont {T{\'o}th}\ and\ \citenamefont
  {G{\"u}hne}(2005)}]{toth_detecting_2005}%
  \BibitemOpen
  \bibfield  {author} {\bibinfo {author} {\bibfnamefont {G.}~\bibnamefont
  {T{\'o}th}}\ and\ \bibinfo {author} {\bibfnamefont {O.}~\bibnamefont
  {G{\"u}hne}},\ }\bibfield  {title} {\bibinfo {title} {Detecting {{Genuine
  Multipartite Entanglement}} with {{Two Local Measurements}}},\ }\href@noop {}
  {\bibfield  {journal} {\bibinfo  {journal} {Phys. Rev. Lett.}\ }\textbf
  {\bibinfo {volume} {94}},\ \bibinfo {pages} {060501} (\bibinfo {year}
  {2005})}\BibitemShut {NoStop}%
\bibitem [{\citenamefont {Lapkiewicz}\ \emph {et~al.}(2011)\citenamefont
  {Lapkiewicz}, \citenamefont {Li}, \citenamefont {Schaeff}, \citenamefont
  {Langford}, \citenamefont {Ramelow}, \citenamefont {Wie{\'s}niak},\ and\
  \citenamefont {Zeilinger}}]{lapkiewicz_experimental_2011}%
  \BibitemOpen
  \bibfield  {author} {\bibinfo {author} {\bibfnamefont {R.}~\bibnamefont
  {Lapkiewicz}}, \bibinfo {author} {\bibfnamefont {P.}~\bibnamefont {Li}},
  \bibinfo {author} {\bibfnamefont {C.}~\bibnamefont {Schaeff}}, \bibinfo
  {author} {\bibfnamefont {N.~K.}\ \bibnamefont {Langford}}, \bibinfo {author}
  {\bibfnamefont {S.}~\bibnamefont {Ramelow}}, \bibinfo {author} {\bibfnamefont
  {M.}~\bibnamefont {Wie{\'s}niak}},\ and\ \bibinfo {author} {\bibfnamefont
  {A.}~\bibnamefont {Zeilinger}},\ }\bibfield  {title} {\bibinfo {title}
  {Experimental non-classicality of an indivisible quantum system},\
  }\href@noop {} {\bibfield  {journal} {\bibinfo  {journal} {Nature}\ }\textbf
  {\bibinfo {volume} {474}},\ \bibinfo {pages} {490} (\bibinfo {year}
  {2011})}\BibitemShut {NoStop}%
\bibitem [{\citenamefont {Chaudhury}\ \emph {et~al.}(2007)\citenamefont
  {Chaudhury}, \citenamefont {Merkel}, \citenamefont {Herr}, \citenamefont
  {Silberfarb}, \citenamefont {Deutsch},\ and\ \citenamefont
  {Jessen}}]{chaudhury_quantum_2007}%
  \BibitemOpen
  \bibfield  {author} {\bibinfo {author} {\bibfnamefont {S.}~\bibnamefont
  {Chaudhury}}, \bibinfo {author} {\bibfnamefont {S.}~\bibnamefont {Merkel}},
  \bibinfo {author} {\bibfnamefont {T.}~\bibnamefont {Herr}}, \bibinfo {author}
  {\bibfnamefont {A.}~\bibnamefont {Silberfarb}}, \bibinfo {author}
  {\bibfnamefont {I.~H.}\ \bibnamefont {Deutsch}},\ and\ \bibinfo {author}
  {\bibfnamefont {P.~S.}\ \bibnamefont {Jessen}},\ }\bibfield  {title}
  {\bibinfo {title} {Quantum {{Control}} of the {{Hyperfine Spin}} of a {{Cs
  Atom Ensemble}}},\ }\href@noop {} {\bibfield  {journal} {\bibinfo  {journal}
  {Phys Rev Lett}\ }\textbf {\bibinfo {volume} {99}},\ \bibinfo {pages}
  {163002} (\bibinfo {year} {2007})}\BibitemShut {NoStop}%
\bibitem [{\citenamefont {Fernholz}\ \emph {et~al.}(2008)\citenamefont
  {Fernholz}, \citenamefont {Krauter}, \citenamefont {Jensen}, \citenamefont
  {Sherson}, \citenamefont {S{\o}rensen},\ and\ \citenamefont
  {Polzik}}]{fernholz_spin_2008}%
  \BibitemOpen
  \bibfield  {author} {\bibinfo {author} {\bibfnamefont {T.}~\bibnamefont
  {Fernholz}}, \bibinfo {author} {\bibfnamefont {H.}~\bibnamefont {Krauter}},
  \bibinfo {author} {\bibfnamefont {K.}~\bibnamefont {Jensen}}, \bibinfo
  {author} {\bibfnamefont {J.~F.}\ \bibnamefont {Sherson}}, \bibinfo {author}
  {\bibfnamefont {A.~S.}\ \bibnamefont {S{\o}rensen}},\ and\ \bibinfo {author}
  {\bibfnamefont {E.~S.}\ \bibnamefont {Polzik}},\ }\bibfield  {title}
  {\bibinfo {title} {Spin {{Squeezing}} of {{Atomic Ensembles}} via
  {{Nuclear}}-{{Electronic Spin Entanglement}}},\ }\href@noop {} {\bibfield
  {journal} {\bibinfo  {journal} {Phys Rev Lett}\ }\textbf {\bibinfo {volume}
  {101}},\ \bibinfo {pages} {073601} (\bibinfo {year} {2008})}\BibitemShut
  {NoStop}%
\bibitem [{\citenamefont {Gatteschi}\ and\ \citenamefont
  {Sessoli}(2003)}]{gatteschi_quantum_2003}%
  \BibitemOpen
  \bibfield  {author} {\bibinfo {author} {\bibfnamefont {D.}~\bibnamefont
  {Gatteschi}}\ and\ \bibinfo {author} {\bibfnamefont {R.}~\bibnamefont
  {Sessoli}},\ }\bibfield  {title} {\bibinfo {title} {Quantum {{Tunneling}} of
  {{Magnetization}} and {{Related Phenomena}} in {{Molecular Materials}}},\
  }\href@noop {} {\bibfield  {journal} {\bibinfo  {journal} {Angew. Chem. Int.
  Ed.}\ }\textbf {\bibinfo {volume} {42}},\ \bibinfo {pages} {268} (\bibinfo
  {year} {2003})}\BibitemShut {NoStop}%
\bibitem [{\citenamefont {Facon}\ \emph {et~al.}(2016)\citenamefont {Facon},
  \citenamefont {Dietsche}, \citenamefont {Grosso}, \citenamefont {Haroche},
  \citenamefont {Raimond}, \citenamefont {Brune},\ and\ \citenamefont
  {Gleyzes}}]{facon_sensitive_2016}%
  \BibitemOpen
  \bibfield  {author} {\bibinfo {author} {\bibfnamefont {A.}~\bibnamefont
  {Facon}}, \bibinfo {author} {\bibfnamefont {E.-K.}\ \bibnamefont {Dietsche}},
  \bibinfo {author} {\bibfnamefont {D.}~\bibnamefont {Grosso}}, \bibinfo
  {author} {\bibfnamefont {S.}~\bibnamefont {Haroche}}, \bibinfo {author}
  {\bibfnamefont {J.-M.}\ \bibnamefont {Raimond}}, \bibinfo {author}
  {\bibfnamefont {M.}~\bibnamefont {Brune}},\ and\ \bibinfo {author}
  {\bibfnamefont {S.}~\bibnamefont {Gleyzes}},\ }\bibfield  {title} {\bibinfo
  {title} {A sensitive electrometer based on a {{Rydberg}} atom in a
  {{Schr\"odinger}}-cat state},\ }\href@noop {} {\bibfield  {journal} {\bibinfo
   {journal} {Nature}\ }\textbf {\bibinfo {volume} {535}},\ \bibinfo {pages}
  {262} (\bibinfo {year} {2016})}\BibitemShut {NoStop}%
\bibitem [{\citenamefont {Majorana}(1932)}]{majorana_atomi_1932}%
  \BibitemOpen
  \bibfield  {author} {\bibinfo {author} {\bibfnamefont {E.}~\bibnamefont
  {Majorana}},\ }\bibfield  {title} {\bibinfo {title} {Atomi orientati in campo
  magnetico variabile},\ }\href@noop {} {\bibfield  {journal} {\bibinfo
  {journal} {Il Nuovo Cimento 1924-1942}\ }\textbf {\bibinfo {volume} {9}},\
  \bibinfo {pages} {43} (\bibinfo {year} {1932})}\BibitemShut {NoStop}%
\bibitem [{Note6()}]{Note6}%
  \BibitemOpen
  \bibinfo {note} {In practice, since the light shift amplitudes strongly vary
  with $m$, we vary the pulse duration in the range $10-\SI {100}{\micro
  \second }$ and the detuning in the range $\SI {50}{\mega \hertz }-\SI
  {1}{\giga \hertz }$, in order to keep similar displacements for all $m$
  states (except the dark states). For the smallest detunings, we take into
  account the corrections to the second-order light shifts. The uncertainties
  in the laser beam waist $w=\SI {40(5)}{\micro \meter }$ and on the excited
  state lifetime $\tau \simeq \SI {11}{\micro \second }$ \cite
  {dzuba_dynamic_2011} lead to a systematic error. We correct an overall 20\%
  error using the constraint $\DOTSB \sum@ \slimits@ _m Q_m=(2J+1)/3$, which
  states that in a completely undetermined state, a symmetric qubit pair has
  1/3 chance to be in $\left | \protect \uparrow \protect \uparrow \right
  >$.}\BibitemShut {Stop}%
\bibitem [{\citenamefont {Dicke}(1954)}]{dicke_coherence_1954}%
  \BibitemOpen
  \bibfield  {author} {\bibinfo {author} {\bibfnamefont {R.~H.}\ \bibnamefont
  {Dicke}},\ }\bibfield  {title} {\bibinfo {title} {Coherence in {{Spontaneous
  Radiation Processes}}},\ }\href@noop {} {\bibfield  {journal} {\bibinfo
  {journal} {Phys. Rev.}\ }\textbf {\bibinfo {volume} {93}},\ \bibinfo {pages}
  {99} (\bibinfo {year} {1954})}\BibitemShut {NoStop}%
\bibitem [{\citenamefont {Choi}\ \emph {et~al.}(2010)\citenamefont {Choi},
  \citenamefont {Goban}, \citenamefont {Papp}, \citenamefont {{van Enk}},\ and\
  \citenamefont {Kimble}}]{choi_entanglement_2010}%
  \BibitemOpen
  \bibfield  {author} {\bibinfo {author} {\bibfnamefont {K.~S.}\ \bibnamefont
  {Choi}}, \bibinfo {author} {\bibfnamefont {A.}~\bibnamefont {Goban}},
  \bibinfo {author} {\bibfnamefont {S.~B.}\ \bibnamefont {Papp}}, \bibinfo
  {author} {\bibfnamefont {S.~J.}\ \bibnamefont {{van Enk}}},\ and\ \bibinfo
  {author} {\bibfnamefont {H.~J.}\ \bibnamefont {Kimble}},\ }\bibfield  {title}
  {\bibinfo {title} {Entanglement of spin waves among four quantum memories},\
  }\href@noop {} {\bibfield  {journal} {\bibinfo  {journal} {Nature}\ }\textbf
  {\bibinfo {volume} {468}},\ \bibinfo {pages} {412} (\bibinfo {year}
  {2010})}\BibitemShut {NoStop}%
\bibitem [{\citenamefont {Haas}\ \emph {et~al.}(2014)\citenamefont {Haas},
  \citenamefont {Volz}, \citenamefont {Gehr}, \citenamefont {Reichel},\ and\
  \citenamefont {Est{\`e}ve}}]{haas_entangled_2014}%
  \BibitemOpen
  \bibfield  {author} {\bibinfo {author} {\bibfnamefont {F.}~\bibnamefont
  {Haas}}, \bibinfo {author} {\bibfnamefont {J.}~\bibnamefont {Volz}}, \bibinfo
  {author} {\bibfnamefont {R.}~\bibnamefont {Gehr}}, \bibinfo {author}
  {\bibfnamefont {J.}~\bibnamefont {Reichel}},\ and\ \bibinfo {author}
  {\bibfnamefont {J.}~\bibnamefont {Est{\`e}ve}},\ }\bibfield  {title}
  {\bibinfo {title} {Entangled {{States}} of {{More Than}} 40 {{Atoms}} in an
  {{Optical Fiber Cavity}}},\ }\href@noop {} {\bibfield  {journal} {\bibinfo
  {journal} {Science}\ }\textbf {\bibinfo {volume} {344}},\ \bibinfo {pages}
  {180} (\bibinfo {year} {2014})}\BibitemShut {NoStop}%
\bibitem [{\citenamefont {McConnell}\ \emph {et~al.}(2015)\citenamefont
  {McConnell}, \citenamefont {Zhang}, \citenamefont {Hu}, \citenamefont
  {{\'C}uk},\ and\ \citenamefont {Vuleti{\'c}}}]{mcconnell_entanglement_2015}%
  \BibitemOpen
  \bibfield  {author} {\bibinfo {author} {\bibfnamefont {R.}~\bibnamefont
  {McConnell}}, \bibinfo {author} {\bibfnamefont {H.}~\bibnamefont {Zhang}},
  \bibinfo {author} {\bibfnamefont {J.}~\bibnamefont {Hu}}, \bibinfo {author}
  {\bibfnamefont {S.}~\bibnamefont {{\'C}uk}},\ and\ \bibinfo {author}
  {\bibfnamefont {V.}~\bibnamefont {Vuleti{\'c}}},\ }\bibfield  {title}
  {\bibinfo {title} {Entanglement with negative {{Wigner}} function of almost
  3,000 atoms heralded by one photon},\ }\href@noop {} {\bibfield  {journal}
  {\bibinfo  {journal} {Nature}\ }\textbf {\bibinfo {volume} {519}},\ \bibinfo
  {pages} {439} (\bibinfo {year} {2015})}\BibitemShut {NoStop}%
\bibitem [{\citenamefont {Ebert}\ \emph {et~al.}(2015)\citenamefont {Ebert},
  \citenamefont {Kwon}, \citenamefont {Walker},\ and\ \citenamefont
  {Saffman}}]{ebert_coherence_2015}%
  \BibitemOpen
  \bibfield  {author} {\bibinfo {author} {\bibfnamefont {M.}~\bibnamefont
  {Ebert}}, \bibinfo {author} {\bibfnamefont {M.}~\bibnamefont {Kwon}},
  \bibinfo {author} {\bibfnamefont {T.~G.}\ \bibnamefont {Walker}},\ and\
  \bibinfo {author} {\bibfnamefont {M.}~\bibnamefont {Saffman}},\ }\bibfield
  {title} {\bibinfo {title} {Coherence and {{Rydberg Blockade}} of {{Atomic
  Ensemble Qubits}}},\ }\href@noop {} {\bibfield  {journal} {\bibinfo
  {journal} {Phys. Rev. Lett.}\ }\textbf {\bibinfo {volume} {115}},\ \bibinfo
  {pages} {093601} (\bibinfo {year} {2015})}\BibitemShut {NoStop}%
\bibitem [{\citenamefont {Zeiher}\ \emph {et~al.}(2015)\citenamefont {Zeiher},
  \citenamefont {Schau{\ss}}, \citenamefont {Hild}, \citenamefont {Macr{\`i}},
  \citenamefont {Bloch},\ and\ \citenamefont
  {Gross}}]{zeiher_microscopic_2015}%
  \BibitemOpen
  \bibfield  {author} {\bibinfo {author} {\bibfnamefont {J.}~\bibnamefont
  {Zeiher}}, \bibinfo {author} {\bibfnamefont {P.}~\bibnamefont {Schau{\ss}}},
  \bibinfo {author} {\bibfnamefont {S.}~\bibnamefont {Hild}}, \bibinfo {author}
  {\bibfnamefont {T.}~\bibnamefont {Macr{\`i}}}, \bibinfo {author}
  {\bibfnamefont {I.}~\bibnamefont {Bloch}},\ and\ \bibinfo {author}
  {\bibfnamefont {C.}~\bibnamefont {Gross}},\ }\bibfield  {title} {\bibinfo
  {title} {Microscopic {{Characterization}} of {{Scalable Coherent Rydberg
  Superatoms}}},\ }\href@noop {} {\bibfield  {journal} {\bibinfo  {journal}
  {Phys. Rev. X}\ }\textbf {\bibinfo {volume} {5}},\ \bibinfo {pages} {031015}
  (\bibinfo {year} {2015})}\BibitemShut {NoStop}%
\bibitem [{\citenamefont {H{\"a}ffner}\ \emph {et~al.}(2005)\citenamefont
  {H{\"a}ffner}, \citenamefont {H{\"a}nsel}, \citenamefont {Roos},
  \citenamefont {Benhelm}, \citenamefont {{Chek-al-kar}}, \citenamefont
  {Chwalla}, \citenamefont {K{\"o}rber}, \citenamefont {Rapol}, \citenamefont
  {Riebe}, \citenamefont {Schmidt}, \citenamefont {Becher}, \citenamefont
  {G{\"u}hne}, \citenamefont {D{\"u}r},\ and\ \citenamefont
  {Blatt}}]{haffner_scalable_2005}%
  \BibitemOpen
  \bibfield  {author} {\bibinfo {author} {\bibfnamefont {H.}~\bibnamefont
  {H{\"a}ffner}}, \bibinfo {author} {\bibfnamefont {W.}~\bibnamefont
  {H{\"a}nsel}}, \bibinfo {author} {\bibfnamefont {C.~F.}\ \bibnamefont
  {Roos}}, \bibinfo {author} {\bibfnamefont {J.}~\bibnamefont {Benhelm}},
  \bibinfo {author} {\bibfnamefont {D.}~\bibnamefont {{Chek-al-kar}}}, \bibinfo
  {author} {\bibfnamefont {M.}~\bibnamefont {Chwalla}}, \bibinfo {author}
  {\bibfnamefont {T.}~\bibnamefont {K{\"o}rber}}, \bibinfo {author}
  {\bibfnamefont {U.~D.}\ \bibnamefont {Rapol}}, \bibinfo {author}
  {\bibfnamefont {M.}~\bibnamefont {Riebe}}, \bibinfo {author} {\bibfnamefont
  {P.~O.}\ \bibnamefont {Schmidt}}, \bibinfo {author} {\bibfnamefont
  {C.}~\bibnamefont {Becher}}, \bibinfo {author} {\bibfnamefont
  {O.}~\bibnamefont {G{\"u}hne}}, \bibinfo {author} {\bibfnamefont
  {W.}~\bibnamefont {D{\"u}r}},\ and\ \bibinfo {author} {\bibfnamefont
  {R.}~\bibnamefont {Blatt}},\ }\bibfield  {title} {\bibinfo {title} {Scalable
  multiparticle entanglement of trapped ions},\ }\href@noop {} {\bibfield
  {journal} {\bibinfo  {journal} {Nature}\ }\textbf {\bibinfo {volume} {438}},\
  \bibinfo {pages} {643} (\bibinfo {year} {2005})}\BibitemShut {NoStop}%
\bibitem [{\citenamefont {Fr{\"o}wis}\ \emph {et~al.}(2017)\citenamefont
  {Fr{\"o}wis}, \citenamefont {Strassmann}, \citenamefont {Tiranov},
  \citenamefont {Gut}, \citenamefont {Lavoie}, \citenamefont {Brunner},
  \citenamefont {Bussi{\`e}res}, \citenamefont {Afzelius},\ and\ \citenamefont
  {Gisin}}]{frowis_experimental_2017}%
  \BibitemOpen
  \bibfield  {author} {\bibinfo {author} {\bibfnamefont {F.}~\bibnamefont
  {Fr{\"o}wis}}, \bibinfo {author} {\bibfnamefont {P.~C.}\ \bibnamefont
  {Strassmann}}, \bibinfo {author} {\bibfnamefont {A.}~\bibnamefont {Tiranov}},
  \bibinfo {author} {\bibfnamefont {C.}~\bibnamefont {Gut}}, \bibinfo {author}
  {\bibfnamefont {J.}~\bibnamefont {Lavoie}}, \bibinfo {author} {\bibfnamefont
  {N.}~\bibnamefont {Brunner}}, \bibinfo {author} {\bibfnamefont
  {F.}~\bibnamefont {Bussi{\`e}res}}, \bibinfo {author} {\bibfnamefont
  {M.}~\bibnamefont {Afzelius}},\ and\ \bibinfo {author} {\bibfnamefont
  {N.}~\bibnamefont {Gisin}},\ }\bibfield  {title} {\bibinfo {title}
  {Experimental certification of millions of genuinely entangled atoms in a
  solid},\ }\href@noop {} {\bibfield  {journal} {\bibinfo  {journal} {Nat.
  Commun.}\ }\textbf {\bibinfo {volume} {8}},\ \bibinfo {pages} {907} (\bibinfo
  {year} {2017})}\BibitemShut {NoStop}%
\bibitem [{\citenamefont {Pu}\ \emph {et~al.}(2018)\citenamefont {Pu},
  \citenamefont {Wu}, \citenamefont {Jiang}, \citenamefont {Chang},
  \citenamefont {Li}, \citenamefont {Zhang},\ and\ \citenamefont
  {Duan}}]{pu_experimental_2018}%
  \BibitemOpen
  \bibfield  {author} {\bibinfo {author} {\bibfnamefont {Y.}~\bibnamefont
  {Pu}}, \bibinfo {author} {\bibfnamefont {Y.}~\bibnamefont {Wu}}, \bibinfo
  {author} {\bibfnamefont {N.}~\bibnamefont {Jiang}}, \bibinfo {author}
  {\bibfnamefont {W.}~\bibnamefont {Chang}}, \bibinfo {author} {\bibfnamefont
  {C.}~\bibnamefont {Li}}, \bibinfo {author} {\bibfnamefont {S.}~\bibnamefont
  {Zhang}},\ and\ \bibinfo {author} {\bibfnamefont {L.}~\bibnamefont {Duan}},\
  }\bibfield  {title} {\bibinfo {title} {Experimental entanglement of 25
  individually accessible atomic quantum interfaces},\ }\href@noop {}
  {\bibfield  {journal} {\bibinfo  {journal} {Sci. Adv.}\ }\textbf {\bibinfo
  {volume} {4}},\ \bibinfo {pages} {3931} (\bibinfo {year} {2018})}\BibitemShut
  {NoStop}%
\bibitem [{Note7()}]{Note7}%
  \BibitemOpen
  \bibinfo {note} {The W-state fidelity is mostly limited by inelastic
  collisions between atoms, which redistribute the spin among neighboring $m$
  states.}\BibitemShut {Stop}%
\bibitem [{\citenamefont {Sudarshan}(1963)}]{sudarshan_equivalence_1963}%
  \BibitemOpen
  \bibfield  {author} {\bibinfo {author} {\bibfnamefont {E.~C.~G.}\
  \bibnamefont {Sudarshan}},\ }\bibfield  {title} {\bibinfo {title}
  {Equivalence of {{Semiclassical}} and {{Quantum Mechanical Descriptions}} of
  {{Statistical Light Beams}}},\ }\href@noop {} {\bibfield  {journal} {\bibinfo
   {journal} {Phys. Rev. Lett.}\ }\textbf {\bibinfo {volume} {10}},\ \bibinfo
  {pages} {277} (\bibinfo {year} {1963})}\BibitemShut {NoStop}%
\bibitem [{\citenamefont {Glauber}(1963)}]{glauber_coherent_1963}%
  \BibitemOpen
  \bibfield  {author} {\bibinfo {author} {\bibfnamefont {R.~J.}\ \bibnamefont
  {Glauber}},\ }\bibfield  {title} {\bibinfo {title} {Coherent and {{Incoherent
  States}} of the {{Radiation Field}}},\ }\href@noop {} {\bibfield  {journal}
  {\bibinfo  {journal} {Phys. Rev.}\ }\textbf {\bibinfo {volume} {131}},\
  \bibinfo {pages} {2766} (\bibinfo {year} {1963})}\BibitemShut {NoStop}%
\bibitem [{\citenamefont {Giraud}\ \emph {et~al.}(2008)\citenamefont {Giraud},
  \citenamefont {Braun},\ and\ \citenamefont
  {Braun}}]{giraud_classicality_2008}%
  \BibitemOpen
  \bibfield  {author} {\bibinfo {author} {\bibfnamefont {O.}~\bibnamefont
  {Giraud}}, \bibinfo {author} {\bibfnamefont {P.}~\bibnamefont {Braun}},\ and\
  \bibinfo {author} {\bibfnamefont {D.}~\bibnamefont {Braun}},\ }\bibfield
  {title} {\bibinfo {title} {Classicality of spin states},\ }\href@noop {}
  {\bibfield  {journal} {\bibinfo  {journal} {Phys. Rev. A}\ }\textbf {\bibinfo
  {volume} {78}},\ \bibinfo {pages} {042112} (\bibinfo {year}
  {2008})}\BibitemShut {NoStop}%
\bibitem [{\citenamefont {Korbicz}\ \emph {et~al.}(2005)\citenamefont
  {Korbicz}, \citenamefont {Cirac},\ and\ \citenamefont
  {Lewenstein}}]{korbicz_spin_2005}%
  \BibitemOpen
  \bibfield  {author} {\bibinfo {author} {\bibfnamefont {J.~K.}\ \bibnamefont
  {Korbicz}}, \bibinfo {author} {\bibfnamefont {J.~I.}\ \bibnamefont {Cirac}},\
  and\ \bibinfo {author} {\bibfnamefont {M.}~\bibnamefont {Lewenstein}},\
  }\bibfield  {title} {\bibinfo {title} {Spin {{Squeezing Inequalities}} and
  {{Entanglement}} of \${{N}}\$ {{Qubit States}}},\ }\href@noop {} {\bibfield
  {journal} {\bibinfo  {journal} {Phys. Rev. Lett.}\ }\textbf {\bibinfo
  {volume} {95}},\ \bibinfo {pages} {120502} (\bibinfo {year}
  {2005})}\BibitemShut {NoStop}%
\bibitem [{Note8()}]{Note8}%
  \BibitemOpen
  \bibinfo {note} {While the W state is rotationally symmetric around $z$, we
  observe in the prepared state a small but significant variation of the
  projection probabilities with the azimuthal angle $\phi $. We show in
  Fig.\protect \tmspace +\thinmuskip {.1667em}\ref {fig_Dicke}b,c,d the $\phi
  $-averaged values of our measurements. The variation with $\phi $ of the
  measured quantities is taken into account in the error bars, and is
  explicitly shown in Appendix \ref {appendix_phi_variation}.}\BibitemShut
  {Stop}%
\bibitem [{\citenamefont {Hillery}(1987)}]{hillery_nonclassical_1987}%
  \BibitemOpen
  \bibfield  {author} {\bibinfo {author} {\bibfnamefont {M.}~\bibnamefont
  {Hillery}},\ }\bibfield  {title} {\bibinfo {title} {Nonclassical distance in
  quantum optics},\ }\href@noop {} {\bibfield  {journal} {\bibinfo  {journal}
  {Phys. Rev. A}\ }\textbf {\bibinfo {volume} {35}},\ \bibinfo {pages} {725}
  (\bibinfo {year} {1987})}\BibitemShut {NoStop}%
\bibitem [{\citenamefont {Wei}\ and\ \citenamefont
  {Goldbart}(2003)}]{wei_geometric_2003}%
  \BibitemOpen
  \bibfield  {author} {\bibinfo {author} {\bibfnamefont {T.-C.}\ \bibnamefont
  {Wei}}\ and\ \bibinfo {author} {\bibfnamefont {P.~M.}\ \bibnamefont
  {Goldbart}},\ }\bibfield  {title} {\bibinfo {title} {Geometric measure of
  entanglement and applications to bipartite and multipartite quantum states},\
  }\href@noop {} {\bibfield  {journal} {\bibinfo  {journal} {Phys. Rev. A}\
  }\textbf {\bibinfo {volume} {68}},\ \bibinfo {pages} {042307} (\bibinfo
  {year} {2003})}\BibitemShut {NoStop}%
\bibitem [{\citenamefont {Vidal}(2006)}]{vidal_concurrence_2006}%
  \BibitemOpen
  \bibfield  {author} {\bibinfo {author} {\bibfnamefont {J.}~\bibnamefont
  {Vidal}},\ }\bibfield  {title} {\bibinfo {title} {Concurrence in collective
  models},\ }\href@noop {} {\bibfield  {journal} {\bibinfo  {journal} {Phys.
  Rev. A}\ }\textbf {\bibinfo {volume} {73}},\ \bibinfo {pages} {062318}
  (\bibinfo {year} {2006})}\BibitemShut {NoStop}%
\bibitem [{\citenamefont {Koashi}\ \emph {et~al.}(2000)\citenamefont {Koashi},
  \citenamefont {Bu{\v z}ek},\ and\ \citenamefont
  {Imoto}}]{koashi_entangled_2000}%
  \BibitemOpen
  \bibfield  {author} {\bibinfo {author} {\bibfnamefont {M.}~\bibnamefont
  {Koashi}}, \bibinfo {author} {\bibfnamefont {V.}~\bibnamefont {Bu{\v z}ek}},\
  and\ \bibinfo {author} {\bibfnamefont {N.}~\bibnamefont {Imoto}},\ }\bibfield
   {title} {\bibinfo {title} {Entangled webs: {{Tight}} bound for symmetric
  sharing of entanglement},\ }\href@noop {} {\bibfield  {journal} {\bibinfo
  {journal} {Phys. Rev. A}\ }\textbf {\bibinfo {volume} {62}},\ \bibinfo
  {pages} {050302} (\bibinfo {year} {2000})}\BibitemShut {NoStop}%
\bibitem [{\citenamefont {Kitagawa}\ and\ \citenamefont
  {Ueda}(1993)}]{kitagawa_squeezed_1993}%
  \BibitemOpen
  \bibfield  {author} {\bibinfo {author} {\bibfnamefont {M.}~\bibnamefont
  {Kitagawa}}\ and\ \bibinfo {author} {\bibfnamefont {M.}~\bibnamefont
  {Ueda}},\ }\bibfield  {title} {\bibinfo {title} {Squeezed spin states},\
  }\href@noop {} {\bibfield  {journal} {\bibinfo  {journal} {Phys Rev A}\
  }\textbf {\bibinfo {volume} {47}},\ \bibinfo {pages} {5138} (\bibinfo {year}
  {1993})}\BibitemShut {NoStop}%
\bibitem [{\citenamefont {Chalopin}\ \emph {et~al.}(2018)\citenamefont
  {Chalopin}, \citenamefont {Bouazza}, \citenamefont {Evrard}, \citenamefont
  {Makhalov}, \citenamefont {Dreon}, \citenamefont {Dalibard}, \citenamefont
  {Sidorenkov},\ and\ \citenamefont
  {Nascimbene}}]{chalopin_quantum-enhanced_2018}%
  \BibitemOpen
  \bibfield  {author} {\bibinfo {author} {\bibfnamefont {T.}~\bibnamefont
  {Chalopin}}, \bibinfo {author} {\bibfnamefont {C.}~\bibnamefont {Bouazza}},
  \bibinfo {author} {\bibfnamefont {A.}~\bibnamefont {Evrard}}, \bibinfo
  {author} {\bibfnamefont {V.}~\bibnamefont {Makhalov}}, \bibinfo {author}
  {\bibfnamefont {D.}~\bibnamefont {Dreon}}, \bibinfo {author} {\bibfnamefont
  {J.}~\bibnamefont {Dalibard}}, \bibinfo {author} {\bibfnamefont {L.~A.}\
  \bibnamefont {Sidorenkov}},\ and\ \bibinfo {author} {\bibfnamefont
  {S.}~\bibnamefont {Nascimbene}},\ }\bibfield  {title} {\bibinfo {title}
  {Quantum-enhanced sensing using non-classical spin states of a highly
  magnetic atom},\ }\href@noop {} {\bibfield  {journal} {\bibinfo  {journal}
  {Nat. Commun.}\ }\textbf {\bibinfo {volume} {9}},\ \bibinfo {pages} {4955}
  (\bibinfo {year} {2018})}\BibitemShut {NoStop}%
\bibitem [{\citenamefont {Wang}\ and\ \citenamefont
  {Sanders}(2003)}]{wang_spin_2003}%
  \BibitemOpen
  \bibfield  {author} {\bibinfo {author} {\bibfnamefont {X.}~\bibnamefont
  {Wang}}\ and\ \bibinfo {author} {\bibfnamefont {B.~C.}\ \bibnamefont
  {Sanders}},\ }\bibfield  {title} {\bibinfo {title} {Spin squeezing and
  pairwise entanglement for symmetric multiqubit states},\ }\href@noop {}
  {\bibfield  {journal} {\bibinfo  {journal} {Phys. Rev. A}\ }\textbf {\bibinfo
  {volume} {68}},\ \bibinfo {pages} {012101} (\bibinfo {year}
  {2003})}\BibitemShut {NoStop}%
\bibitem [{\citenamefont {Horodecki}\ and\ \citenamefont
  {Horodecki}(1996)}]{horodecki_information-theoretic_1996}%
  \BibitemOpen
  \bibfield  {author} {\bibinfo {author} {\bibfnamefont {R.}~\bibnamefont
  {Horodecki}}\ and\ \bibinfo {author} {\bibfnamefont {M.}~\bibnamefont
  {Horodecki}},\ }\bibfield  {title} {\bibinfo {title} {Information-theoretic
  aspects of inseparability of mixed states},\ }\href@noop {} {\bibfield
  {journal} {\bibinfo  {journal} {Phys. Rev. A}\ }\textbf {\bibinfo {volume}
  {54}},\ \bibinfo {pages} {1838} (\bibinfo {year} {1996})}\BibitemShut
  {NoStop}%
\bibitem [{\citenamefont {Konig}\ \emph {et~al.}(2009)\citenamefont {Konig},
  \citenamefont {Renner},\ and\ \citenamefont
  {Schaffner}}]{konig_operational_2009}%
  \BibitemOpen
  \bibfield  {author} {\bibinfo {author} {\bibfnamefont {R.}~\bibnamefont
  {Konig}}, \bibinfo {author} {\bibfnamefont {R.}~\bibnamefont {Renner}},\ and\
  \bibinfo {author} {\bibfnamefont {C.}~\bibnamefont {Schaffner}},\ }\bibfield
  {title} {\bibinfo {title} {The {{Operational Meaning}} of {{Min}}- and
  {{Max}}-{{Entropy}}},\ }\href@noop {} {\bibfield  {journal} {\bibinfo
  {journal} {IEEE Trans. Inf. Theory}\ }\textbf {\bibinfo {volume} {55}},\
  \bibinfo {pages} {4337} (\bibinfo {year} {2009})}\BibitemShut {NoStop}%
\bibitem [{\citenamefont {Man'ko}\ and\ \citenamefont
  {Man'ko}(1997)}]{manko_spin_1997}%
  \BibitemOpen
  \bibfield  {author} {\bibinfo {author} {\bibfnamefont {V.~I.}\ \bibnamefont
  {Man'ko}}\ and\ \bibinfo {author} {\bibfnamefont {O.~V.}\ \bibnamefont
  {Man'ko}},\ }\bibfield  {title} {\bibinfo {title} {Spin state tomography},\
  }\href@noop {} {\bibfield  {journal} {\bibinfo  {journal} {J. Exp. Theor.
  Phys.}\ }\textbf {\bibinfo {volume} {85}},\ \bibinfo {pages} {430} (\bibinfo
  {year} {1997})}\BibitemShut {NoStop}%
\bibitem [{\citenamefont {Monroe}\ \emph {et~al.}(1996)\citenamefont {Monroe},
  \citenamefont {Meekhof}, \citenamefont {King},\ and\ \citenamefont
  {Wineland}}]{monroe_schrodinger_1996}%
  \BibitemOpen
  \bibfield  {author} {\bibinfo {author} {\bibfnamefont {C.}~\bibnamefont
  {Monroe}}, \bibinfo {author} {\bibfnamefont {D.}~\bibnamefont {Meekhof}},
  \bibinfo {author} {\bibfnamefont {B.}~\bibnamefont {King}},\ and\ \bibinfo
  {author} {\bibfnamefont {D.~J.}\ \bibnamefont {Wineland}},\ }\bibfield
  {title} {\bibinfo {title} {A ``{{Schr\"odinger Cat}}'' superposition state of
  an atom},\ }\href@noop {} {\bibfield  {journal} {\bibinfo  {journal}
  {Science}\ }\textbf {\bibinfo {volume} {272}},\ \bibinfo {pages} {1131}
  (\bibinfo {year} {1996})}\BibitemShut {NoStop}%
\bibitem [{\citenamefont {Brune}\ \emph {et~al.}(1996)\citenamefont {Brune},
  \citenamefont {Hagley}, \citenamefont {Dreyer}, \citenamefont {Maitre},
  \citenamefont {Maali}, \citenamefont {Wunderlich}, \citenamefont {Raimond},\
  and\ \citenamefont {Haroche}}]{brune_observing_1996}%
  \BibitemOpen
  \bibfield  {author} {\bibinfo {author} {\bibfnamefont {M.}~\bibnamefont
  {Brune}}, \bibinfo {author} {\bibfnamefont {E.}~\bibnamefont {Hagley}},
  \bibinfo {author} {\bibfnamefont {J.}~\bibnamefont {Dreyer}}, \bibinfo
  {author} {\bibfnamefont {X.}~\bibnamefont {Maitre}}, \bibinfo {author}
  {\bibfnamefont {A.}~\bibnamefont {Maali}}, \bibinfo {author} {\bibfnamefont
  {C.}~\bibnamefont {Wunderlich}}, \bibinfo {author} {\bibfnamefont
  {J.}~\bibnamefont {Raimond}},\ and\ \bibinfo {author} {\bibfnamefont
  {S.}~\bibnamefont {Haroche}},\ }\bibfield  {title} {\bibinfo {title}
  {Observing the progressive decoherence of the ``meter'' in a quantum
  measurement},\ }\href@noop {} {\bibfield  {journal} {\bibinfo  {journal}
  {Phys. Rev. Lett.}\ }\textbf {\bibinfo {volume} {77}},\ \bibinfo {pages}
  {4887} (\bibinfo {year} {1996})}\BibitemShut {NoStop}%
\bibitem [{\citenamefont {Friedman}\ \emph {et~al.}(2000)\citenamefont
  {Friedman}, \citenamefont {Patel}, \citenamefont {Chen}, \citenamefont
  {Tolpygo},\ and\ \citenamefont {Lukens}}]{friedman_quantum_2000}%
  \BibitemOpen
  \bibfield  {author} {\bibinfo {author} {\bibfnamefont {J.~R.}\ \bibnamefont
  {Friedman}}, \bibinfo {author} {\bibfnamefont {V.}~\bibnamefont {Patel}},
  \bibinfo {author} {\bibfnamefont {W.}~\bibnamefont {Chen}}, \bibinfo {author}
  {\bibfnamefont {S.}~\bibnamefont {Tolpygo}},\ and\ \bibinfo {author}
  {\bibfnamefont {J.~E.}\ \bibnamefont {Lukens}},\ }\bibfield  {title}
  {\bibinfo {title} {Quantum superposition of distinct macroscopic states},\
  }\href@noop {} {\bibfield  {journal} {\bibinfo  {journal} {nature}\ }\textbf
  {\bibinfo {volume} {406}},\ \bibinfo {pages} {43} (\bibinfo {year}
  {2000})}\BibitemShut {NoStop}%
\bibitem [{\citenamefont {Sackett}\ \emph {et~al.}(2000)\citenamefont
  {Sackett}, \citenamefont {Kielpinski}, \citenamefont {King}, \citenamefont
  {Langer}, \citenamefont {Meyer}, \citenamefont {Myatt}, \citenamefont {Rowe},
  \citenamefont {Turchette}, \citenamefont {Itano}, \citenamefont {Wineland},\
  and\ \citenamefont {Monroe}}]{sackett_experimental_2000}%
  \BibitemOpen
  \bibfield  {author} {\bibinfo {author} {\bibfnamefont {C.~A.}\ \bibnamefont
  {Sackett}}, \bibinfo {author} {\bibfnamefont {D.}~\bibnamefont {Kielpinski}},
  \bibinfo {author} {\bibfnamefont {B.~E.}\ \bibnamefont {King}}, \bibinfo
  {author} {\bibfnamefont {C.}~\bibnamefont {Langer}}, \bibinfo {author}
  {\bibfnamefont {V.}~\bibnamefont {Meyer}}, \bibinfo {author} {\bibfnamefont
  {C.~J.}\ \bibnamefont {Myatt}}, \bibinfo {author} {\bibfnamefont
  {M.}~\bibnamefont {Rowe}}, \bibinfo {author} {\bibfnamefont {Q.~A.}\
  \bibnamefont {Turchette}}, \bibinfo {author} {\bibfnamefont {W.~M.}\
  \bibnamefont {Itano}}, \bibinfo {author} {\bibfnamefont {D.~J.}\ \bibnamefont
  {Wineland}},\ and\ \bibinfo {author} {\bibfnamefont {C.}~\bibnamefont
  {Monroe}},\ }\bibfield  {title} {\bibinfo {title} {Experimental entanglement
  of four particles},\ }\href@noop {} {\bibfield  {journal} {\bibinfo
  {journal} {Nature}\ }\textbf {\bibinfo {volume} {404}},\ \bibinfo {pages}
  {256} (\bibinfo {year} {2000})}\BibitemShut {NoStop}%
\bibitem [{\citenamefont {Leibfried}\ \emph {et~al.}(2005)\citenamefont
  {Leibfried}, \citenamefont {Knill}, \citenamefont {Seidelin}, \citenamefont
  {Britton}, \citenamefont {Blakestad}, \citenamefont {Chiaverini},
  \citenamefont {Hume}, \citenamefont {Itano}, \citenamefont {Jost},
  \citenamefont {Langer}, \citenamefont {Ozeri}, \citenamefont {Reichle},\ and\
  \citenamefont {Wineland}}]{leibfried_creation_2005}%
  \BibitemOpen
  \bibfield  {author} {\bibinfo {author} {\bibfnamefont {D.}~\bibnamefont
  {Leibfried}}, \bibinfo {author} {\bibfnamefont {E.}~\bibnamefont {Knill}},
  \bibinfo {author} {\bibfnamefont {S.}~\bibnamefont {Seidelin}}, \bibinfo
  {author} {\bibfnamefont {J.}~\bibnamefont {Britton}}, \bibinfo {author}
  {\bibfnamefont {R.~B.}\ \bibnamefont {Blakestad}}, \bibinfo {author}
  {\bibfnamefont {J.}~\bibnamefont {Chiaverini}}, \bibinfo {author}
  {\bibfnamefont {D.~B.}\ \bibnamefont {Hume}}, \bibinfo {author}
  {\bibfnamefont {W.~M.}\ \bibnamefont {Itano}}, \bibinfo {author}
  {\bibfnamefont {J.~D.}\ \bibnamefont {Jost}}, \bibinfo {author}
  {\bibfnamefont {C.}~\bibnamefont {Langer}}, \bibinfo {author} {\bibfnamefont
  {R.}~\bibnamefont {Ozeri}}, \bibinfo {author} {\bibfnamefont
  {R.}~\bibnamefont {Reichle}},\ and\ \bibinfo {author} {\bibfnamefont {D.~J.}\
  \bibnamefont {Wineland}},\ }\bibfield  {title} {\bibinfo {title} {Creation of
  a six-atom `{{Schr\"odinger}} cat' state},\ }\href@noop {} {\bibfield
  {journal} {\bibinfo  {journal} {Nature}\ }\textbf {\bibinfo {volume} {438}},\
  \bibinfo {pages} {639} (\bibinfo {year} {2005})}\BibitemShut {NoStop}%
\bibitem [{\citenamefont {Ourjoumtsev}\ \emph {et~al.}(2006)\citenamefont
  {Ourjoumtsev}, \citenamefont {{Tualle-Brouri}}, \citenamefont {Laurat},\ and\
  \citenamefont {Grangier}}]{ourjoumtsev_generating_2006}%
  \BibitemOpen
  \bibfield  {author} {\bibinfo {author} {\bibfnamefont {A.}~\bibnamefont
  {Ourjoumtsev}}, \bibinfo {author} {\bibfnamefont {R.}~\bibnamefont
  {{Tualle-Brouri}}}, \bibinfo {author} {\bibfnamefont {J.}~\bibnamefont
  {Laurat}},\ and\ \bibinfo {author} {\bibfnamefont {P.}~\bibnamefont
  {Grangier}},\ }\bibfield  {title} {\bibinfo {title} {Generating optical
  {{Schr\"odinger}} kittens for quantum information processing},\ }\href@noop
  {} {\bibfield  {journal} {\bibinfo  {journal} {Science}\ }\textbf {\bibinfo
  {volume} {312}},\ \bibinfo {pages} {83} (\bibinfo {year} {2006})}\BibitemShut
  {NoStop}%
\bibitem [{\citenamefont {{Neergaard-Nielsen}}\ \emph
  {et~al.}(2006)\citenamefont {{Neergaard-Nielsen}}, \citenamefont {Nielsen},
  \citenamefont {Hettich}, \citenamefont {M{\o}lmer},\ and\ \citenamefont
  {Polzik}}]{neergaard-nielsen_generation_2006}%
  \BibitemOpen
  \bibfield  {author} {\bibinfo {author} {\bibfnamefont {J.~S.}\ \bibnamefont
  {{Neergaard-Nielsen}}}, \bibinfo {author} {\bibfnamefont {B.~M.}\
  \bibnamefont {Nielsen}}, \bibinfo {author} {\bibfnamefont {C.}~\bibnamefont
  {Hettich}}, \bibinfo {author} {\bibfnamefont {K.}~\bibnamefont {M{\o}lmer}},\
  and\ \bibinfo {author} {\bibfnamefont {E.~S.}\ \bibnamefont {Polzik}},\
  }\bibfield  {title} {\bibinfo {title} {Generation of a superposition of odd
  photon number states for quantum information networks},\ }\href@noop {}
  {\bibfield  {journal} {\bibinfo  {journal} {Phys. Rev. Lett.}\ }\textbf
  {\bibinfo {volume} {97}},\ \bibinfo {pages} {083604} (\bibinfo {year}
  {2006})}\BibitemShut {NoStop}%
\bibitem [{\citenamefont {Deleglise}\ \emph {et~al.}(2008)\citenamefont
  {Deleglise}, \citenamefont {Dotsenko}, \citenamefont {Sayrin}, \citenamefont
  {Bernu}, \citenamefont {Brune}, \citenamefont {Raimond},\ and\ \citenamefont
  {Haroche}}]{deleglise_reconstruction_2008}%
  \BibitemOpen
  \bibfield  {author} {\bibinfo {author} {\bibfnamefont {S.}~\bibnamefont
  {Deleglise}}, \bibinfo {author} {\bibfnamefont {I.}~\bibnamefont {Dotsenko}},
  \bibinfo {author} {\bibfnamefont {C.}~\bibnamefont {Sayrin}}, \bibinfo
  {author} {\bibfnamefont {J.}~\bibnamefont {Bernu}}, \bibinfo {author}
  {\bibfnamefont {M.}~\bibnamefont {Brune}}, \bibinfo {author} {\bibfnamefont
  {J.-M.}\ \bibnamefont {Raimond}},\ and\ \bibinfo {author} {\bibfnamefont
  {S.}~\bibnamefont {Haroche}},\ }\bibfield  {title} {\bibinfo {title}
  {Reconstruction of non-classical cavity field states with snapshots of their
  decoherence},\ }\href@noop {} {\bibfield  {journal} {\bibinfo  {journal}
  {Nature}\ }\textbf {\bibinfo {volume} {455}},\ \bibinfo {pages} {510}
  (\bibinfo {year} {2008})}\BibitemShut {NoStop}%
\bibitem [{\citenamefont {Monz}\ \emph {et~al.}(2011)\citenamefont {Monz},
  \citenamefont {Schindler}, \citenamefont {Barreiro}, \citenamefont {Chwalla},
  \citenamefont {Nigg}, \citenamefont {Coish}, \citenamefont {Harlander},
  \citenamefont {H{\"a}nsel}, \citenamefont {Hennrich},\ and\ \citenamefont
  {Blatt}}]{monz_14-qubit_2011}%
  \BibitemOpen
  \bibfield  {author} {\bibinfo {author} {\bibfnamefont {T.}~\bibnamefont
  {Monz}}, \bibinfo {author} {\bibfnamefont {P.}~\bibnamefont {Schindler}},
  \bibinfo {author} {\bibfnamefont {J.~T.}\ \bibnamefont {Barreiro}}, \bibinfo
  {author} {\bibfnamefont {M.}~\bibnamefont {Chwalla}}, \bibinfo {author}
  {\bibfnamefont {D.}~\bibnamefont {Nigg}}, \bibinfo {author} {\bibfnamefont
  {W.~A.}\ \bibnamefont {Coish}}, \bibinfo {author} {\bibfnamefont
  {M.}~\bibnamefont {Harlander}}, \bibinfo {author} {\bibfnamefont
  {W.}~\bibnamefont {H{\"a}nsel}}, \bibinfo {author} {\bibfnamefont
  {M.}~\bibnamefont {Hennrich}},\ and\ \bibinfo {author} {\bibfnamefont
  {R.}~\bibnamefont {Blatt}},\ }\bibfield  {title} {\bibinfo {title} {14-qubit
  entanglement: {{Creation}} and coherence},\ }\href@noop {} {\bibfield
  {journal} {\bibinfo  {journal} {Phys. Rev. Lett.}\ }\textbf {\bibinfo
  {volume} {106}},\ \bibinfo {pages} {130506} (\bibinfo {year}
  {2011})}\BibitemShut {NoStop}%
\bibitem [{\citenamefont {Yao}\ \emph {et~al.}(2012)\citenamefont {Yao},
  \citenamefont {Wang}, \citenamefont {Xu}, \citenamefont {Lu}, \citenamefont
  {Pan}, \citenamefont {Bao}, \citenamefont {Peng}, \citenamefont {Lu},
  \citenamefont {Chen},\ and\ \citenamefont {Pan}}]{yao_observation_2012}%
  \BibitemOpen
  \bibfield  {author} {\bibinfo {author} {\bibfnamefont {X.-C.}\ \bibnamefont
  {Yao}}, \bibinfo {author} {\bibfnamefont {T.-X.}\ \bibnamefont {Wang}},
  \bibinfo {author} {\bibfnamefont {P.}~\bibnamefont {Xu}}, \bibinfo {author}
  {\bibfnamefont {H.}~\bibnamefont {Lu}}, \bibinfo {author} {\bibfnamefont
  {G.-S.}\ \bibnamefont {Pan}}, \bibinfo {author} {\bibfnamefont {X.-H.}\
  \bibnamefont {Bao}}, \bibinfo {author} {\bibfnamefont {C.-Z.}\ \bibnamefont
  {Peng}}, \bibinfo {author} {\bibfnamefont {C.-Y.}\ \bibnamefont {Lu}},
  \bibinfo {author} {\bibfnamefont {Y.-A.}\ \bibnamefont {Chen}},\ and\
  \bibinfo {author} {\bibfnamefont {J.-W.}\ \bibnamefont {Pan}},\ }\bibfield
  {title} {\bibinfo {title} {Observation of eight-photon entanglement},\
  }\href@noop {} {\bibfield  {journal} {\bibinfo  {journal} {Nat. Photonics}\
  }\textbf {\bibinfo {volume} {6}},\ \bibinfo {pages} {225} (\bibinfo {year}
  {2012})}\BibitemShut {NoStop}%
\bibitem [{\citenamefont {Kirchmair}\ \emph {et~al.}(2013)\citenamefont
  {Kirchmair}, \citenamefont {Vlastakis}, \citenamefont {Leghtas},
  \citenamefont {Nigg}, \citenamefont {Paik}, \citenamefont {Ginossar},
  \citenamefont {Mirrahimi}, \citenamefont {Frunzio}, \citenamefont {Girvin},\
  and\ \citenamefont {Schoelkopf}}]{kirchmair_observation_2013}%
  \BibitemOpen
  \bibfield  {author} {\bibinfo {author} {\bibfnamefont {G.}~\bibnamefont
  {Kirchmair}}, \bibinfo {author} {\bibfnamefont {B.}~\bibnamefont
  {Vlastakis}}, \bibinfo {author} {\bibfnamefont {Z.}~\bibnamefont {Leghtas}},
  \bibinfo {author} {\bibfnamefont {S.~E.}\ \bibnamefont {Nigg}}, \bibinfo
  {author} {\bibfnamefont {H.}~\bibnamefont {Paik}}, \bibinfo {author}
  {\bibfnamefont {E.}~\bibnamefont {Ginossar}}, \bibinfo {author}
  {\bibfnamefont {M.}~\bibnamefont {Mirrahimi}}, \bibinfo {author}
  {\bibfnamefont {L.}~\bibnamefont {Frunzio}}, \bibinfo {author} {\bibfnamefont
  {S.~M.}\ \bibnamefont {Girvin}},\ and\ \bibinfo {author} {\bibfnamefont
  {R.~J.}\ \bibnamefont {Schoelkopf}},\ }\bibfield  {title} {\bibinfo {title}
  {Observation of quantum state collapse and revival due to the single-photon
  {{Kerr}} effect},\ }\href@noop {} {\bibfield  {journal} {\bibinfo  {journal}
  {Nature}\ }\textbf {\bibinfo {volume} {495}},\ \bibinfo {pages} {205}
  (\bibinfo {year} {2013})}\BibitemShut {NoStop}%
\bibitem [{\citenamefont {Degen}\ \emph {et~al.}(2017)\citenamefont {Degen},
  \citenamefont {Reinhard},\ and\ \citenamefont
  {Cappellaro}}]{degen_quantum_2017}%
  \BibitemOpen
  \bibfield  {author} {\bibinfo {author} {\bibfnamefont {C.~L.}\ \bibnamefont
  {Degen}}, \bibinfo {author} {\bibfnamefont {F.}~\bibnamefont {Reinhard}},\
  and\ \bibinfo {author} {\bibfnamefont {P.}~\bibnamefont {Cappellaro}},\
  }\bibfield  {title} {\bibinfo {title} {Quantum sensing},\ }\href@noop {}
  {\bibfield  {journal} {\bibinfo  {journal} {Rev. Mod. Phys.}\ }\textbf
  {\bibinfo {volume} {89}},\ \bibinfo {pages} {035002} (\bibinfo {year}
  {2017})}\BibitemShut {NoStop}%
\bibitem [{\citenamefont {Wang}\ \emph {et~al.}(2018)\citenamefont {Wang},
  \citenamefont {Luo}, \citenamefont {Huang}, \citenamefont {Chen},
  \citenamefont {Su}, \citenamefont {Liu}, \citenamefont {Chen}, \citenamefont
  {Li}, \citenamefont {Fang}, \citenamefont {Jiang}, \citenamefont {Zhang},
  \citenamefont {Li}, \citenamefont {Liu}, \citenamefont {Lu},\ and\
  \citenamefont {Pan}}]{wang_18-qubit_2018}%
  \BibitemOpen
  \bibfield  {author} {\bibinfo {author} {\bibfnamefont {X.-L.}\ \bibnamefont
  {Wang}}, \bibinfo {author} {\bibfnamefont {Y.-H.}\ \bibnamefont {Luo}},
  \bibinfo {author} {\bibfnamefont {H.-L.}\ \bibnamefont {Huang}}, \bibinfo
  {author} {\bibfnamefont {M.-C.}\ \bibnamefont {Chen}}, \bibinfo {author}
  {\bibfnamefont {Z.-E.}\ \bibnamefont {Su}}, \bibinfo {author} {\bibfnamefont
  {C.}~\bibnamefont {Liu}}, \bibinfo {author} {\bibfnamefont {C.}~\bibnamefont
  {Chen}}, \bibinfo {author} {\bibfnamefont {W.}~\bibnamefont {Li}}, \bibinfo
  {author} {\bibfnamefont {Y.-Q.}\ \bibnamefont {Fang}}, \bibinfo {author}
  {\bibfnamefont {X.}~\bibnamefont {Jiang}}, \bibinfo {author} {\bibfnamefont
  {J.}~\bibnamefont {Zhang}}, \bibinfo {author} {\bibfnamefont
  {L.}~\bibnamefont {Li}}, \bibinfo {author} {\bibfnamefont {N.-L.}\
  \bibnamefont {Liu}}, \bibinfo {author} {\bibfnamefont {C.-Y.}\ \bibnamefont
  {Lu}},\ and\ \bibinfo {author} {\bibfnamefont {J.-W.}\ \bibnamefont {Pan}},\
  }\bibfield  {title} {\bibinfo {title} {18-{{Qubit Entanglement}} with {{Six
  Photons}}' {{Three Degrees}} of {{Freedom}}},\ }\href@noop {} {\bibfield
  {journal} {\bibinfo  {journal} {Phys. Rev. Lett.}\ }\textbf {\bibinfo
  {volume} {120}},\ \bibinfo {pages} {260502} (\bibinfo {year}
  {2018})}\BibitemShut {NoStop}%
\bibitem [{\citenamefont {Dietsche}\ \emph {et~al.}(2019)\citenamefont
  {Dietsche}, \citenamefont {Larrouy}, \citenamefont {Haroche}, \citenamefont
  {Raimond}, \citenamefont {Brune},\ and\ \citenamefont
  {Gleyzes}}]{dietsche_high-sensitivity_2019}%
  \BibitemOpen
  \bibfield  {author} {\bibinfo {author} {\bibfnamefont {E.~K.}\ \bibnamefont
  {Dietsche}}, \bibinfo {author} {\bibfnamefont {A.}~\bibnamefont {Larrouy}},
  \bibinfo {author} {\bibfnamefont {S.}~\bibnamefont {Haroche}}, \bibinfo
  {author} {\bibfnamefont {J.~M.}\ \bibnamefont {Raimond}}, \bibinfo {author}
  {\bibfnamefont {M.}~\bibnamefont {Brune}},\ and\ \bibinfo {author}
  {\bibfnamefont {S.}~\bibnamefont {Gleyzes}},\ }\bibfield  {title} {\bibinfo
  {title} {High-sensitivity magnetometry with a single atom in a superposition
  of two circular {{Rydberg}} states},\ }\href@noop {} {\bibfield  {journal}
  {\bibinfo  {journal} {Nat. Phys.}\ }\textbf {\bibinfo {volume} {15}},\
  \bibinfo {pages} {326} (\bibinfo {year} {2019})}\BibitemShut {NoStop}%
\bibitem [{\citenamefont {Song}\ \emph {et~al.}(2019)\citenamefont {Song},
  \citenamefont {Xu}, \citenamefont {Li}, \citenamefont {Zhang}, \citenamefont
  {Zhang}, \citenamefont {Liu}, \citenamefont {Guo}, \citenamefont {Wang},
  \citenamefont {Ren}, \citenamefont {Hao}, \citenamefont {Feng}, \citenamefont
  {Fan}, \citenamefont {Zheng}, \citenamefont {Wang}, \citenamefont {Wang},\
  and\ \citenamefont {Zhu}}]{song_generation_2019}%
  \BibitemOpen
  \bibfield  {author} {\bibinfo {author} {\bibfnamefont {C.}~\bibnamefont
  {Song}}, \bibinfo {author} {\bibfnamefont {K.}~\bibnamefont {Xu}}, \bibinfo
  {author} {\bibfnamefont {H.}~\bibnamefont {Li}}, \bibinfo {author}
  {\bibfnamefont {Y.-R.}\ \bibnamefont {Zhang}}, \bibinfo {author}
  {\bibfnamefont {X.}~\bibnamefont {Zhang}}, \bibinfo {author} {\bibfnamefont
  {W.}~\bibnamefont {Liu}}, \bibinfo {author} {\bibfnamefont {Q.}~\bibnamefont
  {Guo}}, \bibinfo {author} {\bibfnamefont {Z.}~\bibnamefont {Wang}}, \bibinfo
  {author} {\bibfnamefont {W.}~\bibnamefont {Ren}}, \bibinfo {author}
  {\bibfnamefont {J.}~\bibnamefont {Hao}}, \bibinfo {author} {\bibfnamefont
  {H.}~\bibnamefont {Feng}}, \bibinfo {author} {\bibfnamefont {H.}~\bibnamefont
  {Fan}}, \bibinfo {author} {\bibfnamefont {D.}~\bibnamefont {Zheng}}, \bibinfo
  {author} {\bibfnamefont {D.-W.}\ \bibnamefont {Wang}}, \bibinfo {author}
  {\bibfnamefont {H.}~\bibnamefont {Wang}},\ and\ \bibinfo {author}
  {\bibfnamefont {S.-Y.}\ \bibnamefont {Zhu}},\ }\bibfield  {title} {\bibinfo
  {title} {Generation of multicomponent atomic {{Schr\"odinger}} cat states of
  up to 20 qubits},\ }\href@noop {} {\bibfield  {journal} {\bibinfo  {journal}
  {Science}\ }\textbf {\bibinfo {volume} {365}},\ \bibinfo {pages} {574}
  (\bibinfo {year} {2019})}\BibitemShut {NoStop}%
\bibitem [{\citenamefont {Omran}\ \emph {et~al.}(2019)\citenamefont {Omran},
  \citenamefont {Levine}, \citenamefont {Keesling}, \citenamefont {Semeghini},
  \citenamefont {Wang}, \citenamefont {Ebadi}, \citenamefont {Bernien},
  \citenamefont {Zibrov}, \citenamefont {Pichler}, \citenamefont {Choi},
  \citenamefont {Cui}, \citenamefont {Rossignolo}, \citenamefont {Rembold},
  \citenamefont {Montangero}, \citenamefont {Calarco}, \citenamefont {Endres},
  \citenamefont {Greiner}, \citenamefont {Vuleti{\'c}},\ and\ \citenamefont
  {Lukin}}]{omran_generation_2019}%
  \BibitemOpen
  \bibfield  {author} {\bibinfo {author} {\bibfnamefont {A.}~\bibnamefont
  {Omran}}, \bibinfo {author} {\bibfnamefont {H.}~\bibnamefont {Levine}},
  \bibinfo {author} {\bibfnamefont {A.}~\bibnamefont {Keesling}}, \bibinfo
  {author} {\bibfnamefont {G.}~\bibnamefont {Semeghini}}, \bibinfo {author}
  {\bibfnamefont {T.~T.}\ \bibnamefont {Wang}}, \bibinfo {author}
  {\bibfnamefont {S.}~\bibnamefont {Ebadi}}, \bibinfo {author} {\bibfnamefont
  {H.}~\bibnamefont {Bernien}}, \bibinfo {author} {\bibfnamefont {A.~S.}\
  \bibnamefont {Zibrov}}, \bibinfo {author} {\bibfnamefont {H.}~\bibnamefont
  {Pichler}}, \bibinfo {author} {\bibfnamefont {S.}~\bibnamefont {Choi}},
  \bibinfo {author} {\bibfnamefont {J.}~\bibnamefont {Cui}}, \bibinfo {author}
  {\bibfnamefont {M.}~\bibnamefont {Rossignolo}}, \bibinfo {author}
  {\bibfnamefont {P.}~\bibnamefont {Rembold}}, \bibinfo {author} {\bibfnamefont
  {S.}~\bibnamefont {Montangero}}, \bibinfo {author} {\bibfnamefont
  {T.}~\bibnamefont {Calarco}}, \bibinfo {author} {\bibfnamefont
  {M.}~\bibnamefont {Endres}}, \bibinfo {author} {\bibfnamefont
  {M.}~\bibnamefont {Greiner}}, \bibinfo {author} {\bibfnamefont
  {V.}~\bibnamefont {Vuleti{\'c}}},\ and\ \bibinfo {author} {\bibfnamefont
  {M.~D.}\ \bibnamefont {Lukin}},\ }\bibfield  {title} {\bibinfo {title}
  {Generation and manipulation of {{Schr\"odinger}} cat states in {{Rydberg}}
  atom arrays},\ }\href@noop {} {\bibfield  {journal} {\bibinfo  {journal}
  {Science}\ }\textbf {\bibinfo {volume} {365}},\ \bibinfo {pages} {570}
  (\bibinfo {year} {2019})}\BibitemShut {NoStop}%
\bibitem [{\citenamefont {Wei}\ \emph {et~al.}(2020)\citenamefont {Wei},
  \citenamefont {Lauer}, \citenamefont {Srinivasan}, \citenamefont
  {Sundaresan}, \citenamefont {McClure}, \citenamefont {Toyli}, \citenamefont
  {McKay}, \citenamefont {Gambetta},\ and\ \citenamefont
  {Sheldon}}]{wei_verifying_2020}%
  \BibitemOpen
  \bibfield  {author} {\bibinfo {author} {\bibfnamefont {K.~X.}\ \bibnamefont
  {Wei}}, \bibinfo {author} {\bibfnamefont {I.}~\bibnamefont {Lauer}}, \bibinfo
  {author} {\bibfnamefont {S.}~\bibnamefont {Srinivasan}}, \bibinfo {author}
  {\bibfnamefont {N.}~\bibnamefont {Sundaresan}}, \bibinfo {author}
  {\bibfnamefont {D.~T.}\ \bibnamefont {McClure}}, \bibinfo {author}
  {\bibfnamefont {D.}~\bibnamefont {Toyli}}, \bibinfo {author} {\bibfnamefont
  {D.~C.}\ \bibnamefont {McKay}}, \bibinfo {author} {\bibfnamefont {J.~M.}\
  \bibnamefont {Gambetta}},\ and\ \bibinfo {author} {\bibfnamefont
  {S.}~\bibnamefont {Sheldon}},\ }\bibfield  {title} {\bibinfo {title}
  {Verifying multipartite entangled {{Greenberger}}-{{Horne}}-{{Zeilinger}}
  states via multiple quantum coherences},\ }\href@noop {} {\bibfield
  {journal} {\bibinfo  {journal} {Phys. Rev. A}\ }\textbf {\bibinfo {volume}
  {101}},\ \bibinfo {pages} {032343} (\bibinfo {year} {2020})}\BibitemShut
  {NoStop}%
\bibitem [{\citenamefont {Mermin}(1990)}]{mermin_extreme_1990}%
  \BibitemOpen
  \bibfield  {author} {\bibinfo {author} {\bibfnamefont {N.~D.}\ \bibnamefont
  {Mermin}},\ }\bibfield  {title} {\bibinfo {title} {Extreme quantum
  entanglement in a superposition of macroscopically distinct states},\
  }\href@noop {} {\bibfield  {journal} {\bibinfo  {journal} {Phys. Rev. Lett.}\
  }\textbf {\bibinfo {volume} {65}},\ \bibinfo {pages} {1838} (\bibinfo {year}
  {1990})}\BibitemShut {NoStop}%
\bibitem [{\citenamefont {Bollinger}\ \emph {et~al.}(1996)\citenamefont
  {Bollinger}, \citenamefont {Itano}, \citenamefont {Wineland},\ and\
  \citenamefont {Heinzen}}]{bollinger_optimal_1996}%
  \BibitemOpen
  \bibfield  {author} {\bibinfo {author} {\bibfnamefont {J.~J.}\ \bibnamefont
  {Bollinger}}, \bibinfo {author} {\bibfnamefont {W.~M.}\ \bibnamefont
  {Itano}}, \bibinfo {author} {\bibfnamefont {D.~J.}\ \bibnamefont
  {Wineland}},\ and\ \bibinfo {author} {\bibfnamefont {D.}~\bibnamefont
  {Heinzen}},\ }\bibfield  {title} {\bibinfo {title} {Optimal frequency
  measurements with maximally correlated states},\ }\href@noop {} {\bibfield
  {journal} {\bibinfo  {journal} {Phys. Rev. A}\ }\textbf {\bibinfo {volume}
  {54}},\ \bibinfo {pages} {R4649} (\bibinfo {year} {1996})}\BibitemShut
  {NoStop}%
\bibitem [{Note9()}]{Note9}%
  \BibitemOpen
  \bibinfo {note} {In practice, we observe a small variation of the measured
  $\protect \mathcal {C}_{\protect \mathbf {n}}$ with the azimuthal angle $\phi
  $, discussed in Appendix \ref {appendix_phi_variation}. We show in
  Fig.\protect \tmspace +\thinmuskip {.1667em}\ref {fig_N00N}b the variation of
  $\protect \mathcal {C}_{\protect \mathbf {n}}$ with the polar angle $\theta $
  with a fixed angle $\phi =\SI {3.3(1)}{\radian }$, which maximizes the value
  of $\protect \mathcal {C}_{\protect \mathbf {n}}$.}\BibitemShut {Stop}%
\bibitem [{\citenamefont {Davis}\ \emph {et~al.}(2016)\citenamefont {Davis},
  \citenamefont {Bentsen},\ and\ \citenamefont
  {{Schleier-Smith}}}]{davis_approaching_2016}%
  \BibitemOpen
  \bibfield  {author} {\bibinfo {author} {\bibfnamefont {E.}~\bibnamefont
  {Davis}}, \bibinfo {author} {\bibfnamefont {G.}~\bibnamefont {Bentsen}},\
  and\ \bibinfo {author} {\bibfnamefont {M.}~\bibnamefont {{Schleier-Smith}}},\
  }\bibfield  {title} {\bibinfo {title} {Approaching the {{Heisenberg}} limit
  without single-particle detection},\ }\href@noop {} {\bibfield  {journal}
  {\bibinfo  {journal} {Phys. Rev. Lett.}\ }\textbf {\bibinfo {volume} {116}},\
  \bibinfo {pages} {053601} (\bibinfo {year} {2016})}\BibitemShut {NoStop}%
\bibitem [{\citenamefont {Linnemann}\ \emph {et~al.}(2016)\citenamefont
  {Linnemann}, \citenamefont {Strobel}, \citenamefont {Muessel}, \citenamefont
  {Schulz}, \citenamefont {{Lewis-Swan}}, \citenamefont {Kheruntsyan},\ and\
  \citenamefont {Oberthaler}}]{linnemann_quantum-enhanced_2016}%
  \BibitemOpen
  \bibfield  {author} {\bibinfo {author} {\bibfnamefont {D.}~\bibnamefont
  {Linnemann}}, \bibinfo {author} {\bibfnamefont {H.}~\bibnamefont {Strobel}},
  \bibinfo {author} {\bibfnamefont {W.}~\bibnamefont {Muessel}}, \bibinfo
  {author} {\bibfnamefont {J.}~\bibnamefont {Schulz}}, \bibinfo {author}
  {\bibfnamefont {R.~J.}\ \bibnamefont {{Lewis-Swan}}}, \bibinfo {author}
  {\bibfnamefont {K.~V.}\ \bibnamefont {Kheruntsyan}},\ and\ \bibinfo {author}
  {\bibfnamefont {M.~K.}\ \bibnamefont {Oberthaler}},\ }\bibfield  {title}
  {\bibinfo {title} {Quantum-enhanced sensing based on time reversal of
  nonlinear dynamics},\ }\href@noop {} {\bibfield  {journal} {\bibinfo
  {journal} {Phys. Rev. Lett.}\ }\textbf {\bibinfo {volume} {117}},\ \bibinfo
  {pages} {013001} (\bibinfo {year} {2016})}\BibitemShut {NoStop}%
\bibitem [{\citenamefont {Fr{\"o}wis}\ \emph {et~al.}(2016)\citenamefont
  {Fr{\"o}wis}, \citenamefont {Sekatski},\ and\ \citenamefont
  {D{\"u}r}}]{frowis_detecting_2016}%
  \BibitemOpen
  \bibfield  {author} {\bibinfo {author} {\bibfnamefont {F.}~\bibnamefont
  {Fr{\"o}wis}}, \bibinfo {author} {\bibfnamefont {P.}~\bibnamefont
  {Sekatski}},\ and\ \bibinfo {author} {\bibfnamefont {W.}~\bibnamefont
  {D{\"u}r}},\ }\bibfield  {title} {\bibinfo {title} {Detecting {{Large Quantum
  Fisher Information}} with {{Finite Measurement Precision}}},\ }\href@noop {}
  {\bibfield  {journal} {\bibinfo  {journal} {Phys Rev Lett}\ }\textbf
  {\bibinfo {volume} {116}},\ \bibinfo {pages} {090801} (\bibinfo {year}
  {2016})}\BibitemShut {NoStop}%
\bibitem [{\citenamefont {Macr{\`i}}\ \emph {et~al.}(2016)\citenamefont
  {Macr{\`i}}, \citenamefont {Smerzi},\ and\ \citenamefont
  {Pezz{\`e}}}]{macri_loschmidt_2016}%
  \BibitemOpen
  \bibfield  {author} {\bibinfo {author} {\bibfnamefont {T.}~\bibnamefont
  {Macr{\`i}}}, \bibinfo {author} {\bibfnamefont {A.}~\bibnamefont {Smerzi}},\
  and\ \bibinfo {author} {\bibfnamefont {L.}~\bibnamefont {Pezz{\`e}}},\
  }\bibfield  {title} {\bibinfo {title} {Loschmidt echo for quantum
  metrology},\ }\href@noop {} {\bibfield  {journal} {\bibinfo  {journal} {Phys
  Rev A}\ }\textbf {\bibinfo {volume} {94}},\ \bibinfo {pages} {010102}
  (\bibinfo {year} {2016})}\BibitemShut {NoStop}%
\bibitem [{\citenamefont {Gustavsson}\ \emph {et~al.}(1979)\citenamefont
  {Gustavsson}, \citenamefont {Lundberg}, \citenamefont {Nilsson},\ and\
  \citenamefont {Svanberg}}]{gustavsson_lifetime_1979}%
  \BibitemOpen
  \bibfield  {author} {\bibinfo {author} {\bibfnamefont {M.}~\bibnamefont
  {Gustavsson}}, \bibinfo {author} {\bibfnamefont {H.}~\bibnamefont
  {Lundberg}}, \bibinfo {author} {\bibfnamefont {L.}~\bibnamefont {Nilsson}},\
  and\ \bibinfo {author} {\bibfnamefont {S.}~\bibnamefont {Svanberg}},\
  }\bibfield  {title} {\bibinfo {title} {Lifetime measurements for excited
  states of rare-earth atoms using pulse modulation of a cw dye-laser beam},\
  }\href@noop {} {\bibfield  {journal} {\bibinfo  {journal} {JOSA}\ }\textbf
  {\bibinfo {volume} {69}},\ \bibinfo {pages} {984} (\bibinfo {year}
  {1979})}\BibitemShut {NoStop}%
\bibitem [{\citenamefont {Lee}\ and\ \citenamefont
  {Jeong}(2011)}]{lee_quantification_2011}%
  \BibitemOpen
  \bibfield  {author} {\bibinfo {author} {\bibfnamefont {C.-W.}\ \bibnamefont
  {Lee}}\ and\ \bibinfo {author} {\bibfnamefont {H.}~\bibnamefont {Jeong}},\
  }\bibfield  {title} {\bibinfo {title} {Quantification of {{Macroscopic
  Quantum Superpositions}} within {{Phase Space}}},\ }\href@noop {} {\bibfield
  {journal} {\bibinfo  {journal} {Phys Rev Lett}\ }\textbf {\bibinfo {volume}
  {106}},\ \bibinfo {pages} {220401} (\bibinfo {year} {2011})}\BibitemShut
  {NoStop}%
\bibitem [{Note10()}]{Note10}%
  \BibitemOpen
  \bibinfo {note} {The coupling amplitudes between $\left | m=\pm J \right >$
  and $\left | m'=\pm (J'-2) \right >$ are $1/\protect \sqrt {153}$ smaller
  than the couplings between $\left | m=\pm J \right >$ and $\left | m'=\pm J'
  \right >$. When the population of $\left | m'=\pm J' \right >$ is maximized,
  we expect a residual population of the states $\left | m'=\pm (J'-2) \right
  >$ of 3\% due to these small couplings.}\BibitemShut {Stop}%
\bibitem [{\citenamefont {Wootters}\ and\ \citenamefont
  {Zurek}(1979)}]{wootters_complementarity_1979}%
  \BibitemOpen
  \bibfield  {author} {\bibinfo {author} {\bibfnamefont {W.~K.}\ \bibnamefont
  {Wootters}}\ and\ \bibinfo {author} {\bibfnamefont {W.~H.}\ \bibnamefont
  {Zurek}},\ }\bibfield  {title} {\bibinfo {title} {Complementarity in the
  double-slit experiment: {{Quantum}} nonseparability and a quantitative
  statement of {{Bohr}}'s principle},\ }\href@noop {} {\bibfield  {journal}
  {\bibinfo  {journal} {Phys. Rev. D}\ }\textbf {\bibinfo {volume} {19}},\
  \bibinfo {pages} {473} (\bibinfo {year} {1979})}\BibitemShut {NoStop}%
\bibitem [{\citenamefont {Englert}(1996)}]{englert_fringe_1996}%
  \BibitemOpen
  \bibfield  {author} {\bibinfo {author} {\bibfnamefont {B.-G.}\ \bibnamefont
  {Englert}},\ }\bibfield  {title} {\bibinfo {title} {Fringe {{Visibility}} and
  {{Which}}-{{Way Information}}: {{An Inequality}}},\ }\href@noop {} {\bibfield
   {journal} {\bibinfo  {journal} {Phys. Rev. Lett.}\ }\textbf {\bibinfo
  {volume} {77}},\ \bibinfo {pages} {2154} (\bibinfo {year}
  {1996})}\BibitemShut {NoStop}%
\bibitem [{\citenamefont {Blinov}\ \emph {et~al.}(2004)\citenamefont {Blinov},
  \citenamefont {Moehring}, \citenamefont {Duan},\ and\ \citenamefont
  {Monroe}}]{blinov_observation_2004}%
  \BibitemOpen
  \bibfield  {author} {\bibinfo {author} {\bibfnamefont {B.~B.}\ \bibnamefont
  {Blinov}}, \bibinfo {author} {\bibfnamefont {D.~L.}\ \bibnamefont
  {Moehring}}, \bibinfo {author} {\bibfnamefont {L.-M.}\ \bibnamefont {Duan}},\
  and\ \bibinfo {author} {\bibfnamefont {C.}~\bibnamefont {Monroe}},\
  }\bibfield  {title} {\bibinfo {title} {Observation of entanglement between a
  single trapped atom and a single photon},\ }\href@noop {} {\bibfield
  {journal} {\bibinfo  {journal} {Nature}\ }\textbf {\bibinfo {volume} {428}},\
  \bibinfo {pages} {153} (\bibinfo {year} {2004})}\BibitemShut {NoStop}%
\bibitem [{\citenamefont {Volz}\ \emph {et~al.}(2006)\citenamefont {Volz},
  \citenamefont {Weber}, \citenamefont {Schlenk}, \citenamefont {Rosenfeld},
  \citenamefont {Vrana}, \citenamefont {Saucke}, \citenamefont {Kurtsiefer},\
  and\ \citenamefont {Weinfurter}}]{volz_observation_2006}%
  \BibitemOpen
  \bibfield  {author} {\bibinfo {author} {\bibfnamefont {J.}~\bibnamefont
  {Volz}}, \bibinfo {author} {\bibfnamefont {M.}~\bibnamefont {Weber}},
  \bibinfo {author} {\bibfnamefont {D.}~\bibnamefont {Schlenk}}, \bibinfo
  {author} {\bibfnamefont {W.}~\bibnamefont {Rosenfeld}}, \bibinfo {author}
  {\bibfnamefont {J.}~\bibnamefont {Vrana}}, \bibinfo {author} {\bibfnamefont
  {K.}~\bibnamefont {Saucke}}, \bibinfo {author} {\bibfnamefont
  {C.}~\bibnamefont {Kurtsiefer}},\ and\ \bibinfo {author} {\bibfnamefont
  {H.}~\bibnamefont {Weinfurter}},\ }\bibfield  {title} {\bibinfo {title}
  {Observation of {{Entanglement}} of a {{Single Photon}} with a {{Trapped
  Atom}}},\ }\href@noop {} {\bibfield  {journal} {\bibinfo  {journal} {Phys.
  Rev. Lett.}\ }\textbf {\bibinfo {volume} {96}},\ \bibinfo {pages} {030404}
  (\bibinfo {year} {2006})}\BibitemShut {NoStop}%
\bibitem [{\citenamefont {Wilk}\ \emph {et~al.}(2007)\citenamefont {Wilk},
  \citenamefont {Webster}, \citenamefont {Kuhn},\ and\ \citenamefont
  {Rempe}}]{wilk_single-atom_2007}%
  \BibitemOpen
  \bibfield  {author} {\bibinfo {author} {\bibfnamefont {T.}~\bibnamefont
  {Wilk}}, \bibinfo {author} {\bibfnamefont {S.~C.}\ \bibnamefont {Webster}},
  \bibinfo {author} {\bibfnamefont {A.}~\bibnamefont {Kuhn}},\ and\ \bibinfo
  {author} {\bibfnamefont {G.}~\bibnamefont {Rempe}},\ }\bibfield  {title}
  {\bibinfo {title} {Single-{{Atom Single}}-{{Photon Quantum Interface}}},\
  }\href@noop {} {\bibfield  {journal} {\bibinfo  {journal} {Science}\ }\textbf
  {\bibinfo {volume} {317}},\ \bibinfo {pages} {488} (\bibinfo {year}
  {2007})}\BibitemShut {NoStop}%
\bibitem [{\citenamefont {Togan}\ \emph {et~al.}(2010)\citenamefont {Togan},
  \citenamefont {Chu}, \citenamefont {Trifonov}, \citenamefont {Jiang},
  \citenamefont {Maze}, \citenamefont {Childress}, \citenamefont {Dutt},
  \citenamefont {S{\o}rensen}, \citenamefont {Hemmer}, \citenamefont {Zibrov},\
  and\ \citenamefont {Lukin}}]{togan_quantum_2010}%
  \BibitemOpen
  \bibfield  {author} {\bibinfo {author} {\bibfnamefont {E.}~\bibnamefont
  {Togan}}, \bibinfo {author} {\bibfnamefont {Y.}~\bibnamefont {Chu}}, \bibinfo
  {author} {\bibfnamefont {A.~S.}\ \bibnamefont {Trifonov}}, \bibinfo {author}
  {\bibfnamefont {L.}~\bibnamefont {Jiang}}, \bibinfo {author} {\bibfnamefont
  {J.}~\bibnamefont {Maze}}, \bibinfo {author} {\bibfnamefont {L.}~\bibnamefont
  {Childress}}, \bibinfo {author} {\bibfnamefont {M.~V.~G.}\ \bibnamefont
  {Dutt}}, \bibinfo {author} {\bibfnamefont {A.~S.}\ \bibnamefont
  {S{\o}rensen}}, \bibinfo {author} {\bibfnamefont {P.~R.}\ \bibnamefont
  {Hemmer}}, \bibinfo {author} {\bibfnamefont {A.~S.}\ \bibnamefont {Zibrov}},\
  and\ \bibinfo {author} {\bibfnamefont {M.~D.}\ \bibnamefont {Lukin}},\
  }\bibfield  {title} {\bibinfo {title} {Quantum entanglement between an
  optical photon and a solid-state spin qubit},\ }\href@noop {} {\bibfield
  {journal} {\bibinfo  {journal} {Nature}\ }\textbf {\bibinfo {volume} {466}},\
  \bibinfo {pages} {730} (\bibinfo {year} {2010})}\BibitemShut {NoStop}%
\bibitem [{\citenamefont {Moehring}\ \emph {et~al.}(2007)\citenamefont
  {Moehring}, \citenamefont {Maunz}, \citenamefont {Olmschenk}, \citenamefont
  {Younge}, \citenamefont {Matsukevich}, \citenamefont {Duan},\ and\
  \citenamefont {Monroe}}]{moehring_entanglement_2007}%
  \BibitemOpen
  \bibfield  {author} {\bibinfo {author} {\bibfnamefont {D.~L.}\ \bibnamefont
  {Moehring}}, \bibinfo {author} {\bibfnamefont {P.}~\bibnamefont {Maunz}},
  \bibinfo {author} {\bibfnamefont {S.}~\bibnamefont {Olmschenk}}, \bibinfo
  {author} {\bibfnamefont {K.~C.}\ \bibnamefont {Younge}}, \bibinfo {author}
  {\bibfnamefont {D.~N.}\ \bibnamefont {Matsukevich}}, \bibinfo {author}
  {\bibfnamefont {L.-M.}\ \bibnamefont {Duan}},\ and\ \bibinfo {author}
  {\bibfnamefont {C.}~\bibnamefont {Monroe}},\ }\bibfield  {title} {\bibinfo
  {title} {Entanglement of single-atom quantum bits at a distance},\
  }\href@noop {} {\bibfield  {journal} {\bibinfo  {journal} {Nature}\ }\textbf
  {\bibinfo {volume} {449}},\ \bibinfo {pages} {68} (\bibinfo {year}
  {2007})}\BibitemShut {NoStop}%
\bibitem [{\citenamefont {Cozzolino}\ \emph {et~al.}(2019)\citenamefont
  {Cozzolino}, \citenamefont {Lio}, \citenamefont {Bacco},\ and\ \citenamefont
  {Oxenl{\o}we}}]{cozzolino_high-dimensional_2019}%
  \BibitemOpen
  \bibfield  {author} {\bibinfo {author} {\bibfnamefont {D.}~\bibnamefont
  {Cozzolino}}, \bibinfo {author} {\bibfnamefont {B.~D.}\ \bibnamefont {Lio}},
  \bibinfo {author} {\bibfnamefont {D.}~\bibnamefont {Bacco}},\ and\ \bibinfo
  {author} {\bibfnamefont {L.~K.}\ \bibnamefont {Oxenl{\o}we}},\ }\bibfield
  {title} {\bibinfo {title} {High-{{Dimensional Quantum Communication}}:
  {{Benefits}}, {{Progress}}, and {{Future Challenges}}},\ }\href@noop {}
  {\bibfield  {journal} {\bibinfo  {journal} {Adv. Quantum Technol.}\ }\textbf
  {\bibinfo {volume} {2}},\ \bibinfo {pages} {1900038} (\bibinfo {year}
  {2019})}\BibitemShut {NoStop}%
\bibitem [{\citenamefont {Brahms}\ and\ \citenamefont
  {{Stamper-Kurn}}(2010)}]{brahms_spin_2010}%
  \BibitemOpen
  \bibfield  {author} {\bibinfo {author} {\bibfnamefont {N.}~\bibnamefont
  {Brahms}}\ and\ \bibinfo {author} {\bibfnamefont {D.~M.}\ \bibnamefont
  {{Stamper-Kurn}}},\ }\bibfield  {title} {\bibinfo {title} {Spin optodynamics
  analog of cavity optomechanics},\ }\href@noop {} {\bibfield  {journal}
  {\bibinfo  {journal} {Phys. Rev. A}\ }\textbf {\bibinfo {volume} {82}},\
  \bibinfo {pages} {041804} (\bibinfo {year} {2010})}\BibitemShut {NoStop}%
\bibitem [{\citenamefont {Brennecke}\ \emph {et~al.}(2007)\citenamefont
  {Brennecke}, \citenamefont {Donner}, \citenamefont {Ritter}, \citenamefont
  {Bourdel}, \citenamefont {K{\"o}hl},\ and\ \citenamefont
  {Esslinger}}]{brennecke_cavity_2007}%
  \BibitemOpen
  \bibfield  {author} {\bibinfo {author} {\bibfnamefont {F.}~\bibnamefont
  {Brennecke}}, \bibinfo {author} {\bibfnamefont {T.}~\bibnamefont {Donner}},
  \bibinfo {author} {\bibfnamefont {S.}~\bibnamefont {Ritter}}, \bibinfo
  {author} {\bibfnamefont {T.}~\bibnamefont {Bourdel}}, \bibinfo {author}
  {\bibfnamefont {M.}~\bibnamefont {K{\"o}hl}},\ and\ \bibinfo {author}
  {\bibfnamefont {T.}~\bibnamefont {Esslinger}},\ }\bibfield  {title} {\bibinfo
  {title} {Cavity {{QED}} with a {{Bose}}\textendash{{Einstein}} condensate},\
  }\href@noop {} {\bibfield  {journal} {\bibinfo  {journal} {Nature}\ }\textbf
  {\bibinfo {volume} {450}},\ \bibinfo {pages} {268} (\bibinfo {year}
  {2007})}\BibitemShut {NoStop}%
\bibitem [{\citenamefont {Colombe}\ \emph {et~al.}(2007)\citenamefont
  {Colombe}, \citenamefont {Steinmetz}, \citenamefont {Dubois}, \citenamefont
  {Linke}, \citenamefont {Hunger},\ and\ \citenamefont
  {Reichel}}]{colombe_strong_2007}%
  \BibitemOpen
  \bibfield  {author} {\bibinfo {author} {\bibfnamefont {Y.}~\bibnamefont
  {Colombe}}, \bibinfo {author} {\bibfnamefont {T.}~\bibnamefont {Steinmetz}},
  \bibinfo {author} {\bibfnamefont {G.}~\bibnamefont {Dubois}}, \bibinfo
  {author} {\bibfnamefont {F.}~\bibnamefont {Linke}}, \bibinfo {author}
  {\bibfnamefont {D.}~\bibnamefont {Hunger}},\ and\ \bibinfo {author}
  {\bibfnamefont {J.}~\bibnamefont {Reichel}},\ }\bibfield  {title} {\bibinfo
  {title} {Strong atom\textendash field coupling for
  {{Bose}}\textendash{{Einstein}} condensates in an optical cavity on a chip},\
  }\href@noop {} {\bibfield  {journal} {\bibinfo  {journal} {Nature}\ }\textbf
  {\bibinfo {volume} {450}},\ \bibinfo {pages} {272} (\bibinfo {year}
  {2007})}\BibitemShut {NoStop}%
\bibitem [{\citenamefont {Vitagliano}\ \emph {et~al.}(2014)\citenamefont
  {Vitagliano}, \citenamefont {Apellaniz}, \citenamefont {Egusquiza},\ and\
  \citenamefont {T{\'o}th}}]{vitagliano_spin_2014}%
  \BibitemOpen
  \bibfield  {author} {\bibinfo {author} {\bibfnamefont {G.}~\bibnamefont
  {Vitagliano}}, \bibinfo {author} {\bibfnamefont {I.}~\bibnamefont
  {Apellaniz}}, \bibinfo {author} {\bibfnamefont {I.~L.}\ \bibnamefont
  {Egusquiza}},\ and\ \bibinfo {author} {\bibfnamefont {G.}~\bibnamefont
  {T{\'o}th}},\ }\bibfield  {title} {\bibinfo {title} {Spin squeezing and
  entanglement for an arbitrary spin},\ }\href@noop {} {\bibfield  {journal}
  {\bibinfo  {journal} {Phys. Rev. A}\ }\textbf {\bibinfo {volume} {89}},\
  \bibinfo {pages} {032307} (\bibinfo {year} {2014})}\BibitemShut {NoStop}%
\bibitem [{\citenamefont {Norris}\ \emph {et~al.}(2012)\citenamefont {Norris},
  \citenamefont {Trail}, \citenamefont {Jessen},\ and\ \citenamefont
  {Deutsch}}]{norris_enhanced_2012-1}%
  \BibitemOpen
  \bibfield  {author} {\bibinfo {author} {\bibfnamefont {L.~M.}\ \bibnamefont
  {Norris}}, \bibinfo {author} {\bibfnamefont {C.~M.}\ \bibnamefont {Trail}},
  \bibinfo {author} {\bibfnamefont {P.~S.}\ \bibnamefont {Jessen}},\ and\
  \bibinfo {author} {\bibfnamefont {I.~H.}\ \bibnamefont {Deutsch}},\
  }\bibfield  {title} {\bibinfo {title} {Enhanced {{Squeezing}} of a
  {{Collective Spin}} via {{Control}} of {{Its Qudit Subsystems}}},\
  }\href@noop {} {\bibfield  {journal} {\bibinfo  {journal} {Phys. Rev. Lett.}\
  }\textbf {\bibinfo {volume} {109}},\ \bibinfo {pages} {173603} (\bibinfo
  {year} {2012})}\BibitemShut {NoStop}%
\bibitem [{\citenamefont {Dzuba}\ \emph {et~al.}(2011)\citenamefont {Dzuba},
  \citenamefont {Flambaum},\ and\ \citenamefont {Lev}}]{dzuba_dynamic_2011}%
  \BibitemOpen
  \bibfield  {author} {\bibinfo {author} {\bibfnamefont {V.~A.}\ \bibnamefont
  {Dzuba}}, \bibinfo {author} {\bibfnamefont {V.~V.}\ \bibnamefont
  {Flambaum}},\ and\ \bibinfo {author} {\bibfnamefont {B.~L.}\ \bibnamefont
  {Lev}},\ }\bibfield  {title} {\bibinfo {title} {Dynamic polarizabilities and
  magic wavelengths for dysprosium},\ }\href@noop {} {\bibfield  {journal}
  {\bibinfo  {journal} {Phys. Rev. A}\ }\textbf {\bibinfo {volume} {83}},\
  \bibinfo {pages} {032502} (\bibinfo {year} {2011})}\BibitemShut {NoStop}%
\end{thebibliography}

%

\end{document}